%Paper: hep-th/9308005
%From: candelas@nxth04.cern.ch
%Date: Mon, 2 Aug 93 18:50:10 +0200
%Date (revised): Tue, 3 Aug 93 13:38:38 +0200
%Date (revised): Tue, 3 Aug 93 16:04:23 +0200

%%%%%%%%%%%%%%%%%%%%%%%%%%%
%%                                                                       %%
%%  Periods.tex   by   Per Berglund      berglund@sbitp.ucsb.edu         %%
%%                     Philip Candelas   candelas@cernvm.cern.ch         %%
%%  (harvmac.tex)      Xenia de la Ossa  delaossa@iph.unine.ch           %%
%%       ++            Anamaria Font     afont@conicit.ve                %%
%%    enclosed         Tristan Hubsch    hubsch@reliant.cldc.howard.edu  %%
%%     macros          Dubravka Jancic   jancic@utaphy.bitnet            %%
%%                     Fernando Quevedo  quevedo@iph.unine.ch            %%
%%                                                                       %%
%%%%%%%%%%%%%%%%%%%%%%%%%%%
 %
\expandafter \def \csname CHAPLABELpre\endcsname {1}
\expandafter \def \csname EQLABELXXX\endcsname {1.1?}
\expandafter \def \csname EQLABELXXX\endcsname {1.2?}
\expandafter \def \csname EQLABELXXX\endcsname {1.3?}
\expandafter \def \csname EQLABELXXX\endcsname {1.4?}
\expandafter \def \csname EQLABELBzero\endcsname {1.5?}
\expandafter \def \csname EQLABELXXX\endcsname {1.6?}
\expandafter \def \csname EQLABELvpzero\endcsname {1.7?}
\expandafter \def \csname EQLABELXXX\endcsname {1.8?}
\expandafter \def \csname EQLABELXXX\endcsname {1.9?}
\expandafter \def \csname EQLABELXXX\endcsname {1.10?}
\expandafter \def \csname EQLABELXXX\endcsname {1.11?}
\expandafter \def \csname EQLABELXXX\endcsname {1.12?}
\expandafter \def \csname EQLABELXXX\endcsname {1.13?}
\expandafter \def \csname EQLABELXXX\endcsname {1.14?}
\expandafter \def \csname EQLABELXXX\endcsname {1.15?}
\expandafter \def \csname CHAPLABELOneHyper\endcsname {2}
\expandafter \def \csname EQLABELeCYcond\endcsname {2.1?}
\expandafter \def \csname EQLABELeWCP\endcsname {2.2?}
\expandafter \def \csname EQLABELeDefP\endcsname {2.3?}
\expandafter \def \csname EQLABELeOm\endcsname {2.4?}
\expandafter \def \csname EQLABELeXXX\endcsname {2.5?}
\expandafter \def \csname EQLABELeHVol\endcsname {2.6?}
\expandafter \def \csname EQLABELePiO\endcsname {2.7?}
\expandafter \def \csname EQLABELePOLY\endcsname {2.8?}
\expandafter \def \csname EQLABELeConsA\endcsname {2.9?}
\expandafter \def \csname EQLABELeTheM\endcsname {2.10?}
\expandafter \def \csname EQLABELeConsQ\endcsname {2.11?}
\expandafter \def \csname EQLABELeAij\endcsname {2.12?}
\expandafter \def \csname EQLABELXXX\endcsname {2.13?}
\expandafter \def \csname EQLABELeExp\endcsname {2.14?}
\expandafter \def \csname EQLABELePiEx\endcsname {2.15?}
\expandafter \def \csname EQLABELeXXX\endcsname {2.16?}
\expandafter \def \csname EQLABELeMulti\endcsname {2.17?}
\expandafter \def \csname EQLABELeExP\endcsname {2.18?}
\expandafter \def \csname EQLABELeXXX\endcsname {2.19?}
\expandafter \def \csname EQLABELeNCond\endcsname {2.20?}
\expandafter \def \csname EQLABELePiEXa\endcsname {2.21?}
\expandafter \def \csname EQLABELeSum\endcsname {2.22?}
\expandafter \def \csname EQLABELsumvar\endcsname {2.23?}
\expandafter \def \csname EQLABELdhat\endcsname {2.24?}
\expandafter \def \csname EQLABELrelat\endcsname {2.25?}
\expandafter \def \csname EQLABELeSUM\endcsname {2.26?}
\expandafter \def \csname EQLABELeSia\endcsname {2.27?}
\expandafter \def \csname EQLABELPerM\endcsname {2.28?}
\expandafter \def \csname EQLABELeNewPhi\endcsname {2.29?}
\expandafter \def \csname EQLABELsumvarf\endcsname {2.30?}
\expandafter \def \csname EQLABELpert\endcsname {2.31?}
\expandafter \def \csname EQLABELpertu\endcsname {2.32?}
\expandafter \def \csname EQLABELvp0Im\endcsname {2.33?}
\expandafter \def \csname EQLABELXXX\endcsname {2.34?}
\expandafter \def \csname EQLABELPoch\endcsname {2.35?}
\expandafter \def \csname EQLABELeProdG\endcsname {2.36?}
\expandafter \def \csname EQLABELUnFold\endcsname {2.37?}
\expandafter \def \csname EQLABELIbigz\endcsname {2.38?}
\expandafter \def \csname EQLABELIbigZ\endcsname {2.39?}
\expandafter \def \csname EQLABELMelBar\endcsname {2.40?}
\expandafter \def \csname EQLABELXXX\endcsname {2.41?}
\expandafter \def \csname EQLABELXXX\endcsname {2.42?}
\expandafter \def \csname EQLABELXXX\endcsname {2.43?}
\expandafter \def \csname EQLABELStirling\endcsname {2.44?}
\expandafter \def \csname EQLABELinfinIm\endcsname {2.45?}
\expandafter \def \csname EQLABELBigZ\endcsname {2.46?}
\expandafter \def \csname EQLABELsmallz\endcsname {2.47?}
\expandafter \def \csname EQLABELsumRule\endcsname {2.48?}
\expandafter \def \csname EQLABELSumRule\endcsname {2.49?}
\expandafter \def \csname EQLABELSumRuleP\endcsname {2.50?}
\expandafter \def \csname EQLABELsimpleP\endcsname {2.51?}
\expandafter \def \csname EQLABELImQ\endcsname {2.52?}
\expandafter \def \csname EQLABELvp0Q\endcsname {2.53?}
\expandafter \def \csname EQLABELR2\endcsname {2.54?}
\expandafter \def \csname EQLABELAssume\endcsname {2.55?}
\expandafter \def \csname EQLABELWholeHog\endcsname {2.56?}
\expandafter \def \csname EQLABELvp0Z1\endcsname {2.57?}
\expandafter \def \csname EQLABELvp0Z2\endcsname {2.58?}
\expandafter \def \csname EQLABELvp0Z3\endcsname {2.59?}
\expandafter \def \csname EQLABELonem\endcsname {2.60?}
\expandafter \def \csname EQLABELXXX\endcsname {2.61?}
\expandafter \def \csname EQLABELXXX\endcsname {2.62?}
\expandafter \def \csname EQLABELvp0Wr\endcsname {2.63?}
\expandafter \def \csname EQLABELWr\endcsname {2.64?}
\expandafter \def \csname EQLABELMelBarX\endcsname {2.65?}
\expandafter \def \csname EQLABELzeroWr\endcsname {2.66?}
\expandafter \def \csname EQLABELXXX\endcsname {2.67?}
\expandafter \def \csname EQLABELExpoWr\endcsname {2.68?}
\expandafter \def \csname EQLABELnewnewphi\endcsname {2.69?}
\expandafter \def \csname EQLABELSimpleP\endcsname {2.70?}
\expandafter \def \csname EQLABELvp0WrQ\endcsname {2.71?}
\expandafter \def \csname EQLABELvp0WrZ1\endcsname {2.72?}
\expandafter \def \csname EQLABELvp0WrZ2\endcsname {2.73?}
\expandafter \def \csname EQLABELvp0WrZ3\endcsname {2.74?}
\expandafter \def \csname EQLABELModulG\endcsname {2.75?}
\expandafter \def \csname EQLABELothervp\endcsname {2.76?}
\expandafter \def \csname EQLABELrelall\endcsname {2.77?}
\expandafter \def \csname EQLABELnoRel\endcsname {2.78?}
\expandafter \def \csname EQLABELMoreModG\endcsname {2.79?}
\expandafter \def \csname CHAPLABELCICY manifolds\endcsname {3}
\expandafter \def \csname EQLABELP533Eqs\endcsname {3.1?}
\expandafter \def \csname EQLABELXXX\endcsname {3.2?}
\expandafter \def \csname EQLABELXXX\endcsname {3.3?}
\expandafter \def \csname EQLABELXXX\endcsname {3.4?}
\expandafter \def \csname EQLABELXXX\endcsname {3.5?}
\expandafter \def \csname EQLABELvpP533\endcsname {3.6?}
\expandafter \def \csname EQLABELvpFund\endcsname {3.7?}
\expandafter \def \csname TABLABELeqns\endcsname {3.1?}
\expandafter \def \csname EQLABELeGHgF\endcsname {3.8?}
\expandafter \def \csname EQLABELXXX\endcsname {3.9?}
\expandafter \def \csname EQLABELXXX\endcsname {3.10?}
\expandafter \def \csname EQLABELXXX\endcsname {3.11?}
\expandafter \def \csname EQLABELXXX\endcsname {3.12?}
\expandafter \def \csname EQLABELvpij\endcsname {3.13?}
\expandafter \def \csname EQLABELvpijsym\endcsname {3.14?}
\expandafter \def \csname EQLABELXXX\endcsname {3.15?}
\expandafter \def \csname EQLABELXXX\endcsname {3.16?}
\expandafter \def \csname EQLABELXXX\endcsname {3.17?}
\expandafter \def \csname EQLABELXXX\endcsname {3.18?}
\expandafter \def \csname EQLABELXXX\endcsname {3.19?}
\expandafter \def \csname EQLABELXXX\endcsname {3.20?}
\expandafter \def \csname EQLABELXXX\endcsname {3.21?}
\expandafter \def \csname EQLABELXXX\endcsname {3.22?}
\expandafter \def \csname EQLABELXXX\endcsname {3.23?}
\expandafter \def \csname EQLABELXXX\endcsname {3.24?}
\expandafter \def \csname EQLABELXXX\endcsname {3.25?}
\expandafter \def \csname EQLABELXXX\endcsname {3.26?}
\expandafter \def \csname EQLABELXXX\endcsname {3.27?}
\expandafter \def \csname EQLABELXXX\endcsname {3.28?}
\expandafter \def \csname EQLABELXXX\endcsname {3.29?}
\expandafter \def \csname EQLABELXXX\endcsname {3.30?}
\expandafter \def \csname EQLABELXXX\endcsname {3.31?}
\expandafter \def \csname EQLABELXXX\endcsname {3.32?}
\expandafter \def \csname EQLABELXXX\endcsname {3.33?}
\expandafter \def \csname EQLABELXXX\endcsname {3.34?}
\expandafter \def \csname EQLABELOneCol\endcsname {3.35?}
\expandafter \def \csname EQLABELmnfld\endcsname {3.36?}
\expandafter \def \csname EQLABELXXX\endcsname {3.37?}
\expandafter \def \csname EQLABELNewton\endcsname {3.38?}
\expandafter \def \csname EQLABELXXX\endcsname {3.39?}
\expandafter \def \csname EQLABELXXX\endcsname {3.40?}
\expandafter \def \csname EQLABELeBatyP\endcsname {3.41?}
\expandafter \def \csname EQLABELvpBaty\endcsname {3.42?}
\expandafter \def \csname TABLABELcicys\endcsname {3.2?}
\expandafter \def \csname EQLABELXXX\endcsname {3.43?}
\expandafter \def \csname EQLABELXXX\endcsname {3.44?}
\expandafter \def \csname EQLABELConj\endcsname {3.45?}
\expandafter \def \csname EQLABELcicyR\endcsname {3.46?}
\expandafter \def \csname EQLABELcicyRdef\endcsname {3.47?}
\expandafter \def \csname EQLABELXXX\endcsname {3.48?}
\expandafter \def \csname EQLABELcicyRvp\endcsname {3.49?}
\expandafter \def \csname EQLABELbirat\endcsname {3.50?}
\expandafter \def \csname EQLABELciwcy\endcsname {3.51?}
\expandafter \def \csname CHAPLABELtwoparams\endcsname {4}
\expandafter \def \csname EQLABELtwoexamp\endcsname {4.1?}
\expandafter \def \csname EQLABELpfer\endcsname {4.2?}
\expandafter \def \csname EQLABELemp\endcsname {4.3?}
\expandafter \def \csname EQLABELpifer\endcsname {4.4?}
\expandafter \def \csname EQLABELpiU\endcsname {4.5?}
\expandafter \def \csname EQLABELUfun\endcsname {4.6?}
\expandafter \def \csname EQLABELUhyp\endcsname {4.7?}
\expandafter \def \csname EQLABELUzero\endcsname {4.8?}
\expandafter \def \csname EQLABELpismall\endcsname {4.9?}
\expandafter \def \csname EQLABELXXX\endcsname {4.10?}
\expandafter \def \csname EQLABELXXX\endcsname {4.11?}
\expandafter \def \csname EQLABELeP\endcsname {4.12?}
\expandafter \def \csname EQLABELePdef\endcsname {4.13?}
\expandafter \def \csname EQLABELeOONE\endcsname {4.14?}
\expandafter \def \csname EQLABELQSym\endcsname {4.15?}
\expandafter \def \csname EQLABELGSymA\endcsname {4.16?}
\expandafter \def \csname EQLABELGSymB\endcsname {4.17?}
\expandafter \def \csname EQLABELFeMatP\endcsname {4.18?}
\expandafter \def \csname EQLABELeEYPiO\endcsname {4.19?}
\expandafter \def \csname EQLABELXXX\endcsname {4.20?}
\expandafter \def \csname EQLABELpi0U\endcsname {4.21?}
\expandafter \def \csname EQLABELUl\endcsname {4.22?}
\expandafter \def \csname EQLABELpi0sf\endcsname {4.23?}
\expandafter \def \csname EQLABELXXX\endcsname {4.24?}
\expandafter \def \csname EQLABELXXX\endcsname {4.25?}
\expandafter \def \csname EQLABELeTrP\endcsname {4.26?}
\expandafter \def \csname EQLABELeGofW\endcsname {4.27?}
\expandafter \def \csname EQLABELXXX\endcsname {4.28?}
\expandafter \def \csname EQLABELXXX\endcsname {4.29?}
\expandafter \def \csname EQLABELXXX\endcsname {4.30?}
\expandafter \def \csname CHAPLABELMore\endcsname {5}
\expandafter \def \csname EQLABELePo\endcsname {5.1?}
\expandafter \def \csname EQLABELePoF\endcsname {5.2?}
\expandafter \def \csname EQLABELvpMore\endcsname {5.3?}
\expandafter \def \csname EQLABELNMparams\endcsname {5.4?}
\expandafter \def \csname EQLABELnewphiNM\endcsname {5.5?}
\expandafter \def \csname EQLABELeFoP\endcsname {5.6?}
\expandafter \def \csname EQLABELMelBarNM\endcsname {5.7?}
\expandafter \def \csname EQLABELNMvp0Wr\endcsname {5.8?}
\expandafter \def \csname EQLABELWrNM\endcsname {5.9?}
\expandafter \def \csname EQLABELWrNMghf\endcsname {5.10?}
\expandafter \def \csname EQLABELsmallNMvp\endcsname {5.11?}
\expandafter \def \csname EQLABELsNMvp\endcsname {5.12?}
\expandafter \def \csname EQLABELA\endcsname {5.13?}
\expandafter \def \csname EQLABELB\endcsname {5.14?}
\expandafter \def \csname EQLABELC\endcsname {5.15?}
\expandafter \def \csname EQLABELD\endcsname {5.16?}
\expandafter \def \csname EQLABEL3103\endcsname {5.17?}
\expandafter \def \csname EQLABELRadical\endcsname {5.18?}
\expandafter \def \csname EQLABELCICYR\endcsname {5.19?}
\expandafter \def \csname EQLABELCICYRdef\endcsname {5.20?}
\expandafter \def \csname EQLABELXXX\endcsname {5.21?}
\expandafter \def \csname EQLABELCICYRvp\endcsname {5.22?}
\expandafter \def \csname EQLABELRpi\endcsname {5.23?}
\expandafter \def \csname EQLABELstd\endcsname {5.24?}
\expandafter \def \csname EQLABELREqs\endcsname {5.25?}
\expandafter \def \csname EQLABELXXX\endcsname {5.26?}
\expandafter \def \csname EQLABELXXX\endcsname {5.27?}
\expandafter \def \csname EQLABELInvG\endcsname {5.28?}
\expandafter \def \csname EQLABELRvp\endcsname {5.29?}
\expandafter \def \csname EQLABEL4102\endcsname {5.30?}
\expandafter \def \csname EQLABELthe30\endcsname {5.31?}
\expandafter \def \csname EQLABEL4102def\endcsname {5.32?}
\expandafter \def \csname EQLABELXXX\endcsname {5.33?}
\expandafter \def \csname EQLABELeCYGcond\endcsname {5.34?}
\expandafter \def \csname EQLABELXXX\endcsname {5.35?}
\expandafter \def \csname EQLABELXXX\endcsname {5.36?}
\expandafter \def \csname EQLABELXXX\endcsname {5.37?}
\expandafter \def \csname EQLABELXXX\endcsname {5.38?}
\expandafter \def \csname EQLABELXXX\endcsname {5.39?}
\expandafter \def \csname EQLABELXXX\endcsname {5.40?}
\expandafter \def \csname EQLABELInt0\endcsname {5.41?}
\expandafter \def \csname EQLABELvp0\endcsname {5.42?}
\expandafter \def \csname EQLABELIntj\endcsname {5.43?}
\expandafter \def \csname EQLABELvpj\endcsname {5.44?}
\expandafter \def \csname CHAPLABELconcl\endcsname {6}
%%%%%%%%%%%%%%%%%%%%%%%%%%%
%%%[ TeXnicalities ]%%%%%%%
%%%%%%%%%%%%%%%%%%%%%%%%%%%
\input harvmac
 %
%%%%%%%%%%%%%%%%%%%%%%%%%%%
%%%[ More TeXnicalities ]%%%%%%%%%%%%%%%%%%%%%%
%%%%%%%%%%%%%%%%%%%%%%%%%%%
%\input zip
%%%%%%%%%%%%%%%%%%%%%%%%
%                                                                      %
%       "zip.tex", a set of macros to be used with "harvmac.tex"       %
%            Latest change: 27. VII '92.   (Tristan Hubsch)            %
%                                                                      %
%%%%%%%%%%%%%%%%%%%%%%%%
 %
\catcode`@=11
\def\rlx{\relax\leavevmode}                  % Guess what this is for...
 %
 %
%%%%%%%%%%%%%%%%%%%%%%%%
%%%****FONTS****FONTS****FONTS****FONTS****FONTS****FONTS****FONTS****%%
 % That is, where fonts may not be available...
 %
 % Poor man's boldface; use in lieu of proper boldface
\def\BM#1{\relax\leavevmode\setbox0=\hbox{$#1$}
           \kern-.025em\copy0\kern-\wd0
            \kern.05em\copy0\kern-\wd0
             \kern-.025em\raise.0433em\copy0\kern-\wd0
              \raise.0144em\box0 }
 %
 % Some basic black-board bold (capital) letters
 % ...should work rather well in sub- and super-scripts also...
 %
\def\inbar{\vrule height1.5ex width.4pt depth0pt}
\def\sinbar{\vrule height1ex width.35pt depth0pt}
\def\ssinbar{\vrule height.7ex width.3pt depth0pt}
\font\cmss=cmss10
\font\cmsss=cmss10 at 7pt
\def\ZZ{\rlx\leavevmode
             \ifmmode\mathchoice
                    {\hbox{\cmss Z\kern-.4em Z}}
                    {\hbox{\cmss Z\kern-.4em Z}}
                    {\lower.9pt\hbox{\cmsss Z\kern-.36em Z}}
                    {\lower1.2pt\hbox{\cmsss Z\kern-.36em Z}}
               \else{\cmss Z\kern-.4em Z}\fi}
\def\Ik{\rlx{\rm I\kern-.18em k}}  % Yes, I know. This ain't capital.
\def\IC{\rlx\leavevmode
             \ifmmode\mathchoice
                    {\hbox{\kern.33em\inbar\kern-.3em{\rm C}}}
                    {\hbox{\kern.33em\inbar\kern-.3em{\rm C}}}
                    {\hbox{\kern.28em\sinbar\kern-.25em{\sevenrm C}}}
                    {\hbox{\kern.25em\ssinbar\kern-.22em{\fiverm C}}}
             \else{\hbox{\kern.3em\inbar\kern-.3em{\rm C}}}\fi}
\def\IP{\rlx{\rm I\kern-.18em P}}
\def\IR{\rlx{\rm I\kern-.18em R}}
\def\Ione{\rlx{\rm 1\kern-2.7pt l}}
 %
%%%%%%%%%%%%%%%%%%%%%%%%
%%%****SHAPE****SHAPE****SHAPE****SHAPE****SHAPE****SHAPE****SHAPE****%%
 %
 % Get in shape, Man, (never mind the content)!
\def\boxit#1{\vbox{\hrule\hbox{\vrule\kern3pt
     \vbox{\kern3pt#1\kern3pt}\kern3pt\vrule}\hrule}}

\def\intem#1{\par\leavevmode%
              \llap{\hbox to\parindent{\hss{#1}\hfill~}}\ignorespaces}
 %
 % Indents #1 lines by width of #2 and puts #2 in the "niche".

 % Similar to "niche" :

 %
 % Math...
 % My version of \eqalign, \eqalignno ...
\newskip\humongous \humongous=0pt plus 1000pt minus 1000pt   % isn't it?
\def\caja{\mathsurround=0pt}
\newif\ifdtup
 %
 % display pattern: [         a &= b        \cr]
\def\eqalign#1{\,\vcenter{\openup2\jot \caja
     \ialign{\strut \hfil$\displaystyle{##}$&$
      \displaystyle{{}##}$\hfil\crcr#1\crcr}}\,}
 %
 % display pattern: [   a &= b  &  c &= d   \cr]
\def\twoeqsalign#1{\,\vcenter{\openup2\jot \caja
     \ialign{\strut \hfil$\displaystyle{##}$&$
      \displaystyle{{}##}$\hfil&\hfill$\displaystyle{##}$&$
       \displaystyle{{}##}$\hfil\crcr#1\crcr}}\,}
 %
 % display to full hsize, numbered at far right
\def\panorama{\global\dtuptrue \openup2\jot \caja
     \everycr{\noalign{\ifdtup \global\dtupfalse
      \vskip-\lineskiplimit \vskip\normallineskiplimit
      \else \penalty\interdisplaylinepenalty \fi}}}
 %
 % display pattern: [         a &= b        &(*) \cr]
\def\eqalignno#1{\panorama \tabskip=\humongous
     \halign to\displaywidth{\hfil$\displaystyle{##}$
      \tabskip=0pt&$\displaystyle{{}##}$\hfil
       \tabskip=\humongous&\llap{$##$}\tabskip=0pt\crcr#1\crcr}}
 %
 % display pattern: [      a &= b &= c      &(*) \cr]

 %
 % display pattern: [   a &= b  &  c &= d   &(*) \cr]

 %
 % display pattern: [    a &= b &= c &= d   &(*) \cr]

 %
 % For extra v-space between rows of a matrix or eqn-alignment,
 % use "\noalign{\vskip2mm}". In the above equation alignments,
 % "\openup2mm" does the same, but has no effect in "\matrix".
 %
%%%%%%%%%%%%%%%%%%%%%%%%
%%%****SHORT****SHORT****SHORT****SHORT****SHORT****SHORT****SHORT****%%
 %
 % Redefinitions of TeX's commands :
 %
          % Polish l-slash, L-slash
        % Scandinavian o-slash, O-slash
          % P-mirror, double-S
        % tie-after, cedilla
\let\ii=\i          % include mathmode !!!
          % under-bar, under-dot
\def\,{\hskip1.5pt}           % why only in math-mode?
 %
 % Some abbreviations that save typing :
\let\a=\alpha
\let\b=\beta
\let\c=\chi
\let\d=\delta       \let\vd=\partial             \let\D=\Delta
\let\e=\epsilon     
\let\f=\phi         \let\vf=\varphi              
\let\g=\gamma                                    \let\G=\Gamma
\let\h=\eta
\let\i=\iota
\let\j=\psi                                      \let\J=\Psi
\let\k=\kappa
\let\l=\lambda                                   
\let\m=\mu
\let\n=\nu
\let\p=\pi          \let\vp=\varpi               \let\P=\Pi
\let\q=\theta                   
\let\r=\rho         
\let\s=\sigma                   \let\S=\Sigma

\let\w=\omega                                    \let\W=\Omega
\let\x=\xi                                       
                                  
\let\z=\zeta
 %
 % Additional math symbols
 %
\def\Box{\sqcap\llap{$\sqcup$}}
\def\lapp{\lower.4ex\hbox{\rlap{$\sim$}} \raise.4ex\hbox{$<$}}
\def\gapp{\lower.4ex\hbox{\rlap{$\sim$}} \raise.4ex\hbox{$>$}}
\def\con{\ifmmode\raise.1ex\hbox{\bf*}
          \else\raise.1ex\hbox{\bf*}\fi}
\def\bo{{\raise.15ex\hbox{\large$\Box\kern-.39em$}}}
  %  a very fat nothing

\def\Ree{\mathop{\Re e}}

\def\dual{\relax\leavevmode\lower.9ex\hbox{\titlerms*}}
\def\define{\buildrel\rm def\over =}

\let\8=\otimes
 %
 %
%%%%%%%%%%%%%%%%%%%%%%%%
%%****MACROS***MACROS***MACROS***MACROS***MACROS***MACROS***MACROS****%%
 %
 % Math macros
 %

\let\2=\underline

\let\Tw=\widetilde
 %
 % Let's take arguments...
%  Use:  "A\like{B}".
\def\dt#1{{\buildrel{\smash{\lower1pt\hbox{.}}}\over{#1}}}
\def\pd#1#2{{\partial#1\over\partial#2}}

\font\eightrm=cmr8
\def\6(#1){\relax\leavevmode\hbox{\eightrm(}#1\hbox{\eightrm)}}
\def\0#1{\relax\ifmmode\mathaccent"7017{#1}     % a little circle atop,
                \else\accent23#1\relax\fi}      % as a halo of a saint
\def\7#1#2{{\mathop{\null#2}\limits^{#1}}}      % puts #1 atop #2
\def\5#1#2{{\mathop{\null#2}\limits_{#1}}}      % puts #1 beneath #2
 %
 % Will grow vertically with size of argument

 %
 % Vertical arrows with labels

 %
 % For vert. arrows to grow, say "\bigg\down\crlap{...}", using the
 % side-script: a label for vertical delimiters

 %

 %
 % Horizontal arrows that can grow
\newbox\t@b@x
\def\rightarrowfill{$\m@th \mathord- \mkern-6mu
     \cleaders\hbox{$\mkern-2mu \mathord- \mkern-2mu$}\hfill
      \mkern-6mu \mathord\rightarrow$}
\def\tooo#1{\setbox\t@b@x=\hbox{$\scriptstyle#1$}%
             \mathrel{\mathop{\hbox to\wd\t@b@x{\rightarrowfill}}%
              \limits^{#1}}\,}
\def\leftarrowfill{$\m@th \mathord\leftarrow \mkern-6mu
     \cleaders\hbox{$\mkern-2mu \mathord- \mkern-2mu$}\hfill
      \mkern-6mu \mathord-$}
\def\froo#1{\setbox\t@b@x=\hbox{$\scriptstyle#1$}%
             \mathrel{\mathop{\hbox to\wd\t@b@x{\leftarrowfill}}%
              \limits^{#1}}\,}
 %
 % fractions
\def\frac#1#2{{#1\over#2}}
\def\frc#1#2{\relax\ifmmode{\textstyle{#1\over#2}} % A small fraction,
                    \else$#1\over#2$\fi}           % good in text.
                            % Like {1\over{#1}}
 %
 % The basic Theorem-like macro, uses equation numbers
 % Use:  \Claim\cLABEL{Theorem}{This is a theorem.}
\def\Claim#1#2#3{\bigskip\begingroup%
                  \xdef #1{\secsym\the\meqno}%
                   \writedef{#1\leftbracket#1}%
                    \global\advance\meqno by1\wrlabeL#1%
                     \noindent{\bf#2}\,#1{}\,:~\sl#3\vskip1mm\endgroup}

\def\QED{\rlx\hfill$\Box$\kern-7pt\raise3pt\hbox{$\surd$}\bigskip}
 %
 % Math miscellanea
 %

\def\K#1#2{\relax\def\normalbaselines{\baselineskip12pt\lineskip3pt
                                       \lineskiplimit3pt}
             \left[\matrix{#1}\right.\!\left\|\,\matrix{#2}\right]}
\def\muthstrut{\vphantom1}
\def\mutrix#1{\null\,\vcenter{\normalbaselines\m@th
        \ialign{\hfil$##$\hfil&&~\hfil$##$\hfill\crcr
            \muthstrut\crcr\noalign{\kern-\baselineskip}
            #1\crcr\muthstrut\crcr\noalign{\kern-\baselineskip}}}\,}

 %
 % Young tableaux: use "\Box" for a box and "\Z" for newline
 % The 2-1 stair-tableau:  \YT{\Box\Box}{\Box\Box\Z\Box}
 % Note: the first argument sets the width!
\def\YT#1#2{\vcenter{\hbox{\vbox{\baselineskip=\normalbaselineskip%
             \def\Box{$\sqcap\llap{$\sqcup$}$\kern-1.2pt}%
              \def\Z{\hfil\vskip-8.8pt}%
               \setbox0=\hbox{#1}\hsize\wd0\parindent=0pt#2}\,}}}
\def\EU{\rlx\ifmmode \c_{{}_E} \else$\c_{{}_E}$\fi}
\def\TM{\rlx\ifmmode {\cal T_M} \else$\cal T_M$\fi}
\def\TW{\rlx\ifmmode {\cal T_W} \else$\cal T_W$\fi}
\def\CM{\rlx\ifmmode {\cal T\rlap{\bf*}\!\!_M}
             \else$\cal T\rlap{\bf*}\!\!_M$\fi}
\def\hm#1#2{\rlx\ifmmode H^{#1}({\cal M},{#2})
                 \else$H^{#1}({\cal M},{#2})$\fi}
\def\CP#1{\rlx\ifmmode\IP^{#1}\else\IP$^{#1}$\fi}
\def\cP#1{\rlx\ifmmode\IC{\rm P}^{#1}\else$\IC{\rm P}^{#1}$\fi}

\def\sll#1{\rlx\rlap{\,\raise1pt\hbox{/}}{#1}}
\def\Sll#1{\rlx\rlap{\,\kern.6pt\raise1pt\hbox{/}}{#1}\kern-.6pt}

\let\SSS=\scriptstyle
\let\ttt=\textstyle
\let\ddd=\displaystyle
 %
 % Text miscellanea
 %
\def\ie{\hbox{\it i.e.}}        % By Knuth, use commas: ..., \ie, ... !

\def\CY{Calabi-\kern-.2em Yau}
\def\LGO{Landau-Ginzburg orbifold}
\def\3{\ifmmode\ldots\else$\ldots$\fi}
\def\Z{\hfil\break\rlx\hbox{}\quad}
\def\3{\ifmmode\ldots\else$\ldots$\fi}
\def\?{d\kern-.3em\raise.64ex\hbox{-}}           % d-dash
\def\9{\raise.43ex\hbox{-}\kern-.37em D}         % D-Dash

 %
 % References
 %

 %

 %

\def\MPL#1{{\it Mod.\,Phys.\,Lett.\,}{\bf#1\,}}

\def\CQG#1{{\it Class.\,Quant.\,Grav.\,}{\bf#1\,}}

 %
 %
%%%%%%%%%%%%%%%%%%%%%%%%
%%============>>>>            SAVE  TIMBER            <<<<============%%
 %
\baselineskip=13.0861pt plus2pt minus1pt
\parskip=\medskipamount
\let\ft=\foot
\noblackbox
\def\SaveTimber{\abovedisplayskip=1.5ex plus.3ex minus.5ex
                \belowdisplayskip=1.5ex plus.3ex minus.5ex
                \abovedisplayshortskip=.2ex plus.2ex minus.4ex
                \belowdisplayshortskip=1.5ex plus.2ex minus.4ex
                \baselineskip=12pt plus1pt minus.5pt
 \parskip=\smallskipamount
 \def\ft##1{\unskip\,\begingroup\footskip9pt plus1pt minus1pt\setbox%
             \strutbox=\hbox{\vrule height6pt depth4.5pt width0pt}%
              \global\advance\ftno by1\footnote{$^{\the\ftno)}$}{##1}%
               \endgroup}
 \def\listrefs{\footatend\vfill\immediate\closeout\rfile%
                \writestoppt\baselineskip=10pt%
                 \centerline{{\bf References}}%
                  \bigskip{\frenchspacing\parindent=20pt\escapechar=` %
                   \rightskip=0pt plus4em\spaceskip=.3333em%
                    \input refs.tmp\vfill\eject}\nonfrenchspacing}}
 %
 % For European standard
\def\Afour{\ifx\answ\bigans
            \hsize=16.5truecm\vsize=24.7truecm
             \else
              \hsize=24.7truecm\vsize=16.5truecm
               \fi}
\catcode`@=12
%%%%%%%%%%%%%%%%%%%%%%%%%
%                                                                       %
%                          End of  "zip.tex".                           %
%                        May the article begin!                         %
%                                                                       %
%%%%%%%%%%%%%%%%%%%%%%%%%
%\input xenia.mac
%%%
%Fonts%%%%%%%%%%%%%%%%%%%%%%%%%
%%%

\font\eightrm=cmr8 at 8pt

\font\seventeenrm=cmr17 at 17pt
\font\twentyonerm=cmr17 at 21pt

\font\ss=cmss10

\font\csc=cmcsc10

\font\twelvecal=cmsy10 at 12pt

\font\twelvemath=cmmi12

\font\seventeenbold=cmbx7 at 17pt

\font\fively=lasy5
\font\sevenly=lasy7
\font\tenly=lasy10

\textfont10=\tenly
\scriptfont10=\sevenly
\scriptscriptfont10=\fively
%%%
%Formatting%%%%%%%%%%%%%%%%%%%%
%%%
\magnification=1200
\parskip=10pt
\parindent=20pt
\def\today{\ifcase\month\or January\or February\or March\or April\or May\or
June
       \or July\or August\or September\or October\or November\or December\fi
       \space\number\day, \number\year}

\def\title#1{\footline={\ifnum\pageno<2\hfil
       \else\hss\tenrm\folio\hss\fi}\vskip1truein\centerline{{#1}
       \footnote{\raise1ex\hbox{*}}{\eightrm Supported in part
       by the Robert A. Welch Foundation and N.S.F. Grants
       PHY-880637 and\break PHY-8605978.}}}

\def\Z{\hfill\break}
\def\newpage{\vfill\eject}
\def\abstract#1{\centerline{\bf ABSTRACT}\vskip.2truein{\narrower\noindent#1
       \smallskip}}

\def\runninghead#1#2{\voffset=2\baselineskip\nopagenumbers
       \headline={\ifodd\pageno\rightheadline\else \leftheadline\fi}
       \def\rightheadline{{\sl#1}\hfill{\rm\folio}}
       \def\leftheadline{{\rm\folio}\hfill{\sl#2}}}

\newcount\footnoteno
\def\Footnote#1{\advance\footnoteno by 1
                \let\SF=\empty
                \ifhmode\edef\SF{\spacefactor=\the\spacefactor}\/\fi
                $^{\the\footnoteno}$\ignorespaces
                \SF\vfootnote{$^{\the\footnoteno}$}{#1}}

\def\place#1#2#3{\vbox to0pt{\kern-\parskip\kern-7pt
                             \kern-#2truein\hbox{\kern#1truein #3}
                             \vss}\nointerlineskip}
\def\figurecaption#1#2{\kern.75truein\vbox{\hsize=5truein\noindent{\bf Figure
    \figlabel{#1}:} #2}}
\def\tablecaption#1#2{\kern.75truein\lower12truept\hbox{\vbox{\hsize=5truein
    \noindent{\bf Table\hskip5truept\tablabel{#1}:} #2}}}
\def\boxed#1{\lower3pt\hbox{
                       \vbox{\hrule\hbox{\vrule
                        \vbox{\kern2pt\hbox{\kern3pt#1\kern3pt}\kern3pt}\vrule}
                         \hrule}}}

%%%
%Greek characters%%%%%%%%%%%%%%
%%%
\def\a{\alpha}
\def\b{\beta}
\def\g{\gamma}\def\G{\Gamma}
\def\d{\delta}\def\D{\Delta}
\def\e{\epsilon}
\def\z{\zeta}

\def\k{\kappa}
\def\l{\lambda}
\def\m{\mu}
\def\n{\nu}
\def\x{\xi}

\def\p{\pi}\def\P{\Pi}\def\vp{\varpi}
\def\r{\rho}
\def\s{\sigma}\def\S{\Sigma}

\def\O{\Omega}

%%%
%Calligraphic capitals%%%%%%%%%
%%%
\def\ca#1{\relax\ifmmode {{\cal #1}}\else $\cal #1$\fi}

\def\calb{{\cal B}}

\def\calm{{\cal M}}

%%%
% Poor man's Blackboard Bold%%%
%%%
\def\inbar{\vrule height1.5ex width.4pt depth0pt}
\def\IB{\relax{\rm I\kern-.18em B}}
\def\IC{\relax\hbox{\kern.25em$\inbar\kern-.3em{\rm C}$}}
\def\ID{\relax{\rm I\kern-.18em D}}
\def\IE{\relax{\rm I\kern-.18em E}}
\def\IF{\relax{\rm I\kern-.18em F}}
\def\IG{\relax\hbox{\kern.25em$\inbar\kern-.3em{\rm G}$}}
\def\IH{\relax{\rm I\kern-.18em H}}
\def\II{\relax{\rm I\kern-.18em I}}
\def\IK{\relax{\rm I\kern-.18em K}}
\def\IL{\relax{\rm I\kern-.18em L}}
\def\IM{\relax{\rm I\kern-.18em M}}
\def\IN{\relax{\rm I\kern-.18em N}}
\def\IO{\relax\hbox{\kern.25em$\inbar\kern-.3em{\rm O}$}}
\def\IP{\relax{\rm I\kern-.18em P}}
\def\IQ{\relax\hbox{\kern.25em$\inbar\kern-.3em{\rm Q}$}}
\def\IR{\relax{\rm I\kern-.18em R}}
\def\IZ{\relax\ifmmode\hbox{\ss Z\kern-.4em Z}\else{\ss Z\kern-.4em Z}\fi}
\def\IGa{\relax{\rm I}\kern-.18em\Gamma}
\def\IPi{\relax{\rm I}\kern-.18em\Pi}
\def\ITh{\relax\hbox{\kern.25em$\inbar\kern-.3em\Theta$}}
\def\IOm{\relax\thinspace\inbar\kern1.95pt\inbar\kern-5.525pt\Omega}

%Papers, Lecture Notes on Complex Manifolds etc.

\def\ie{{\it i.e.\ \/}}

\def\noblackboxes{\overfullrule=0pt}
\def\define{\buildrel\rm def\over =}

\def\cy{Calabi--Yau}
\def\cym{Calabi--Yau manifold}
\def\cys{Calabi--Yau manifolds}
\def\cicy{CICY\ }

\def\cicys{CICY manifolds}
\def\K{K\"ahler}

\def\H#1#2{\relax\ifmmode {H^{#1#2}}\else $H^{#1 #2}$\fi}
\def\M{\relax\ifmmode{\calm}\else $\calm$\fi}

\def\Bigcheck{\lower3.8pt\hbox{\smash{\hbox{{\twentyonerm \v{}}}}}}
\def\bigboldcheck{\smash{\hbox{{\seventeenbold\v{}}}}}

\def\Bighat{\lower3.8pt\hbox{\smash{\hbox{{\twentyonerm \^{}}}}}}

\def\Msharp{\relax\ifmmode{\calm^\sharp}\else $\smash{\calm^\sharp}$\fi}
\def\Mflat{\relax\ifmmode{\calm^\flat}\else $\smash{\calm^\flat}$\fi}
\def\preMcheck{\kern2pt\hbox{\Bigcheck\kern-12pt{$\cal M$}}}
\def\Mcheck{\relax\ifmmode\preMcheck\else $\preMcheck$\fi}
\def\preMhat{\kern2pt\hbox{\Bighat\kern-12pt{$\cal M$}}}
\def\Mhat{\relax\ifmmode\preMhat\else $\preMhat$\fi}

\def\Bsharp{\relax\ifmmode{\calb^\sharp}\else $\calb^\sharp$\fi}
\def\Bflat{\relax\ifmmode{\calb^\flat}\else $\calb^\flat$ \fi}
\def\preBcheck{\hbox{\Bigcheck\kern-9pt{$\cal B$}}}
\def\Bcheck{\relax\ifmmode\preBcheck\else $\preBcheck$\fi}
\def\preBhat{\hbox{\Bighat\kern-9pt{$\cal B$}}}
\def\Bhat{\relax\ifmmode\preBhat\else $\preBhat$\fi}

\def\figBcheck{\kern3pt\hbox{\raise1pt\hbox{\bigboldcheck}\kern-11pt
    {\twelvecal B}}}
\def\figBsharp{{\twelvecal B}\raise5pt\hbox{$\twelvemath\sharp$}}
\def\figBflat{{\twelvecal B}\raise5pt\hbox{$\twelvemath\flat$}}

\def\gcheck{\hbox{\lower2.5pt\hbox{\Bigcheck}\kern-8pt$\g$}}
\def\lhat{\hbox{\raise.5pt\hbox{\Bighat}\kern-8pt$\l$}}

\def\Fcheck{\kern2pt\hbox{\raise1pt\hbox{\Bigcheck}\kern-10pt{$\cal F$}}}
\def\Fhat{\kern2pt\hbox{\raise1pt\hbox{\Bighat}\kern-10pt{$\cal F$}}}

\def\cp#1{\relax\ifmmode {\IP\kern-2pt{}_{#1}}\else $\IP\kern-2pt{}_{#1}$\fi}
\def\h#1#2{\relax\ifmmode {b_{#1#2}}\else $b_{#1#2}$\fi}
\def\Z{\hfill\break}
\def\imag{\Im m}
\def\half{{1\over 2}}

\def\frac#1#2{{#1\over #2}}

\def\pd#1#2{{\partial #1\over\partial #2}}

\def\cone{\relax\thinspace\hbox{$<\kern-.8em{)}$}}
\mathchardef\mho"0A30

\def\asymp{\sim}
\def\-{\hphantom{-}}

%References

\def\npb#1{Nucl.\ Phys.\ {\bf B#1}}

\def\cmp#1{Commun. Math. Phys. {\bf #1}}
\def\plb#1{Phys. Lett. {\bf #1B}}

\def\preprint#1{University of #1 Report}

% Pictures

\def\picture #1 by #2 (#3){\vbox to #2{\hrule width #1 height 0pt depth 0pt
                                       \vfill\special{picture #3}}}
\def\scaledpicture #1 by #2 (#3 scaled #4){{\dimen0=#1 \dimen1=#2
           \divide\dimen0 by 1000 \multiply\dimen0 by #4
            \divide\dimen1 by 1000 \multiply\dimen1 by #4
            \picture \dimen0 by \dimen1 (#3 scaled #4)}}
\def\illustration #1 by #2 (#3){\vbox to #2%
{\hrule width #1 height 0pt depth0pt\vfill\special{illustration #3}}}

\def\scaledillustration #1 by #2 (#3 scaled #4){{\dimen0=#1 \dimen1=#2
           \divide\dimen0 by 1000 \multiply\dimen0 by #4
            \divide\dimen1 by 1000 \multiply\dimen1 by #4
            \illustration \dimen0 by \dimen1 (#3 scaled #4)}}

%Letters, Letters of Recommendation, Referee's Reports, Itineraries, etc.

\def\delaOssa{\nobreak\vskip1truein\hbox to\hsize
       {\hskip 4truein Xenia de la Ossa\hfill}}

\def\hoy{\number\day\space de \ifcase\month\or enero\or febrero\or marzo\or
       abril\or mayo\or junio\or julio\or agosto\or septiembre\or octubre\or
       noviembre\or diciembre\fi\space de \number\year}

%%%
%Equation macros%%%%%%%%%%%%%%%
%%%

\newif\ifproofmode
\proofmodefalse

\newif\ifforwardreference
\forwardreferencefalse

\newif\ifchapternumbers
\chapternumbersfalse

\newif\ifcontinuousnumbering
\continuousnumberingfalse

\newif\iffigurechapternumbers
\figurechapternumbersfalse

\newif\ifcontinuousfigurenumbering
\continuousfigurenumberingfalse

\newif\iftablechapternumbers
\tablechapternumbersfalse

\newif\ifcontinuoustablenumbering
\continuoustablenumberingfalse

\font\eqsixrm=cmr6

\def\marginstyle{\eqsixrm}

\newtoks\chapletter
\newcount\chapno
\newcount\eqlabelno
\newcount\figureno
\newcount\tableno

\chapno=0
\eqlabelno=0
\figureno=0
\tableno=0

\def\chapfolio{\ifnum\chapno>0 \the\chapno\else\the\chapletter\fi}

\def\bumpchapno{\ifnum\chapno>-1 \global\advance\chapno by 1
\else\global\advance\chapno by -1 \setletter\chapno\fi
\ifcontinuousnumbering\else\global\eqlabelno=0 \fi
\ifcontinuousfigurenumbering\else\global\figureno=0 \fi
\ifcontinuoustablenumbering\else\global\tableno=0 \fi}

\def\setletter#1{\ifcase-#1{}\or{}%
\or\global\chapletter={A}%
\or\global\chapletter={B}%
\or\global\chapletter={C}%
\or\global\chapletter={D}%
\or\global\chapletter={E}%
\or\global\chapletter={F}%
\or\global\chapletter={G}%
\or\global\chapletter={H}%
\or\global\chapletter={I}%
\or\global\chapletter={J}%
\or\global\chapletter={K}%
\or\global\chapletter={L}%
\or\global\chapletter={M}%
\or\global\chapletter={N}%
\or\global\chapletter={O}%
\or\global\chapletter={P}%
\or\global\chapletter={Q}%
\or\global\chapletter={R}%
\or\global\chapletter={S}%
\or\global\chapletter={T}%
\or\global\chapletter={U}%
\or\global\chapletter={V}%
\or\global\chapletter={W}%
\or\global\chapletter={X}%
\or\global\chapletter={Y}%
\or\global\chapletter={Z}\fi}

\def\tempsetletter#1{\ifcase-#1{}\or{}%
\or\global\chapletter={A}%
\or\global\chapletter={B}%
\or\global\chapletter={C}%
\or\global\chapletter={D}%
\or\global\chapletter={E}%
\or\global\chapletter={F}%
\or\global\chapletter={G}%
\or\global\chapletter={H}%
\or\global\chapletter={I}%
\or\global\chapletter={J}%
\or\global\chapletter={K}%
\or\global\chapletter={L}%
\or\global\chapletter={M}%
\or\global\chapletter={N}%
\or\global\chapletter={O}%
\or\global\chapletter={P}%
\or\global\chapletter={Q}%
\or\global\chapletter={R}%
\or\global\chapletter={S}%
\or\global\chapletter={T}%
\or\global\chapletter={U}%
\or\global\chapletter={V}%
\or\global\chapletter={W}%
\or\global\chapletter={X}%
\or\global\chapletter={Y}%
\or\global\chapletter={Z}\fi}

\def\chapshow#1{\ifnum#1>0 \relax#1%
\else{\tempsetletter{\number#1}\chapno=#1\chapfolio}\fi}

\def\chaplabel#1{\bumpchapno\ifproofmode\ifforwardreference
\immediate\write1{\noexpand\expandafter\noexpand\def
\noexpand\csname CHAPLABEL#1\endcsname{\the\chapno}}\fi\fi
\global\expandafter\edef\csname CHAPLABEL#1\endcsname
{\the\chapno}\ifproofmode\llap{\hbox{\marginstyle #1\ }}\fi\chapfolio}

\def\chapref#1{\ifundefined{CHAPLABEL#1}??\ifproofmode\ifforwardreference%
\else\write16{ ***Undefined Chapter Reference #1*** }\fi
\else\write16{ ***Undefined Chapter Reference #1*** }\fi
\else\edef\LABxx{\getlabel{CHAPLABEL#1}}\chapshow\LABxx\fi
\ifproofmode\write2{Chapter #1}\fi}

\def\eqnum{\global\advance\eqlabelno by 1
\eqno(\ifchapternumbers\chapfolio.\fi\the\eqlabelno)}

\def\eqlabel#1{\global\advance\eqlabelno by 1 \ifproofmode\ifforwardreference
\immediate\write1{\noexpand\expandafter\noexpand\def
\noexpand\csname EQLABEL#1\endcsname{\the\chapno.\the\eqlabelno?}}\fi\fi
\global\expandafter\edef\csname EQLABEL#1\endcsname
{\the\chapno.\the\eqlabelno?}\eqno(\ifchapternumbers\chapfolio.\fi
\the\eqlabelno)\ifproofmode\rlap{\hbox{\marginstyle #1}}\fi}

\def\eqalignnum{\global\advance\eqlabelno by 1
&(\ifchapternumbers\chapfolio.\fi\the\eqlabelno)}

\def\eqalignlabel#1{\global\advance\eqlabelno by 1 \ifproofmode
\ifforwardreference\immediate\write1{\noexpand\expandafter\noexpand\def
\noexpand\csname EQLABEL#1\endcsname{\the\chapno.\the\eqlabelno?}}\fi\fi
\global\expandafter\edef\csname EQLABEL#1\endcsname
{\the\chapno.\the\eqlabelno?}&(\ifchapternumbers\chapfolio.\fi
\the\eqlabelno)\ifproofmode\rlap{\hbox{\marginstyle #1}}\fi}

\def\eqref#1{\hbox{(\ifundefined{EQLABEL#1}***)\ifproofmode\ifforwardreference%
\else\write16{ ***Undefined Equation Reference #1*** }\fi
\else\write16{ ***Undefined Equation Reference #1*** }\fi
\else\edef\LABxx{\getlabel{EQLABEL#1}}%
\def\LAByy{\expandafter\stripchap\LABxx}\ifchapternumbers%
\chapshow{\LAByy}.\expandafter\stripeq\LABxx%
\else\ifnum\number\LAByy=\chapno\relax\expandafter\stripeq\LABxx%
\else\chapshow{\LAByy}.\expandafter\stripeq\LABxx\fi\fi)\fi}%
\ifproofmode\write2{Equation #1}\fi}

\def\fignum{\global\advance\figureno by 1
\relax\iffigurechapternumbers\chapfolio.\fi\the\figureno}

\def\figlabel#1{\global\advance\figureno by 1
\relax\ifproofmode\ifforwardreference
\immediate\write1{\noexpand\expandafter\noexpand\def
\noexpand\csname FIGLABEL#1\endcsname{\the\chapno.\the\figureno?}}\fi\fi
\global\expandafter\edef\csname FIGLABEL#1\endcsname
{\the\chapno.\the\figureno?}\iffigurechapternumbers\chapfolio.\fi
\ifproofmode\llap{\hbox{\marginstyle#1
\kern1.2truein}}\relax\fi\the\figureno}

\def\figref#1{\hbox{\ifundefined{FIGLABEL#1}!!!\ifproofmode\ifforwardreference%
\else\write16{ ***Undefined Figure Reference #1*** }\fi
\else\write16{ ***Undefined Figure Reference #1*** }\fi
\else\edef\LABxx{\getlabel{FIGLABEL#1}}%
\def\LAByy{\expandafter\stripchap\LABxx}\iffigurechapternumbers%
\chapshow{\LAByy}.\expandafter\stripeq\LABxx%
\else\ifnum \number\LAByy=\chapno\relax\expandafter\stripeq\LABxx%
\else\chapshow{\LAByy}.\expandafter\stripeq\LABxx\fi\fi\fi}%
\ifproofmode\write2{Figure #1}\fi}

\def\tabnum{\global\advance\tableno by 1
\relax\iftablechapternumbers\chapfolio.\fi\the\tableno}

\def\tablabel#1{\global\advance\tableno by 1
\relax\ifproofmode\ifforwardreference
\immediate\write1{\noexpand\expandafter\noexpand\def
\noexpand\csname TABLABEL#1\endcsname{\the\chapno.\the\tableno?}}\fi\fi
\global\expandafter\edef\csname TABLABEL#1\endcsname
{\the\chapno.\the\tableno?}\iftablechapternumbers\chapfolio.\fi
\ifproofmode\llap{\hbox{\marginstyle#1
\kern1.2truein}}\relax\fi\the\tableno}

\def\tabref#1{\hbox{\ifundefined{TABLABEL#1}!!!\ifproofmode\ifforwardreference%
\else\write16{ ***Undefined Table Reference #1*** }\fi
\else\write16{ ***Undefined Table Reference #1*** }\fi
\else\edef\LABtt{\getlabel{TABLABEL#1}}%
\def\LABTT{\expandafter\stripchap\LABtt}\iftablechapternumbers%
\chapshow{\LABTT}.\expandafter\stripeq\LABtt%
\else\ifnum\number\LABTT=\chapno\relax\expandafter\stripeq\LABtt%
\else\chapshow{\LABTT}.\expandafter\stripeq\LABtt\fi\fi\fi}%
\ifproofmode\write2{Table#1}\fi}

\newdimen\sectionskip     \sectionskip=20truept
\newcount\sectno
\def\section#1#2{\sectno=0 \null\vskip\sectionskip
    \centerline{\chaplabel{#1}.~~{\bf#2}}\nobreak\vskip.2truein
    \noindent\ignorespaces}

\def\advancesectno{\global\advance\sectno by 1}
\def\sectfolio{\number\sectno}
\def\subsection#1{\goodbreak\advancesectno\null\vskip10pt
                  \noindent\chapfolio.~\sectfolio.~{\bf #1}
                  \nobreak\vskip.05truein\noindent\ignorespaces}

\def\uttg#1{\null\vskip.1truein
    \ifproofmode \line{\hfill{\bf Draft}:
    UTTG--{#1}--\number\year}\line{\hfill\today}
    \else \line{\hfill UTTG--{#1}--\number\year}
    \line{\hfill\ifcase\month\or January\or February\or March\or April%
    \or May\or June\or July\or August\or September\or October\or November%
    \or December\fi
    \space\number\year}\fi}

\def\contents{\noindent
   {\bf Contents\Z}\nobreak\vskip.05truein\noindent\ignorespaces}

\def\getlabel#1{\csname#1\endcsname}
\def\ifundefined#1{\expandafter\ifx\csname#1\endcsname\relax}
\def\stripchap#1.#2?{#1}
\def\stripeq#1.#2?{#2}

%%%%%%%%%%%%%%%%%%%%%%%%%%%%%%
%Reference macros%%%%%%%%%%%%%
%%%%%%%%%%%%%%%%%%%%%%%%%%%%%%
%
\catcode`@=11 % This allows us to modify PLAIN macros.
\def\space@ver#1{\let\@sf=\empty\ifmmode#1\else\ifhmode%
\edef\@sf{\spacefactor=\the\spacefactor}\unskip${}#1$\relax\fi\fi}
\newcount\referencecount     \referencecount=0
\newif\ifreferenceopen       \newwrite\referencewrite
\newtoks\rw@toks
\def\refmark#1{\relax[#1]}
\def\refend{\refmark{\number\referencecount}}
\newcount\lastrefsbegincount \lastrefsbegincount=0
\def\refsend{\refmark{\count255=\referencecount%
\advance\count255 by -\lastrefsbegincount%
\ifcase\count255 \number\referencecount%
\or\number\lastrefsbegincount,\number\referencecount%
\else\number\lastrefsbegincount-\number\referencecount\fi}}
\def\refch@ck{\chardef\rw@write=\referencewrite
\ifreferenceopen\else\referenceopentrue
\immediate\openout\referencewrite=referenc.texauxil \fi}
%
% In \obeyendofline, we say `\let^^M=\relax
{\catcode`\^^M=\active % these lines must end with %
  \gdef\obeyendofline{\catcode`\^^M\active \let^^M\ }}%
%
% In \ignoreendofline, we say `\let^^M=\relax
{\catcode`\^^M=\active % these lines must end with %
  \gdef\ignoreendofline{\catcode`\^^M=5}}
{\obeyendofline\gdef\rw@start#1{\def\t@st{#1}\ifx\t@st\blankend%
\endgroup\@sf\relax\else\ifx\t@st\bl@nkend\endgroup\@sf\relax%
\else\rw@begin#1
\backtotext
\fi\fi}}
{\obeyendofline\gdef\rw@begin#1
{\def\n@xt{#1}\rw@toks={#1}\relax%
\rw@next}}
\def\blankend{}
{\obeylines\gdef\bl@nkend{
}}
\newif\iffirstrefline  \firstreflinetrue
\def\rwr@teswitch{\ifx\n@xt\blankend\let\n@xt=\rw@begin%
\else\iffirstrefline\global\firstreflinefalse%
\immediate\write\rw@write{\noexpand\obeyendofline\the\rw@toks}%
\let\n@xt=\rw@begin%
\else\ifx\n@xt\rw@@d \def\n@xt{\immediate\write\rw@write{%
\noexpand\ignoreendofline}\endgroup\@sf}%
\else\immediate\write\rw@write{\the\rw@toks}%
\let\n@xt=\rw@begin\fi\fi\fi}
\def\rw@next{\rwr@teswitch\n@xt}
\def\rw@@d{\backtotext} \let\rw@end=\relax
\let\backtotext=\relax

\newdimen\refindent     \refindent=30pt
\def\Textindent#1{\noindent\llap{#1\enspace}\ignorespaces}
\def\refitem#1{\par\hangafter=0 \hangindent=\refindent\Textindent{#1}}
\def\REFNUM#1{\space@ver{}\refch@ck\firstreflinetrue%
\global\advance\referencecount by 1 \xdef#1{\the\referencecount}}
\def\refnum#1{\space@ver{}\refch@ck\firstreflinetrue%
\global\advance\referencecount by 1\xdef#1{\the\referencecount}\refend}

\def\REF#1{\REFNUM#1%
\immediate\write\referencewrite{%
\noexpand\refitem{#1.}}%
\begingroup\obeyendofline\rw@start}
\def\ref{\refnum\?%
\immediate\write\referencewrite{\noexpand\refitem{\?.}}%
\begingroup\obeyendofline\rw@start}
\def\Ref#1{\refnum#1%
\immediate\write\referencewrite{\noexpand\refitem{#1.}}%
\begingroup\obeyendofline\rw@start}
\def\REFS#1{\REFNUM#1\global\lastrefsbegincount=\referencecount%
\immediate\write\referencewrite{\noexpand\refitem{#1.}}%
\begingroup\obeyendofline\rw@start}

\def\REFSCON#1{\REF#1}

\def\cite#1{\refmark#1}
\catcode`@=12 % at signs are no longer letters
%
%%%
%%%%%%%%%%%%%%%%%%%%%%%%%%%
%%%[ Yet More TeXnicalities ]%%%%%%%%%%%%%%%%%%
%%%%%%%%%%%%%%%%%%%%%%%%%%%
\def\hourandminute{\count255=\time\divide\count255 by
60\xdef\hour{\number\count255}
\multiply\count255 by-60\advance\count255 by\time
\hour:\ifnum\count255<10 0\fi\the\count255}

\def\cite#1{{\refmark#1}}

\def\cropen#1{\crcr\noalign{\vskip #1}}%For use with \eqalign
\def\contents{\line{{\bf Contents}\hfill}\nobreak\vskip.2truein\noindent%
              \ignorespaces}
\def\bar{\overline}

\def\rd{{\rm d}}
\def\\{\hfill\break}
\def\:{\kern-1.5truept}
\def\LG{Landau-Ginzburg}
\def\LGO#1{Landau-Ginzburg orbifold{#1}}
\def\cicy{\hbox{CICY}}
\def\MPL#1{Mod.\ Phys.\ Lett.\ {\bf #1}}
\def\CQG#1{Class.\ Quantum\ Grav.\ {\bf #1}}
 % For CYCI configurations
 % ... in displayed math-mode:
\def\Cnfg#1#2{\matrix{#1}\mkern-6mu\left[\matrix{#2}\right]}
 % and in-line, in text-mode:

\def\subsection#1{\goodbreak\advancesectno\null%\vskip10pt
                  \noindent{\it \chapfolio.\sectfolio.~#1}
                  \nobreak\vskip.05truein\noindent\ignorespaces}
 % display pattern: [   a &= b  &  c &= d  &  e &= f   &(*) \cr]
\def\threeqsalignno#1{\panorama \tabskip=\humongous
     \halign to\displaywidth{\hfil$\displaystyle{##}$
      \tabskip=0pt&$\displaystyle{{}##}$\hfil
       \tabskip=0pt&\hfil$\displaystyle{##}$
        \tabskip=0pt&$\displaystyle{{}##}$\hfil
         \tabskip=0pt&\hfil$\displaystyle{##}$
          \tabskip=0pt&$\displaystyle{{}##}$\hfil
           \tabskip=\humongous&\llap{$##$}\tabskip=0pt\crcr#1\crcr}}
%%%%%%%%%%%%%%%%%%%%%%%%%%%
%%%%%%%%%%%%%%%%%%%%%%%%%%%
%%%%%%%%%%%%%%%%%%%%%%%%%%%
\proofmodefalse
\baselineskip=13pt
\parskip=2pt
\chapternumberstrue
\figurechapternumberstrue
\tablechapternumberstrue
%\forwardreferencetrue
\ifproofmode
\immediate\openout2=allcrossreferfile \fi
\ifforwardreference\input labelfile
\ifproofmode\immediate\openout1=labelfile \fi\fi
\noblackboxes
\hfuzz=1pt
\vfuzz=2pt
%
%%%%%%%%%%%%%
\nopagenumbers\pageno=-1
\null\vskip-40pt
\rightline{\eightrm CERN-TH.~6865/93, HUPAPP-93/3, NEIP 93-004}\vskip-3pt
\rightline{\eightrm NSF-ITP-93-96, UTTG-13-93}\vskip-3pt
\rightline{\eightrm hepth 9308005}\vskip-3pt
\rightline{\eightrm 2 August 1993}
\vskip .6truein
\centerline{\seventeenrm PERIODS FOR CALABI--YAU AND}
\vskip .2truein
\centerline{\seventeenrm LANDAU--GINZBURG VACUA}
\bigskip \bigskip \bigskip \bigskip
\centerline{{\csc Per~Berglund}$^{1,2}$\footnote{$^{\sharp}$}
            {\eightrm After Sept.~15th: Institute for Advanced Study,
              Olden Lane, Princeton, NJ 08540, USA.},\quad
            {\csc Philip~Candelas}$^{2,3}$,\quad
            {\csc Xenia~de la Ossa}$^{4\,\sharp}$,\quad
            {\csc Anamar{\'\ii}a~Font}$^5$,}
\vskip3mm
\centerline{{\csc Tristan~H\"ubsch}$^6$\footnote{$^{\flat}$}
            {\eightrm On leave from the Institute ``Rudjer
              Bo\v{s}kovi\'c'', Zagreb, Croatia.},\quad
            {\csc Dubravka~Jan\v{c}i\'c}$^2$,\quad
            {\csc Fernando~Quevedo}$^4$}
\bigskip \bigskip \bigskip \bigskip
\line{
\vtop{\baselineskip=11pt\hsize = 2.0truein
\centerline{$^1$\it Institute\:\ for\:\ Theoretical\:\ Physics}
\centerline{\it University of California}
\centerline{\it Santa Barbara}
\centerline{\it CA 93106, USA}}\hfill
\vtop{\baselineskip=11pt\hsize = 2.0truein
\centerline{$^2$\it Theory Group}
\centerline{\it Department of Physics}
\centerline{\it University of Texas}
\centerline{\it Austin, TX 78712, USA}}\hfill
\vtop{\baselineskip=11pt\hsize = 2.0truein
\centerline {$^3$\it Theory Division}
\centerline {\it  CERN}
\centerline {\it CH 1211 Geneva 23}
\centerline {\it Switzerland} }}
\vskip.2truein
\line{
\vtop{\baselineskip=11pt\hsize = 2.0truein
\centerline{$^4$\it Institut de Physique\qquad}
\centerline{\it Universit\'e\:\ de\:\ Neuch\^atel\qquad}
\centerline{\it CH-2000 Neuch\^atel\qquad}
\centerline{\it Switzerland\qquad}}\hfill
\vtop{\baselineskip=11pt\hsize = 2.0truein
\centerline{$^5$\it Departamento de F{\'\ii}sica}
\centerline{\it Universidad\:\ Central\:\ de\:\ Venezuela}
\centerline{\it A.P.~20513, Caracas 1020-A}
\centerline{\it Venezuela}}\hfill
\vtop{\baselineskip=11pt\hsize = 2.0truein
\centerline{$^6$\it Department of Physics}
\centerline{\it Howard University}
\centerline{\it Washington}
\centerline{\it DC 20059, USA} }}
\bigskip \bigskip \bigskip \bigskip
\vbox{\narrower\centerline{\bf ABSTRACT}
\vskip.15truein
\vbox{\baselineskip 11pt\noindent  The complete structure of the moduli
space of \cys\ and the associated Landau-Ginzburg theories, and hence also
of the corresponding low-energy effective theory that results from (2,2)
superstring compactification, may be determined in terms of certain
holomorphic functions called periods. These periods are shown to be
readily calculable for a great many such models. We illustrate this by
computing the periods explicitly for a number of classes of \cys. We also
point out that it is possible to read off from the periods certain
important information relating to the mirror manifolds.
 \bigskip
\leftline{\eightrm CERN-TH.~6865/93}%\vskip-3pt
\leftline{\eightrm 2 August 1993}}}
\newpage
\headline={\ifproofmode\hfil\eightrm draft:\ \today\
\hourandminute\else\hfil\fi}
%
%%%%%%%%%%%%%%%%%%%%%%%%
%%%[ The Article! ]%%%%%
%%%%%%%%%%%%%%%%%%%%%%%%
\contents

\item{1.~}Preamble\bigskip
\item{2.~}Hypersurfaces in Weighted $\CP{N}$
\itemitem{\it 2.1~}{\it Generalities}
\itemitem{\it 2.2~}{\it The fundamental expansion}
\itemitem{\it 2.3~}{\it Convergence and analytic continuation:
                        small $\vf_\a$}
\itemitem{\it 2.4~}{\it Convergence and analytic continuation:
                        general $\vf_\a$}
\itemitem{\it 2.5~}{\it The remaining periods}\bigskip
\item{3.~}\cicy\ Manifolds
\itemitem{\it 3.1~}{\it Complete intersections in a single
                        projective space}
\itemitem{\it 3.2~}{\it Calabi-Yau hypersurfaces in products of
                        projective spaces}
\itemitem{\it 3.3~}{\it A natural conjecture}\bigskip
\item{4.~}Hypersurfaces with Two Parameters
\itemitem{\it 4.1~}{\it Fermat hypersurfaces}
\itemitem{\it 4.2~}{\it A non-Fermat example}
\itemitem{\it 4.3~}{\it The Picard--Fuchs equations}\bigskip
\item{5.~}Other Important Generalizations
\itemitem{\it 5.1~}{\it Manifolds with no known mirror}
\itemitem{\it 5.2~}{\it Twisted moduli}
\itemitem{\it 5.3~}{\it A new look at the contours}\bigskip
\item{6.~}Concluding Remarks
\vfill
\newpage
\pageno=1\footline={\rm\hfil\folio\hfil}
\section{pre}{Preamble}
Superstring vacua with (2,2) worldsheet supersymmetry depend on two classes of
parameters: the complex structure moduli and the K\"ahler moduli,
which are in a 1--1 correspondence with
the $E_6$-charged
matter fields in the low energy effective theory. The dynamics of these
matter fields is determined by the geometry on these two components of
the parameter space, the kinetic terms being determined by the metric and
the Yukawa couplings by a certain natural third rank tensor.
A key feature of the geometry is that both the metric and the
Yukawa coupling on each of the two sectors of the moduli space are
determined from a single holomorphic object that varies over the moduli
space. For the space of complex structures this is the holomorphic 3-form
$\O$, or equivalently, the vector formed from its periods,
$\vp_i\define \oint_{\g^i}\O$,
with the $\g^i$ a basis of homology cycles. It is reasonable to think of
the $\vp_i$ as the components of $\O$ and from these periods much of the
structure of the moduli space and of the low energy effective theory is
readily available.

In virtue of mirror symmetry the moduli space of the K\"ahler class
parameters of a manifold, \M, is identified with the space of complex
structures of the mirror manifold, \ca{W}, of \M. Thus whenever the mirror
of a given manifold is known, a study of the complex structure periods of
the mirror, \ca{W}, allows one to solve for the structure of the space of
K\"ahler class parameters of \M. For most constructions, it is easier to
determine the complex structure periods $\vp_i$ and so their K\"ahler
class equivalents are determined via the mirror map.

A procedure of explicitly solving for the structure of the {\it complete}
moduli space of a given family of vacua, or equivalently of finding both
the kinetic terms and the Yukawa couplings of the corresponding low energy
theory, factors roughly speaking into two stages.
  (i) Finding a complete basis of (complex structure) periods $\vp$ and
      $\hat\vp$ for the manifold and its mirror and
 (ii) finding a symplectic basis $\P_i = m_i{}^j\vp_j$, with $m$ a constant
      matrix, such that the cycles corresponding to $\P$ form an integral
      symplectic homology basis.
In this basis, the K\"ahler potential for the moduli space and
the Yukawa couplings for the effective theory are given by the
expressions~
\REFS{\AS}{A.~Strominger, \cmp{133} (1990) 163.}
\REFSCON{\Cd}{P.~Candelas and X.~C.~de~la~Ossa, \npb{355} (1991) 455.}
\refsend:
$$ \eqalign{
   e^{-K}   &= -i\int_\M \O\wedge\bar\O = -i\P^T\S\,\bar\P~, \cropen{5pt}
 y_{\a\b\g} &=\hphantom{-i}\int_\M \O\wedge\vd_\a\vd_\b\vd_\g\O
             =\P^T\S\,\vd_\a\vd_\b\vd_\g\P                      \cr}
  \eqlabel{XXX}
$$
where $\S$ is the matrix $\left({\- 0~{\bf 1}\atop -{\bf 1}~0}\right)$. Of
course, in the cases where a construction of the mirror models is not known,
this procedure still provides a description of the complex structure sector
of the moduli space and this application is logically independent of mirror
symmetry.

This article addresses step (i) of the process outlined above. We present a
widely applicable method of calculating the complex structure periods
$\vp_i$ for a large number of \cys\ and \LGO{s}. Through the mirror map,
the considerations
of this article are useful also for purposes of enumerative geometry.
Such applications will, however, be developed elsewhere.
There is as yet no general method for implementing step (ii), but
worked examples may be found in Refs.~
\REFS\rCdGP{P.~Candelas, X.~ de la Ossa, P.~Green and L.~Parkes,
       \npb{359} (1991) 21.}
\REFSCON\rMorrison{D.~Morrison, ``Picard-Fuchs Equations and Mirror Maps
       for Hypersurfaces'', in {\it Essays on Mirror Symmetry}, ed.\
       S.-T.~Yau (Intl. Press, Hong Kong, 1992).}
\REFSCON\rFont{A.~Font, \npb{391} (1993) 358.}
\REFSCON\rKT{A.~Klemm and S.~Theisen, \npb{389} (1993) 153.}
\REFSCON\rCdFKM{P.~Candelas, X.~de~la~Ossa, A.~Font, S.~Katz and
       D.~Morrison, in preparation.}
\refsend.
More precisely the present work is primarily concerned with two
observations: (a) that the periods may be calculated for a wide range of
models and (b) that although this fact is logically separate from the
existence of mirror symmetry nevertheless it is the case that the periods
encode important information about the mirror. Thus it is, for example,
that for the class of \cys\ that can be realised as hypersurfaces in weighted
\cp4's the periods are written most naturally in terms of the weights of
the mirror manifolds.

Before embarking on generalizations let us first recall the computation of
the complex structure periods for a one-parameter family of mirrors of
quintic threefolds~\cite{\rCdGP}. These mirror manifolds, $\cal W$, are
defined as the locus, ${\cal M}/G$
where \M\ is the zero set of the polynomial
$$
  p(x;\j)~ = ~\sum_{k=1}^5 x_k^5 - 5\j\, x_1 x_2 x_3 x_4 x_5~,
  \eqlabel{XXX}
$$
and the coordinates $x_k$ identified under the action of a group, $G$,
which is abstractly $\ZZ_5^{~3}$ and which is generated by\Footnote{We
will use the notation $(\ZZ_k:r_1,r_2,r_3,r_4,r_5)$ for a $\ZZ_k$ symmetry
with the action\hfil\break $(x_1,x_2,x_3,x_4,x_5) \to
(\w^{r_1} x_1,\w^{r_2} x_2, \w^{r_3} x_3,\w^{r_4} x_4, \w^{r_5} x_5)$,
where $\w^k = 1$.}
$$ \eqalign{ g_0~ &= ~(\ZZ_5 : 1,0,0,0,4)~,  \cr
             g_1~ &= ~(\ZZ_5 : 0,1,0,0,4)~,  \cr
             g_2~ &= ~(\ZZ_5 : 0,0,1,0,4)~.  \cr}
  \eqlabel{XXX}
$$
Now $b_{1,1}({\cal M}) =
b_{2,1}({\cal W}) = 1$, and $\j$ parametrizes both the complex structures
of $\cal W$ and the variations of the K\"ahler class of $\cal M$; hence the
two-fold application of such calculations\Footnote{In fact, $\j$ also
parametrizes a 1-dimensional subspace of the 101-dimensional space of
complex structures on $\cal M$ and its mirror analogue of this
1-dimensional slice in the 101-dimensional space of variations of the
K\"ahler class on $\cal W$.}.

The holomorphic three-form is chosen to be
$$
  \O(\j) = 5\j\, {5^3\over (2\p i)^3}\,
                   {x_5\, \rd x_1\wedge \rd x_2\wedge \rd x_3
                    \over \displaystyle{\pd{p(x;\j)}{x_4}}}~,
  \eqlabel{XXX}
$$
with a prefactor chosen to simplify later expressions. The $5^3$
corresponds to the order of $G$ and the factors of $2\p i$ relate to a
residue calculation that follows shortly.

We define also a `fundamental cycle' $B_0$
$$ {\eqalign{
   B_0=\Big\{ x_k &\big|~x_5~ = ~1,\qquad|x_1|=|x_2|=|x_3|=\d~,
                                                             \cropen{-8pt}
   & \quad x_4~\hbox{given by the solution to $p(x)=0$ that tends to zero
                     as $\j\to\infty$}~\Big\}\cr}} \eqlabel{Bzero}
$$
and a fundamental period
$$
  \vp_0(\j)~ = ~\int_{B_0}\O(\j)~.
  \eqlabel{XXX}
$$
The evaluation of the fundamental period proceeds by noting that
$$ \eqalignno{ \vp_0(\j)~
    &= ~-5\j\, {1\over (2\p i)^4}\int_{\g_1\times\cdots\times\g_4}
         {x_5 \rd x_1 \rd x_2 \rd x_3 \rd x_4\over p(x;\j)}\cropen{5pt}
    &= ~-5\j\, {1\over (2\p i)^5}\int_{\g_1\times\cdots\times\g_5}
         {\rd x_1 \rd x_2 \rd x_3 \rd x_4 \rd x_5\over p(x;\j)}~.
\eqalignlabel{vpzero}\cr}
$$
In these relations $\g_i$ denote the circles $|x_i|=\d$. The first equality
follows by noting that in virtue of the definition of $B_0$ the value of
$x_4$ for which $p$ vanishes lies inside the circle $\g_4$ for
sufficiently large $\j$ and that the residue of $1/p$ is $1/\pd{p}{x_4}$.
The minus sign reflects a choice of orientation for $B_0$ and a factor of
$5^3$ has been absorbed into the contour by taking this to be
$\g_1{\times}\cdots{\times}\g_4$ rather than
$\g_1{\times}\cdots{\times}\g_4/G$. The second equality follows by noting
that on account of the homogeneity properties of the integrand under the
scaling $x_k\to\l x_k$ the integral is, despite appearances, independent
of $x_5$. We have therefore introduced unity in the guise of
${1\over 2\p i}\int{dx_5\over x_5}$. The last expression will be easy to
generalize and we may in fact consider the residue~\eqref{vpzero} as the
definition of $\vp_0$.

The next maneuver is to expand $1/p$ in powers of $\j$
$$
  \vp_0(\j) = {1\over (2\p i)^5}
   \sum_{m=0}^\infty {1\over (5\j)^m} \int_{\g_1\times\cdots\times\g_5}
   {dx_1 dx_2 dx_3 dx_4 dx_5\over x_1 x_2 x_3 x_4 x_5}
   {(x_1^5{+}x_2^5{+}x_3^5{+}x_4^5{+}x_5^5)^m \over  (x_1 x_2 x_3 x_4 x_5)^m}~.
  \eqlabel{XXX}
$$
The integrals may be evaluated by residues. The only terms that contribute
are the terms in the expansion of $(x_1^5+\cdots+x_5^5)^m$ that cancel the
factor of $(x_1 \cdots x_5)^m$ in the denominator. Such terms arise only
when $m=5n$ in which case
$(x_1^5+\cdots+x_5^5)^{5n}$ contains
$x_1^{5n} x_2^{5n} x_3^{5n} x_4^{5n} x_5^{5n}$
with coefficient $(5n)!/(n!)^5$. Thus
$$
  \vp_0(\j)~ = ~\sum_{n=0}^\infty {(5n)!\over (n!)^5\, (5\j)^{5n}}~,\qquad
  |\j|\ge 1~,\quad 0<\arg(\j)<{2\p\over 5}~.
  \eqlabel{XXX}
$$
The restriction on $\arg(\j)$ is due to the fact that $\vp_0(\j)$ has
branch points when $\j^5=1$ and we take the cuts to run out radially from
these points.

The fundamental period may be continued analytically to the region
$|\j|<1$ with the result that
$$
  \vp_0(\j)~ = ~-{1\over 5}\sum_{m=1}^\infty {\G({m\over 5})(5\a^2\j)^m
  \over \G(m)\G^4(1-{m\over 5})}~,\qquad |\j|<1~.
  \eqlabel{XXX}
$$
The functions $\vp_j(\j)\define \vp_0(\a^j\j)$ are a set of five periods
any four of which are linearly independent. Owing to the zeros from
$\G(1-{m\over5})$ in the denominator, the $\vp_j$ satisfy
the single constraint
$$
  \sum_{j=1}^5 \vp_j(\j)=0~.
  \eqlabel{XXX}
$$
Our first observation is that the process outlined above for the simple
case of the one-parameter family of mirrors of quintic threefolds admits
significant generalizations.

There are large classes of Calabi-Yau and Landau-Ginzburg vacua that have
been investigated during the past few years and to which we would like to
extend the calculation of the periods. Recently, a complete listing of
non-degenerate Landau-Ginzburg potentials leading to $N=2$ string vacua
with $c=9$ has been achieved~
\REFS\rMaxSkI{M.~Kreuzer and H.~Skarke, \npb{388} (1992) 113.}
\REFSCON\rMaxSkII{M.~Kreuzer and H.~Skarke, \cmp{150} (1992) 137.}
\REFSCON\rAR{A.~Klemm and R.~Schimmrigk, ``Landau-Ginzburg Vacua'',
       CERN preprint CERN-TH-6459/92, Universit\"at Heidelberg report
       HD-THEP-92-13.}
\refsend.
7,555 (about $3\over 4$) of these models admit a
standard geometrical interpretation in terms of
hypersurfaces defined from the vanishing of a polynomial in a weighted
\CP4. These models are Calabi-Yau compactifications and it is for this
large class of models that we find the fundamental period in Section~2.
Our ability to do this is due to the fact that for all these models, there
exists generically a contour
$\g_1{\times}\g_2{\times}\g_3{\times}\g_4{\times}\g_5$ as in
\eqref{vpzero}. For some $2\over3$ of these \cy\ compactifications, the
mirror models are known in virtue of the construction of Ref.~
\Ref\rBH{P.~Berglund and T.~H\"ubsch, \npb{393} (1993) 377, also in
       {\sl Essays on Mirror Manifolds}, p.388, ed.\ S.-T.~Yau,
       (Intl.\ Press, Hong Kong, 1992).}.
This construction applies to deformation classes of hypersurfaces for
which a smooth representative may be found corresponding to a polynomial
which is the sum of five monomials, that is, the number of monomials is
equal to the number of projective coordinates. The prescription for
finding the polynomial that corresponds to the mirror of a given model
proceeds by transposing the matrix of the exponents of the original
polynomial and then forming the quotient by a suitable group%
\REF\rBriRon{B.R.~Greene and M.R.~Plesser: ``Mirror Manifolds, A Brief
Review and Progress Report'', Cornell and Yale University preprints CLNS
91-1109, YCTP-P32-91.}%
\Footnote{The  class of Fermat surfaces for which $p=\sum_i{x_i^{a_i}}$,
corresponding to the $A$ series of products of minimal models, are
invariant under transposition and the  prescription to find the mirrors of
these models by orbifoldizing the original manifold~\cite{\rBriRon}\ is
recovered as a particular case.}. As we will see, the fundamental period
for this class of models can be nicely written in terms of the weights of
the mirror manifold.  This is clearly a reflection of mirror symmetry.
Thus, these periods can, in some sense, be thought of as periods for the
\K\ class parameters of the mirror manifold.  As particular cases of the
general expression, a more explicit form of the period for Fermat
hypersurfaces is given which allows us to reproduce the known results for
one and two moduli examples that have been studied.  We also present here a
method to find the other periods by analytically continuing the fundamental
period and using the geometrical
symmetries of the model.

Another class of \cy\ models that has been studied is the class of complete
intersection Calabi-Yau (CICY) models~
\REFS\rTristan{T.~H\"ubsch, \cmp{108} (1987) 291.}
\REFSCON\rGH{P.~Green and T.~H\"ubsch, \cmp{109} (1987) 99.}
\REFSCON\rCLS{P.~Candelas, A.M.~Dale, C.~A.~L\"utken and R.~Schimmrigk,\Z
       \npb{298} (1988) 493.}
\REFSCON\rBeast{T.~H\"ubsch: {\sl \cy\ Manifolds---A Bestiary for
       Physicists}\Z (World Scientific, Singapore, 1992).}
\refsend.
These spaces correspond to non-singular intersection of hypersurfaces on
products of (weighted) projective spaces.
The prototypical\vadjust{\newpage} example of such a
model is provided by the Tian--Yau manifold~
\Ref\rTY{S.-T.~Yau: in Proc. of {\sl Symposium on Anomalies, Geometry,
       Topology\/}, pp.395, esp. the Appendix, by G.~Tian and S.-T.~Yau,
       pp.402-405, eds. W.A.~Bardeen and A.R.~White (World Scientific,
       Singapore, 1985).}
$$
   \Cnfg{\CP3\cr \CP3\cr}{3&0&1\cr 0&3&1\cr}~.
  \eqlabel{XXX}
$$
The notation denotes that the manifold is realised in $\CP3\times\CP3$ by
three polynomials $p_\a$. The columns of the matrix give the degrees of
the $p_\a$ in the variables of each of the projective spaces. More
generally, there can be $N$ polynomials $p_\a$ and $F$ factor spaces such
that the total dimension of the factor spaces is $N+3$. In favourable
cases, such models correspond to \LGO{s}~
\REFS\rGVW{B.R.~Greene, C.~Vafa and N.P.~Warner, \npb{324} (1989) 371.}
\REFSCON\rSplit{T.~H\"ubsch, \CQG{8} (1991) L31.}
\REFSCON\rCQG{P.~Berglund, B.R.~Greene and T.~H\"ubsch, \MPL{A7} (1992)
       1885.}
\refsend,
and we address the question of finding the periods for these models
starting with Section~3.

For complete intersection manifolds, explicit constructions of mirror
models are less well known and it is harder to write a general expression
for the K\"ahler class periods explicitly. We proceed by first analyzing
two simple classes of \cicys. Those with $N=1$ or $F=1$. The mirror
manifolds of the \cicys\ with $N=1$ are the manifolds studied by Libgober
and Teitelbaum~~\Ref{\rLT}{A.~Libgober and J.~Teitelbaum, ``Lines on
\cy\ Complete Intersections, Mirror Symmetry and Picard Fuchs Equations'',
\preprint{Illinois} (1992).}. These are the mirrors of the five manifolds
$$\vbox{\hskip0pt
 {\hsize=1in \parindent=0pt
 \valign{#\hfill\strut\cr
 \CP4[5]~~,\cr
 \CP5[3,3]~~,\cr
 \CP5[2,4]~~,\cr
 \CP6[2,2,3]~~,\cr
 \CP7[2,2,2,2]~~,\cr}}}
  \eqlabel{XXX}
$$
which may be realised by transverse polynomials in a single projective
space. These manifolds, which are the simplest one modulus \cicys, are of
interest also owing to the fact that the structure of their parameter
spaces is somewhat different from the one parameter spaces that have been
studied hitherto. We also employ the construction of
Batyrev~~\Ref\Baty{V. Batyrev, {\it Duke Math. J.} {\bf 69} (1993) 349.}\
to compute the fundamental periods for the mirrors of the five manifolds
that may be realised by a single transverse polynomial in a product of
projective spaces
$$\vbox{
 {\hsize=1in \parindent=0pt
 \valign{$$#$$\hfill\strut\vfil\cr
 \Cnfg{\CP4\cr}{5\cr}~,                                        \cr
 \Cnfg{\CP2\cr \CP2\cr}{3\cr 3\cr}~,                           \cr
 \Cnfg{\CP1\cr \CP3\cr}{2\cr 4\cr}~,                           \cr
 \Cnfg{\CP1\cr \CP1\cr \CP2\cr}{2\cr 2\cr 3\cr}~,              \cr
 \Cnfg{\CP1\cr \CP1\cr \CP1\cr \CP1\cr}{2\cr 2\cr 2\cr 2\cr}~. \cr}}
      \vskip-20pt}
  \eqlabel{XXX}
$$
Having computed the fundamental period for these simple \cicys\ we
conjecture the form of the fundamental period for a general \cicy\ and for
a general \cicy\ in weighted projective spaces for the
K\"ahler class parameters of the embedding spaces.
In Section~4, we study some two-parameter examples to illustrate the
analysis.

Section~5 is devoted to other important generalizations.
In subsection~{\it5.1} we show that the calculations of Section~2 apply
straightforwardly even to \cys\ which are hypersurfaces in weighted
$\CP4$ spaces and for which there is not a known construction for its
mirror.
In subsection~{\it5.2}, we
point out that the method is not limited to the parameters associated to
the \K-classes of the embedding spaces and as an illustration of this we
compute the dependence of the periods on the `twisted parameters' for the
manifolds
$$
  \Cnfg{\CP3\cr \CP2\cr}{3&1\cr 0&3\cr}~~~\hbox{and}~~~
  \Cnfg{\CP4\cr \CP1\cr}{4&1\cr 0&2\cr}~. \eqlabel{XXX}
$$
This is of interest since it requires the introduction of
`non-polynomial' deformations of the defining polynomials.
\Footnote{The first case~~\Ref\rROLF{R.~Schimmrigk: \plb{193} (1987) 175.}\
is of potential phenomenological interest since a particular quotient leads to
a three-generation model. Knowledge of the structure of the moduli space will
be useful when  determining the low-energy effective action for this model.}
 We find also that it is necessary to introduce a new set of contours in order
to evaluate the periods.  In connection with this we return in
subsection~{\it5.3} to a consideration of the periods for the manifold
$\CP4[5]$. We introduce a set of alternative contours and argue that these
contours are actually the truly fundamental contours in the sense that they
seem
to exist in general for any Landau-Ginzburg potential associated to a general
weighted complete intersection \cym. In Section~6, we discuss our results and
the limitations of the method.

While completing this work we received a copy of an interesting paper by
Batyrev and Van~Straten~
\Ref\rBatnew{V. Batyrev and D. Van Straten, ``Generalized Hypergeometric
Functions and Rational Curves on Calabi--Yau Complete Intersections in Toric
Varieties'', Universit\"at Essen report in preparation.}\
which has substantial overlap with the present work.

\newpage
\section{OneHyper}{Hypersurfaces in Weighted $\CP{N}$}
\vskip-10pt
\subsection{Generalities}
Our aim, in this section, is to derive an explicit expression for the
fundamental period for any \cym\ that can be realised as a hypersurface in
a weighted \CP{N}. We shall show also that this period is most naturally
written in terms of data that pertain to the mirror manifold.

We consider a hypersurface ${\cal M}$, defined with reference to the
zero-set of a defining polynomial $P$ of degree $d$ in a weighted
projective space $\CP{N}_{(k_0,\ldots,k_N)}$. Of course, the case of
immediate interest is $N=4$, but the results of this section are valid for
general $N$. The relation $$
   d~ = ~\sum_{i=0}^N k_i~,\qquad d~ = ~\deg(P)~, \eqlabel{eCYcond}
$$
ensures that the hypersurface $\cal M$ is a \cym. In the corresponding
\LGO{} with $N{+}1$ chiral superfields $X_i$, $P$ is the superpotential.
It will be shown in later sections how to extend the results to the more
general case of (at least many of the) intersections of hypersurfaces in
products of projective spaces and the corresponding \LGO{s}.

The defining polynomial, $P$, is usually required to be non-degenerate
so that the defined hypersurface $\cal M$ is smooth. That is the
derivative, $\rd P$, must not vanish simultaneously with $P$ in the
(weighted) projective space $\CP{N}_{(k_0,\ldots,k_N)}$. In the affine
space, $\IC^{N{+}1}_{(k_0,\ldots,k_N)}$, $\rd P$ and $P$ must
simultaneously vanish precisely at the origin $x_i=0$, $i=0,\ldots,N$.
This condition seems to be required for the interpretation of \LGO{s}:
only such non-degenerate models are well understood. The weighted
homogeneous coordinates $x_i$ have weights $k_i$; that is,
$$
  x_i~ \approx ~\l^{k_i}x_i,\qquad i=0,\ldots, N~. \eqlabel{eWCP}
$$
The basic strategy is to write $P$ as a deformation of a suitable defining
polynomial, $P_0$, and calculate the periods in terms of a power series
expansion in the moduli $\f_\a$, $\a=0,\ldots,M{-}1$, defined by
$$
  P(x_i;\f_\a) ~~ \define ~~P_0(x_i)~ - ~\f_0\,\prod_{i=0}^N x_i~
                                + ~\sum_{\a=1}^{M-1} \f_\a M^\a(x_i)~.
				 \eqlabel{eDefP}
$$
To compare with the simpler case of Ref.~\cite{\rCdGP} recounted in the
introduction, write $\f_0=d\j$. The fundamental period is defined
initially for sufficiently large-$\f_0$ and is then extended over the
whole parameter space by analytic continuation.

The fundamental monomial $M^0(x_i) \define \prod_{i} x_i$ is a
nontrivial deformation at generic points of the moduli
space~
\Ref\rHY{T.~H\"ubsch and S.-T.~Yau, \MPL{A7} (1992) 3277.},
and has here been isolated among other deformations, $M^\a(x_i)$, for
future convenience.
In favourable circumstances, the complete list of degree-$d$ monomials
$M^\a(x_i)$, together with the fundamental monomial $\prod_i x_i$ spans
the whole space of deformations.
In general, however, there are other, non-polynomial, deformations~
\REFS\rPDM{P.~Green and T.~H\"ubsch, \cmp{113} (1987) 505.}
\REFSCON\rCfC{P.~Berglund and T.~H\"ubsch, ``Couplings for
       Compactification'', Howard University preprint HUPAPP-93/2,
       \npb{~}(in press).}
\refsend\
and our present methods seem to imply a restriction to the sector
representable by polynomial deformations.  It is tempting to identify
these deformations with the `untwisted sector' of \LGO{s} or even
superconformal field theories, but this is simply false in
general~\cite{\rCQG}; hence the use of the names `polynomial' and
`non-polynomial' deformations and sectors. We do not have a complete
understanding of how to represent in general the parameters of the twisted
sectors. However it is sometimes possible to represent certain twisted
parameters by certain `non-polynomial' deformations of the defining
polynomial that also involve radicals of polynomials.
\Footnote{In certain situations some of the states in the twisted sector
of \LGO{s} may be represented as rational functions of the coordinates~
\Ref\rLS{M.~Lynker and R.~Schimmrigk, \plb{249}(1990) 237.}. Thus, one may
guess that in general the fractional powers appearing in the `non-polynomial'
defomormations will be negative as well as positive (and zero).}

 For these cases
there appears to be no difficulty in calculating periods for these
parameters. We present two examples of this procedure in
subsection~{\it5.2}.

The present methods apply also to finite quotients of
\cy\ hypersurfaces since the `allowable symmetries' leave $\prod_i x_i$
invariant~
\Ref\rGP{B.R.~Greene and R.~Plesser, in {\sl Essays on Mirror
      Symmetry}, p.1, ed.\ S.-T.~Yau\Z (Intl.\ Press, Hong Kong, 1992).};
the collection of admissible deformations $M^\a (x_i)$ in this case will
of course become smaller.
This straightforwardly covers practically all known constructions of
mirror models~\cite{\rBH} and thereby provides a method for calculating
also the K\"ahler class periods for all these models. \bigskip

The workhorse of the present article is the residue
formula~
\REFS\rABG{M.~Atiyah, R.~Bott and L.~G{\aa}rding, Acta Math.~{\bf 131}
        (1973) 145.}
\REFSCON\rPhilip{P.~Candelas, \npb{298} (1988) 458.}
\refsend\
for the holomorphic $(N{-}1)$-form
$$
  \O(\f_\a)~~ \define
            ~~{\rm Res}_{\cal M}\bigg[ {\mu\over P(x_i;\f_\a)}\bigg]~,
	 \eqlabel{eOm}
$$
for the case where each $\cal M$ (for each choice of $\f_\a$) is the
hypersurface $P(x_i;\f_\a)=0$ in a (weighted) projective space
$\CP{N}_{(k_0,\ldots,k_N)}$. The residue is evaluated point by point for
each $x \in {\cal M}$. The resulting $\O$ is the holomorphic
$(N{-}1)$-form on the manifold ${\cal M}$ parametrized by the $\f_\a$.
Here, $$
    \mu ~~ \define ~~{1\over (N{+}1)!} \e^{i_0 i_1\cdots i_N}
        k_{i_0} x_{i_0} \rd x_{i_1} \cdots \rd x_{i_N}~,
  \eqlabel{eXXX}
$$
is the natural holomorphic volume-like form\Footnote{Actually, this is not
a section of the canonical bundle ${\cal K}$ on the weighted projective
space, and so is {\it not} a holomorphic volume form. Instead, it is a
section of the product bundle ${\cal K} \otimes {\cal O}(d)$. However, the
integrand in \eqref{eOm}\ is homogeneous of degree zero and {\it is} a
section of the canonical bundle. This provides for the canonicity of
\eqref{eOm}.} on $\CP{N}_{(k_0,\ldots,k_N)}$. Back in the affine space,
$\IC_{N{+}1}^{(k_0,\ldots,k_N)}$, we may take $\m$ to be given by
$$
   \mu ~~ = ~~\prod_{i=0}^N {\rm d} x_i~.
  \eqlabel{eHVol}
$$
Most of the time, the affine case will be implied and the context should
make it clear if the projective case is understood instead; we therefore
make no notational distinction between these two. The manifold $\cal M$ is
smooth if the defining polynomial $P(x_i;\f_\a)$ is non degenerate and
this ensures that the residue \eqref{eOm}\ is well defined.

The fundamental period of $\O$ is then {\it defined} to be
$$
 \vp_0(\f_\a)~~ \define
   ~~ - \f_0\, \oint_{B_0} \O(\f_\a)~~ =
   ~~ - \f_0\, {C\over(2\p i)^{N+1}} \int_\G {\prod_{i=0}^N {\rm d} x_i
                               \over P(x_i;\f_\a)}~,
  \eqlabel{ePiO}
$$
where the integral has been re-expressed in the affine space
$\IC_{N{+}1}^{(k_0,\ldots,k_N)}$ as in Section~\chapref{pre}. Here
$\G\subset \IC^{N{+}1}_{(k_0,\ldots,k_N)}$ is chosen so as to reproduce
the integral of the residue \eqref{eOm}\ over the cycle $B_0$ in
$\cal M$ and $C$ is a convenient normalization constant. The
explicit prefactor $\f_0$ and the overall sign are again merely for later
convenience.

\subsection{The fundamental expansion}
Bearing in mind the classification of transverse polynomials of
Refs.~\cite{{\rBH,\rMaxSkII,\rAR}},
we choose the reference polynomial $P_0(x_i)$ in Eq.~\eqref{eDefP}\ to be
given by the sum of $N$ monomials
$$
  P_0(x_i)~ = ~\sum_{j=0}^N \prod_{i=0}^N x_i^{a_{ij}}~.
  \eqlabel{ePOLY}
$$
Thus $P_0$ is specified by the $(N{+}1)\times(N{+}1)$ matrix  of exponents
$a_{ij}$ satisfying
$$
  \sum_{i=0}^N k_i a_{ij}~ = ~d~,\qquad {\rm for~all~}j~. \eqlabel{eConsA}
$$
For $M^\a(x_i)$ we take the most general set of monomials which we write as
$$
  M^\a~ = ~\prod_{i}x_{i}^{q_{i}^{\a}}\qquad \a=1,\cdots, M-1~,
   \eqlabel{eTheM}
$$
for some matrix of exponents $q_i^\a$ that is subject to the requirement
$$
  \sum_{i=0}^N  k_i q_{i}^{\a}~ = ~d~,\qquad {\rm for~all~}\a~.
  \eqlabel{eConsQ}
$$

This covers all the types of non-degenerate polynomials listed in
Ref.~\cite{\rBH}. For all these, the exponent matrices are block-diagonal,
with blocks which are either themselves diagonal or which take one of the
following two forms:
$$
  \left[\matrix{a_1 &  1  &  \cdots  &  0  &  0  \cr
                          0  & a_2 &  \cdots  &  0  &  0  \cr
                  \vdots  & \vdots &  \ddots  &  1  &  0  \cr
                          0  &  0  &  \cdots  & a_{r-1} &  1  \cr
                          0  &  0  &  \cdots  &  0  & a_r \cr}\right]~,
    \qquad \left[\matrix{a_1 &  1  &  \cdots  &  0  &  0  \cr
                          0  & a_2 &  \cdots  &  0  &  0  \cr
                  \vdots  & \vdots &  \ddots  &  1  &  0  \cr
                          0  &  0  &  \cdots  & a_{s-1} &  1  \cr
                          1  &  0  &  \cdots  &  0  & a_s \cr}\right]~.
  \eqlabel{eAij}
$$
This class includes, as well as polynomials of Fermat type, also
polynomials such as, say,
$$
  x_0^7 + x_1^7 x_3 +  x_3^3 + x_2^7 x_4 + x_4^3~,  \eqlabel{XXX}
$$
whose exponent matrix has a $1\times 1$ Fermat block corresponding to
$x_0$ and two blocks of the type of the first of the matrices above.
The remaining types of non-degenerate polynomials in the
complete lists of Refs.~\cite{{\rMaxSkII,\rAR}} have more monomials than
coordinates. These polynomials are recovered in Eq.~\eqref{eDefP}\ by
including such additional monomials in the collection of $M_\a(x_i)$.
The limit in which the parameter associated to this monomial is zero
is however a singular point in the moduli space.

We now turn to the calculation of the fundamental period
$\vp_0(\f_\a)$. Rewriting the defining polynomial in Eq.~\eqref{eDefP}\ as
$P = p -\f_0\prod_{i}x_i$, we first factor out the fundamental monomial
$$
  P~ = ~-\f_0 \prod_j x_j\>
  \left(1-{p\over{\f_0\prod_i x_i}}\right)~,
  \eqlabel{eExp}
$$
and then make a large-$\f_0$ expansion of Eq.~\eqref{ePiO}:
$$
  \vp_0(\f_\a)~
 = ~{C\over {(2\p i)^{N+1}}} \oint_\G \prod_{j=0}^N {\rd x_j \over x_j}\>
       \sum_{n=0}^{\infty} { p^n \over {\f_0^n \prod_i x_i^n} }~.
  \eqlabel{ePiEx}
$$
This is valid regardless of the smoothness (non-degeneracy) of $P_0$,
since the large-$\f_0$ expansion is equivalent to a small-${1\over\f_0}$
expansion around the very singular model defined by
$P_\infty={-}\f_0\prod_{i=0}^N x_i$. At the same time however, the
parameters $\f_\a$, $\a=0,\ldots,M-1$, parametrize a local neighborhood of
the zero-set of $P_0$, so the choice of $P_0$ defines the
$\f_\a$-parametrization. That is, different choices of $P_0$ provide
different $\f_\a$-coordinate patches on the moduli space.

To continue with the calculation started in \eqref{ePiEx}, focus now on
the expansion of
$$
  p^n ~ = ~ \left[\sum_{\ell=0}^N \prod_{i=0}^N x_i^{a_{i\ell}}~
     + ~\sum_{\a=1}^{M-1} \f_\a \prod_{i=0}^N x_i^{q_{i}^{\a}}\right]^n~.
  \eqlabel{eXXX}
$$
Using the multinomial formula,
$$
  \bigg(\sum_{j=1}^r X_j\bigg)^n~
 = ~\sum_{\scriptstyle{n_i} \atop \scriptstyle \sum n_i=n}
                   {n! \over {n_1! n_2!\cdots n_r!}}
                    X_1^{n_1} X_2^{n_2} \cdots X_r^{n_r}~,
  \eqlabel{eMulti}
$$
one finds an explicit expression for $p^n$,
$$ p^n~
   = ~\sum_{\scriptstyle {n_i,m_\a} \atop
            \scriptstyle \sum_\ell n_\ell+\sum_\a m_\a=n}\mkern-40mu
      \Big({n! \over {\prod n_i! \prod m_\a!}}\Big)
       \prod_{j=0}^N x_j^{(\sum_\ell a_{j\ell}n_\ell
                          + \sum_\a q_{j}^{\a}m_\a)}
        \Big(\prod_{\a=1}^{M-1} \f_\a^{m_\a}\Big)~.
  \eqlabel{eExP}
$$
Inserting~\eqref{eExP} back into the expression~\eqref{ePiEx} for
$\vp_0(\f_\a)$, each summand contains a product of $N{+}1$ integrals
$$
  \prod_{j=0}^N {1\over2\p i} \oint {\rd x_j \over x_j}~
    x_j^{(\sum_\ell a_{j\ell}n_\ell + \sum_\a q_{j}^{\a}m_\a - n)}~.
    \eqlabel{eXXX}
$$
This $N+1$-fold integral is nonzero precisely if the $N+1$ relations
$$
  \sum_\ell a_{j\ell} n_\ell + \sum_\a q_{j}^{\a}m_\a~ = ~n~,
              \qquad j=0,\ldots,N~, \eqlabel{eNCond}
$$
are satisfied and each integral just picks out the residue at $x_i=0$.
Multiplying~\eqref{eNCond} by $k_j$, summing over $j=0,\ldots,N$ and
using Eqs.~\eqref{eConsA}, \eqref{eConsQ}\ and \eqref{eCYcond}, we obtain
$\sum_i n_i +  \sum_\a m_\a~ =  n~$,
whence the condition imposed on $n_\ell,m_\a$ in the summations of
Eq.~\eqref{eExP}\ is automatically satisfied for \CY\ hypersurfaces.
Thus we find that the fundamental period is then given by by the
remarkably simple expression:
$$
  \vp_0(\f_\a) = \sum_{n_i,m_\a}
{\G(n{+}1) \over \prod_\a\G(m_\a{+}1) \prod_i\G(n_i{+}1)}\,
{\prod_{\a=1}^{M-1} {\f_\a}^{m_\a}\over \f_0^n}
  \eqlabel{ePiEXa}
$$
where
$$
  n = \sum_i n_i + \sum_\a m_\a~,  \eqlabel{eSum}
$$
and the sums over $n_i$ and $m_\a$ are restricted by Eq.~\eqref{eNCond},
which provides $N{+}1$ conditions for $N{+}M{+}1$ unknowns $n_i,
m_\a,n$. We are left with $M$ free summation variables.
There are of course a number of different ways of choosing which variables
to sum over. In what follows we will solve for the $n_i$.

We now would like to rewrite the above expression for $\vp_0$ in a form
which realizes mirror symmetry and which also
will turn out to be useful for the analytic continuation to small-$\f_0$ and
for obtaining the other periods.
The crucial observation here is that the summation variables $n_i$ and $n$
can be easily eliminated in \eqref{ePiEXa}, through \eqref{eNCond} and
\eqref{eSum}. The $n_i$ and $n$ may be expressed
in terms of the $M$ independent summation variables
 $\Tw{m}_\a, \a=1,\ldots, M{-}1$,
and a new variable $r$ by setting
$$ \matrix{
  n = \hat d r +  t{\cdot}\Tw{m}~,\qquad
  n_i= \hat k_i r + s_i{\cdot}\Tw{m}~,                  \cr
  \noalign{\vglue3mm}
  m_\a = p^\a{\Tw{m}}_\a  \quad(\hbox{no sum over}~\a)~,   \cr}
  \eqlabel{sumvar}
$$
where the dot-product denotes summation over $\a=1,\ldots,M{-}1$ and
where the coefficients $\hat d$, $\hat k_i$, $t^{\a}$ and $s_i{}^\a$
satisfy
$$\eqalignno{
       \sum_{j=0}^N \hat k_j a_{ij} &= \hat d~,\qquad
                       \hbox{for each}~i~,   \eqalignlabel{dhat}\cr
       \sum_{j=0}^N a_{ij} s_j{}^\a + q_i^\a p^\a &= t^\a~,\qquad
                       \hbox{for each $i$ and each $\a$},
                                            \eqalignlabel{relat}\cr}
$$
for the constraints \eqref{eNCond} to hold. Note also
that \eqref{eSum} implies
$$
 \hat d = \sum_{i=0}^N \hat k_i \qquad\hbox{and}\qquad
   t^\a= \sum_{i=0}^N s_i{}^\a + p^\a~.      \eqlabel{eSUM}
$$
Also, Eqs.~\eqref{sumvar} imply that the new variable $r$ and the
quantities $t{\cdot}\Tw{m}$ and $s_i{\cdot}\Tw{m}$, for all $i$, are
integers. The $p^\a$ are chosen so that all $t^\a$ and $s_i^\a$, and if
possible also all $\Tw{m}_\a$, are integers; we do not know if this is
always possible, but have not been able to find a counterexample.

{}From these relations we have that
$$
   \sum_{j=0}^N (a_{ij}-1) s_j{}^\a~ = ~p^\a(1 - q_i^\a)~,\qquad
                    \hbox{for each $i$ and each $\a$}. \eqlabel{eSia}
$$
Finally, with the above choice of $a_{ij}$, diagonal or as in~\eqref{eAij},
the matrix $[a_{ij}-1]$ is invertible in general\Footnote{Amusingly,
$[a_{ij}{-}1]$ does have one vanishing eigenvalue for $\IP^4[5]$, owing to
$k_i=1$, which means that the $s_i^\a$ can be determined only up to a
scaling factor per each $\a$---which still suffices our present need.}
and the system of equations~\eqref{eSia} can be solved for $s_i{}^\a$ in
terms of $p^\a, q_i^\a$ and $a_{ij}$. The $p^\a$ may then be chosen so as
to obtain integral $s_i{}^\a$'s. Thereafter, the $t^\a$ are determined
through~\eqref{eSUM}. As promised, Eqs.~\eqref{sumvar} then eliminate the
$N{+}1$ $n_i$'s and replace $n$ by $r$, leaving $M$ summation
variables $r, \tilde m_\a$, $\a=1,\ldots,M{-}1$.

With these substitutions, the expression for the fundamental
period~\eqref{ePiEXa} becomes
$$
  \vp_0 (\f_0,\f_\a) =
\sum_{r,m_\a}
  {\G(\hat dr{+}1{+}\sum_\a t^\a m_\a)
    \over \prod_{i=0}^N \G(\hat k_i r{+}1{+}\sum_\a s_i{}^\a m_\a)}\,
  {1\over\f_0^{\hat dr}}
  \prod_{\a=1}^{M-1} {{\vf_\a}^{m_\a}\over(p^\a m_\a)!}~;
   \eqlabel{PerM}
$$
here and hereafter, we drop the tilde over the $m$'s.
We have also defined the new variables
$$
   \vf_\a\define(\f_\a^{\>p^\a}/\f_0^{t^\a})~. \eqlabel{eNewPhi}
$$

Equation \eqref{PerM} is the main result of this section. It gives an
explicit form of the fundamental period for all models
defined as transverse hypersurfaces in weighted $\CP4$, once we have
solved equations \eqref{dhat} and \eqref{relat} for the coefficients
$\hat d$, $\hat k_i$, $t^\a$ and $s_i^\a$. This is straightforward in all
the cases. Most remarkably, upon rewriting the  condition~\eqref{dhat} as
$~\sum_j \hat k_j a_{ji}^T=\hat d~$, comparison with Eq.~\eqref{eConsA}
makes it clear that the $\hat k_i$ {\it are the weights associated to the
transposed polynomial $P_0^T$ of degree $\hat d$ that is
used to construct the mirror manifold\/} in the class of models
of Ref.~\cite{\rBH}.

We now illustrate this for the subclass of Fermat hypersurfaces. In this
case  the matrix $a_{ij}$
is diagonal $a_{ij}=\delta_{ij}d/k_i$  and $\hat d=d$,
$\hat k_i=k_i$. The independent summation variables are
$m_\a$ and $r$. We find,
$s_i^\a= k_i(t^\a-q_i^\a p^\a)/d$ and \eqref{sumvar}
become
$$
  n=d~r+Q,\qquad  n_i=r~k_i+{k_i\over d}(Q - Q_i)~,
  \eqlabel{sumvarf}
$$
where $Q \equiv m\cdot t^{\a}$,
$Q_i \equiv pm \cdot q_i$. The period then becomes
$$
  \vp_0(\f_{0}, \vf_\a)~ =\sum_{r,m_\a}~{(d r{+}Q)!~
   \prod_{\a}\vf_\a^{m_\a}\over
    \prod_{j=0}^N\big(r k_j{+}{k_j\over d} (Q{-}Q_j)\big) !~
     \prod_{\a} (p^\a m_\a !~(\f_{0})^{d r} }.
  \eqlabel{pert}
$$

The case of a single modulus corresponds to setting $M=0$ or equivalently
$m_\a =0$ in Eq.~\eqref{PerM}. In this situation we find that
Eq.~\eqref{PerM} simplifies further to
$$
   \vp_0(\f_0) =\sum_r {\G(\hat dr+1)
    \over {\prod_j \G(\hat k_j r+1)\, \f_0^{\hat dr}}}~. \eqlabel{pertu}
$$
This agrees with Batyrev's results obtained by means of toric geometry
\cite{\Baty}. It is  also in agreement with earlier studies restricted to
one-modulus models~\cite{{\rCdGP--\rKT}} and in particular with the
quintic discussed in the introduction.  Also, from \eqref{pert}
we can reproduce the periods for the two moduli Fermat examples studied in
Ref.~\cite\rCdFKM. Applications of Eq.~\eqref{PerM} in the case of
two-parameter models, including non-Fermat hypersurfaces, will be given in
Section~4.

\subsection{Convergence and analytic continuation: small $\vf_\a$}
As should be clear from the general expression~\eqref{PerM}, the
large-$\f_0$ series expansion of $\vp_0$ will be of the form of a
generalized hypergeometric series. The analytic continuation of such
series is studied by means of integral representations of
Mellin--Barnes type~
\REF\Erdelyi{A.~Erd\'elyi, F.~Oberhettinger, W.~Magnus and
         F.~G.~Tricomi, {\sl Higher Transcendental Functions, 3 Vols.},
         (McGraw--Hill, New York, 1953).}
\cite{{\Erdelyi,\rCdGP}}.
For the cases we consider this provides the analytic continuation of the
series~\eqref{PerM} into
the small-$\f_0$ regime.

\bigskip
\noindent$\underline{\hbox{\it The large-$\f_0$ expansion}}\>$:~~%
Let us start by writing \eqref{PerM} as
$$
  \vp_0 (\f_0,\vf_\a) = \sum_{m_\a} {\cal I}_m (\f_0)\>
   {\G(t{\cdot}m{+}1)\over {\prod_{i=0}^N \G(s_i{\cdot}m{+}1)}}
  \,\prod_{\a=1}^{M{-}1} {\vf_\a^{m_\a}\over {(p^\a m_\a) !}}~,
  \eqlabel{vp0Im}
$$
where
$$
  {\cal I}_m (\f_0)~ \define ~\sum_{r=0}^{\infty} \f_0^{-\hat d r}
     {(t{\cdot}m{+}1)_{\hat d r}\over
     {\prod_{i=0}^N(s_i{\cdot}m{+}1)_{\hat k_i r}}}~,\eqlabel{XXX}
$$
and we have used the Pochhammer's symbol
$$
  \big(a\big)_r~ \define ~{\G(a{+}r)\over\G(a)}~.
  \eqlabel{Poch}
$$
For sufficiently small $\vf_\a$, the series~\eqref{vp0Im} converges
if the function ${\cal I}_m(\f_0)$ does, and we now turn to study
${\cal I}_m(\f_0)$.

In fact, ${\cal I}_m (\f_0)$ is a generalized hypergeometric
function ${}_{\hat d{+}1}F_{\hat d}$. To see this, we will need
the multiplication formula for the
$\G$-function~\cite{\Erdelyi}(Vol.~I):
$$
  \G(kz)~ = ~(2\p)^{1-k\over2} k^{kz-{1\over2}}
                         \prod_{q=0}^{k-1}\G(z+{q\over k})~,
\eqlabel{eProdG}
$$
whereby
$$
  \G(kr{+}z)~ = ~k^{kr}\G(z)\prod_{q=0}^{k-1}\Big({z+q\over k}\Big)_r~,
   \qquad
 \big(z\big)_{kr}~ = ~k^{kr}\prod_{q=0}^{k-1}\Big({z+q\over k}\Big)_r~.
    \eqlabel{UnFold}
$$
Thus, we obtain
$$ \eqalign{
  {\cal I}_m (\f_0) &= \sum_{r=0}^{\infty} {z^{-r}\over r!}
   {{\G(r{+}1)
     \prod_{p=1}^{\hat d}
        \left({t{\cdot}m{+}p}\over{\hat d}\right)_r }\over
   {\prod_{i=0}^N \prod_{q_i=1}^{{\hat k}_i}
     \left({s_i{\cdot}m{+} q_i}\over{\hat k}_i\right)_r}}\cr
  &= {}_{{\hat d}{+}1}F_{\hat d}\left(
     \matrix{ a_1,\ldots,a_{{\hat d}{+}1}\cropen{3pt}
              c_0{}^1,\ldots,c_N^{{\hat k}_N}};~z^{-1}\right)
       ,\qquad |z|>1~,\cr}\eqlabel{Ibigz}
$$
where
$$ \eqalign{
   z &= \f_0^{\>\hat d}\, {\prod_{i=0}^N {\hat k}_i^{{\hat k}_i}
         \over {\hat d}^{\hat d}}~,\cr
   a_p &= {{t{\cdot}m + p} \over {\hat d}}~,\quad p=1,\ldots,{\hat d}~,
          \qquad {\rm and}~a_{{\hat d}{+}1} = 1~,\cr
   c_i{}^{q_i} &= {{s_i{\cdot}m + q_i}\over{\hat k}_i}~,
         \quad i=0,\ldots,N,\quad q_i = 1,\ldots,{\hat k}_i~.\cr}
  \eqlabel{IbigZ}
$$

The function ${\cal I}_m (\f_0)$ also has a Mellin--Barnes integral
representation
$$
  {\cal I}_m (\f_0) = - {1\over {2\p i}}\, {\prod_{i=0}^N
   \G(s_i{\cdot}m{+}1)\over\G(t{\cdot}m{+}1)}\,
    \int_\g\,\rd\r\, e^{i\p\r}\, \f_0^{-\hat d\r}\, {\p\over{\sin\p\r}}\,
  {\G(\hat d\r{+}t{\cdot}m{+}1)\over
      {\prod_{i=0}^N\G(\hat k_i \r{+}s_i{\cdot}m{+}1)}}~,
  \eqlabel{MelBar}
$$
where the contour $\g$ is parallel to the imaginary axis: $\r=-\e{+}iy$,
with $\e$ small, real and positive. This integral converges
{\it for all} $m$ as long as $0<\arg\f_0<{{2\pi}\over{\hat d}}$.
This can straightforwardly  be checked~\cite{\Erdelyi}(Vol. I, p.49) by using
$$
 \lim_{|y|\to\infty} |\G(x{+}iy)|=
    (2\p)^{1\over 2} |y|^{-{1\over 2}{+}x} e^{-{{\p |y|}\over 2}}~,
    \quad x~{\rm and}~y~{\rm real}~,\eqlabel{XXX}
$$
and taking $\r = x{+}iy$, we find that in the limit $|y|\to\infty$ the
integrand in~\eqref{MelBar}\ behaves as
$$
  \Big| e^{i\p\r}\, \f_0^{-\hat d\r}\, {\p\over{\sin\p\r}}\,
    {\G(\hat d\r{+}t{\cdot}m{+}1)\over
      \prod_{i=0}^N\G(\hat k_i \r{+}s_i{\cdot}m{+}1)}\Big|~
 \asymp ~{\k\over 2\p}\> |z|^{{-}x}\>
   |y|^{p{\cdot}m{-}{N\over2}}\> e^{-|y|[\p{\pm}(\p{-}{\hat d}\q)]}~,
  \eqlabel{XXX}
$$
where we use the notation~\eqref{IbigZ}, $\q\define\arg\f_0$ and
$$
  \k~\define~(2\p)^{4-N\over2}\bigg[{{\hat d}{}^{{1\over2}{+}t{\cdot}m}
   \over \prod_{i=0}^N {\hat k}_i{}^{{1\over2}{+}s_i{\cdot}m}}\bigg]~.
  \eqlabel{XXX}
$$

It is easy to see that the integral representation~\eqref{MelBar}\
reproduces the series expansion for the
fundamental period~\eqref{PerM}\ by closing
the contour of integration in~\eqref{MelBar}\ to the right
with a semicircle of infinite radius and summing over
the residues at the poles enclosed by this contour.  The only poles
that this contour encloses occur at $\r = r~,~r=0,1,2,\ldots,$ due to
the factor $1/{\sin\pi\rho}$ in the integrand. It remains to check, however,
that the integral over the infinite semicircle vanishes.
Using Stirling's formula
$$
  \G(z) \asymp (2\pi)^{\half}\, e^{-z}\, z^{z{-}\half} \eqlabel{Stirling}
$$
we find that the integrand on the semicircle has the following behaviour
as $|\r|\to\infty$
$$\eqalignno{
  \Big| e^{i\p\r}\, \f_0^{-\hat d\r}\, {\p\over{\sin\p\r}}\,
 &  {\G(\hat d\r{+}t\cdot m{+}1)\over
     {\prod_{i=0}^N\G(\hat k_i \r{+}s_i{\cdot}m{+}1)}}\Big|      \cr
 &\hskip12mm\asymp ~{\k\over{2\p}}\>
   |z|^{{-}|\r||\cos\chi}\> |\r|^{p{\cdot}m{-}{N\over2}}\>
      e^{-|\r||\sin\chi|[\p{\pm}(\p{-}{\hat d}\q)]}~,
                                          \eqalignlabel{infinIm} \cr}
$$
where $\chi=\arg\rho$, with $|\chi|<{\p\over2}$,
and again $0<\q<2\p/{\hat d}$.
Thus, the infinite semicircle contribution vanishes {\it for any} $m_\a$
as long as
$$
  |z|>1~,\qquad\hbox{that is},\qquad
  |\f_0|>{\hat d}\big(\prod_{i=0}^N
                      {\hat k_i}^{\hat k_i}\big)^{-1/{\hat d}}~,
  \eqlabel{BigZ}
$$
which is precisely the region of convergence of ${\cal I}_m$ in
\eqref{Ibigz}.

\bigskip
\noindent$\underline{\hbox{\it The small-$\f_0$ expansion}}\>$:~~%
By closing the contour $\g$ to the left instead, where $|\chi|>{\p\over2}$,
a series representation for the fundamental period is obtained
in the complementary region
$$
  |z|<1~, \qquad\hbox{that is},\qquad
  |\f_0|<{\hat d}\big(\prod_{i=0}^N
                      {\hat k_i}^{\hat k_i}\big)^{-1/{\hat d}}~.
  \eqlabel{smallz}
$$
In the region now enclosed by the contour, the integrand has
\item{1.} poles at $\r = -(t{\cdot}m{+}S)/{\hat d}$, $S=1,2,\ldots$,
    from $\G({\hat d}\r{+}t{\cdot}m+1)$;
\item{2.} poles at $\r = -R$, $R=1,2,\ldots$, from $1/\sin\p\r$;
\item{3.} zeros at $\r = -(s_i{\cdot}m{+}S_i)/{\hat k_i}$,
    $S_i=1,2,\ldots$, from $\G({\hat k_i}\r{+}s_i{\cdot}m+1)$,
    for each $i=0,\ldots,N$.

\noindent
We recall that $t{\cdot}m, s_i{\cdot}m \in \ZZ$, $i=0,\ldots,N$ by
virtue of their definition~\eqref{sumvar}. Therefore, the poles in item~1.\
include all integral $\r\leq{-}(t{\cdot}m+1)/{\hat d}$, which coincide
with poles from item~2.; these will then be double poles unless canceled
by the zeros from item~3. Indeed, the zeros in
item~3.\ include all integral $\r\leq{-}(s_i{\cdot}m+1)/{\hat k_i}$,
$i=0,\ldots,N$.
Thus, when $\r\in\ZZ$ and $\r\leq{-}(s_i{\cdot}m+1)/{\hat k_i}$ for one or
two $i$'s, integrand in~\eqref{MelBar} has simple poles or is analytic,
respectively, rather than having double poles.

 This possible appearance of double poles is in fact a standard
characteristic of hypergeometric functions~\cite{\Erdelyi}
and could be anticipated already from the form~\eqref{Ibigz}:
since $t{\cdot}m\in\ZZ$, one of $a_p$ in~\eqref{IbigZ} must differ from
$a_{\hat d+1}=1$ by an integer.
 Similarly, their (partial) cancellation is indicated by some of the
$c_i{}^{q_i}$'s being integral---which happens once for every
$i=0,\ldots,N$ because $s_i{\cdot}m$ are all integers.

\midinsert
\vskip10pt
\boxit{0pt}{\vbox{\parskip=0pt
 \vglue2in\hrule height0pt depth0pt width\hsize %for the outer box
  %
  % the coordinate system, label and data
 \place{.23}{-.33}{\vrule height.3pt width5in}
 \place{4.66}{2.03}{\vrule height3.2in width.3pt}
 \place{5.73}{2.0}{$\r$-plane}
 \place{4.8}{1.4}{In this illustration:}
 \place{4.9}{1.1}{${\hat d}=11~,\quad{\hat k_0}=2$,}
 \place{4.9}{0.8}{$t{\cdot}m=7~,~s_0{\cdot}m=2$,}
 \place{4.9}{0.55}{for some fixed $m_\a$.}
  %
  % the vertical part of the contour
 \place{4.55}{2.13}{\vrule height3.4in width.8pt}
 \place{3.28}{2.0}{the vertical part}
 \place{3.44}{1.8}{of the contour}
  %
  % the poles of $\G({\hat d}\r + t{\cdot}m + 1)$:
 \place{.5}{1.8}{$\bullet$~: poles of $\G({\hat d}\r+t{\cdot}m+1)$}
 \place{0.42}{-.3}{$\bullet$}
 \place{0.54727}{-.3}{$\bullet$}
 \place{0.67455}{-.3}{$\bullet$}
 \place{0.80182}{-.3}{$\bullet$}
 \place{0.92909}{-.3}{$\bullet$}
 \place{1.05636}{-.3}{$\bullet$}
 \place{1.18364}{-.3}{$\bullet$}
 \place{1.31091}{-.3}{$\bullet$}
 \place{1.43818}{-.3}{$\bullet$}
 \place{1.56545}{-.3}{$\bullet$}
 \place{1.69273}{-.3}{$\bullet$}
 \place{1.82}{-.3}{$\bullet$}
 \place{1.94727}{-.3}{$\bullet$}
 \place{2.07455}{-.3}{$\bullet$}
 \place{2.20182}{-.3}{$\bullet$}
 \place{2.32909}{-.3}{$\bullet$}
 \place{2.45636}{-.3}{$\bullet$}
 \place{2.58364}{-.3}{$\bullet$}
 \place{2.71091}{-.3}{$\bullet$}
 \place{2.83818}{-.3}{$\bullet$}
 \place{2.96545}{-.3}{$\bullet$}
 \place{3.09273}{-.3}{$\bullet$}
 \place{3.22}{-.3}{$\bullet$}
 \place{3.34727}{-.3}{$\bullet$}
 \place{3.47455}{-.3}{$\bullet$}
 \place{3.60182}{-.3}{$\bullet$}
 \place{3.585}{.05}{$\Big\downarrow$}
 \place{3.215}{.5}{$\ddd-{t{\cdot}m+1\over{\hat d}}$}
  %
  % the poles of $1/\sin\p\r$:
 \place{.475}{1.5}{$\odot$~: poles of $1/\sin\p\r$}
 \place{0.4}{-.275}{$\odot$} \place{0.35}{-.5}{$-3$}
 \place{1.8}{-.275}{$\odot$} \place{1.75}{-.5}{$-2$}
 \place{3.2}{-.275}{$\odot$} \place{3.15}{-.5}{$-1$}
 \place{3.21}{-.7}{$\Big\uparrow$}
 \place{2}{-1.1}{the only double pole}
 \place{4.6}{-.275}{$\odot$} \place{4.75}{-.5}{$0$}
 \place{6.0}{-.275}{$\odot$} \place{5.95}{-.5}{$+1$}
  %
  % the zeros of $\G({\hat k_0}\r + s_0{\cdot}m + 1)$:
 \place{.422}{1.28}{{\titlesy\char"02}~:
                 zeros of $\G({\hat k_0}\r+s_0{\cdot}m+1)$}
 \place{0.355}{-.235}{\titlesy\char"02}
 \place{1.055}{-.235}{\titlesy\char"02}
 \place{1.755}{-.235}{\titlesy\char"02}
 \place{2.455}{-.235}{\titlesy\char"02}
 \place{2.51}{.05}{$\Big\downarrow$}
 \place{2.1}{.5}{$\ddd-{s_0{\cdot}m+1\over{\hat k_0}}$}
 \vglue2truein
}}
\vskip10pt
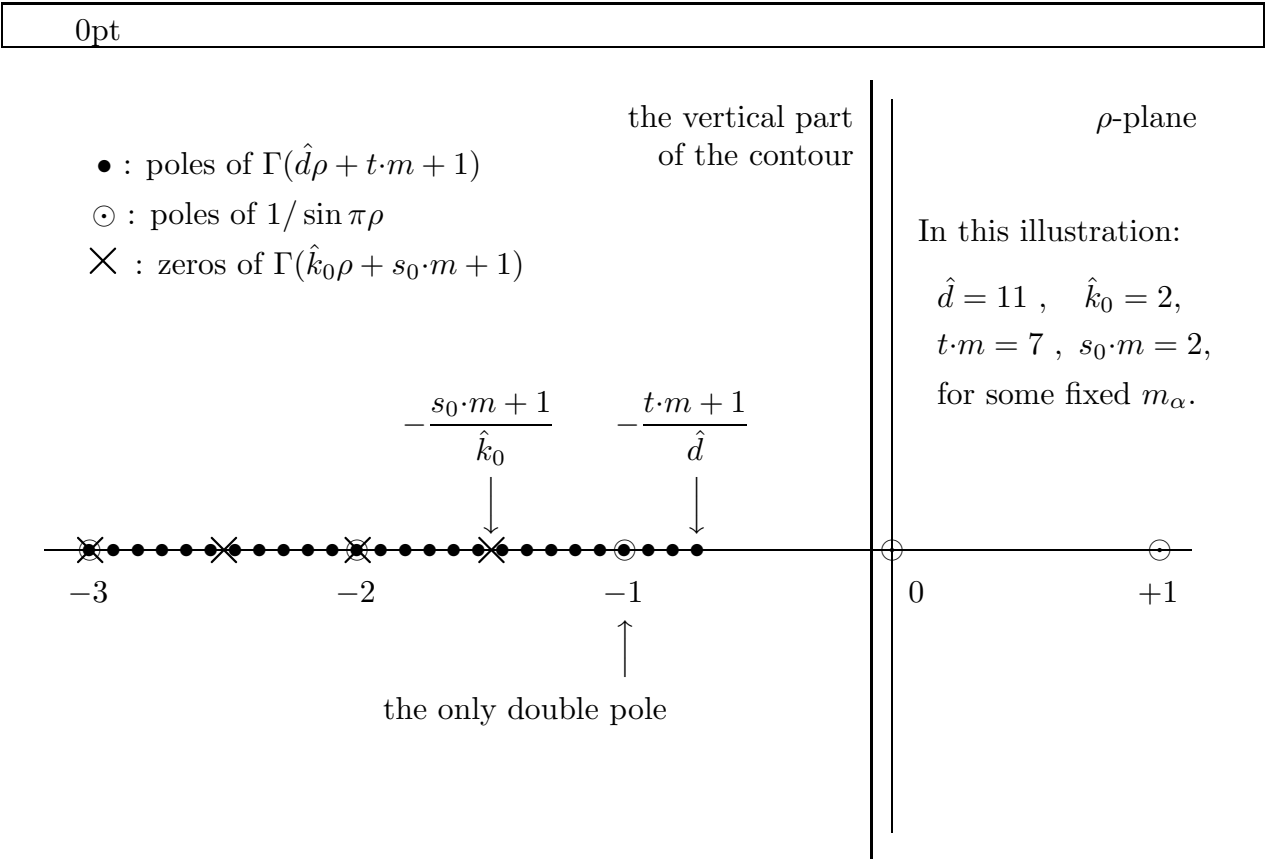
\centerline{{\bf Figure~1.}~~A sketch of a possible pole structure for
${\cal I}_m(\f_0)$ in Eq.~\eqref{MelBar}.}
\vskip20pt
\endinsert

This dependence of the pole structure on the relative values of $t{\cdot}m$
and $s_i{\cdot}m$ prevents us from writing a general expression in closed form
for the small-$\f_0$ expansion of $\vp_0(\f_0,\vf_\a)$. The following brief
discussion should help a case by case analysis; recall that
Eqs.~\eqref{eSia} determine the $s_i^\a$, whereupon
Eqs.~\eqref{eSUM} determine the $t^\a$.

{}From the items~1.--3.\ above, the following is clear:
\nobreak\vglue3mm
\vbox{\narrower\narrower\parskip0pt\noindent
      For fixed integers $m_\a\geq0$ and $R>0$, the extra pole at $\r=-R$:
\item{a.} remains uncancelled if
      $(1+s_i{\cdot}m-{\hat k}_iR)>0$ for all $i=0,\ldots,N$, and
\item{b.} is a double pole if also $(1+t{\cdot}m-{\hat d}R)\leq0$.}
\bigskip

The condition~a.\ implies the ``sum rule''
$$
   \sum_{i=0}^N s_i{\cdot}m + N + 1~ > ~{\hat d}R~,
    \eqlabel{sumRule}
$$
for the pole from $1/\sin\p\r$ at $\r=-R$ to remain.
Combined with condition~b., this yields
$$
   \sum_{i=0}^N s_i{\cdot}m + N~ > ~t{\cdot}m~,\qquad
   \hbox{for double poles to occur.} \eqlabel{SumRule}
$$
Finally, using the second of Eqs.~\eqref{eSUM}, we obtain
$$
   N~ > ~p{\cdot}m~,\qquad
   \hbox{for double poles to occur.} \eqlabel{SumRuleP}
$$
Therefore, there are at most a few---if any---double poles.
Further conditions may be derived from Eqs.~\eqref{sumvar} by recalling
that $n, n_i, m_\a\geq0$, but these do not appear very helpful in the
general case.

To describe the small-$\f_0$ evaluation of the integral~\eqref{MelBar},
focus first on the nonintegral-$\r$ poles:
$$ {\eqalign{ \hbox{simple poles}~:~
 & \r^\sharp = {-}(R{+}{r\over\hat d})~,\qquad R=0,1,2\ldots~, \cr
   \noalign{\vglue3mm}
 & r=t{\cdot}m,\, t{\cdot}m{+}1,\ldots, t{\cdot}m{+}{\hat d}{-}1,\quad
         r\neq0\,({\rm mod}\,\hat d)~, \cr}}
 \eqlabel{simpleP}
$$
which stem solely from $\G(\hat d\r{+}t{\cdot}m{+}1)$. This
`rational-$\r$' part of ${\cal I}_m$ is straightforwardly\vadjust{\newpage}
obtained
$$ {\eqalign{
   {\cal I}^{(Q)}_m(\f_0) = -{(-1)^{p{\cdot}m}\over\hat d\p^N}
   {\prod_{i=0}^N \G(s_i{\cdot}m{+}1)\over\G(t{\cdot}m{+}1)}
   \sum_{{\SSS r=t{\cdot}m\atop\SSS r\neq0({\rm mod}\,\hat d)}}
       ^{t{\cdot}m{+}\hat d{-}1}
 & e^{{-}{i\p r\over\hat d}}
        {\prod_{j=0}^N \sin({\hat k_jr\p\over\hat d})
         \over \sin({\p r\over\hat d})}(-\f_0)^r            \cropen{5pt}
 & \mkern-60mu\times\sum_{R=0}^\infty
   {\prod_{j=0}^N \G(\hat k_j(R{+}{r\over\hat d}){-}s_j{\cdot}m)
     \over\G(\hat dR{+}r{-}t{\cdot}m)}\>\f_0^{\hat dR}~.    \cr}}
   \eqlabel{ImQ}
$$
Correspondingly,
$$ {\eqalign{
   \vp_0^{(Q)} = -{1\over\hat d\p^N} \sum_{m_\a} (-1)^{p\cdot m}
   \sum_{{\SSS r=t{\cdot}m\atop\SSS r\neq0({\rm mod}\,\hat d)}}
       ^{t{\cdot}m{+}\hat d{-}1}
 & e^{{-}{i\p r\over\hat d}}
        {\prod_{j=0}^N \sin({\hat k_jr\p\over\hat d})
         \over \sin({\p r\over\hat d})}(-\f_0)^r              \cropen{5pt}
 & \mkern-90mu\times\sum_{R=0}^\infty
   {\prod_{j=0}^N \G(\hat k_j(R{+}{r\over\hat d}){-}s_j{\cdot}m)
     \over\G(\hat dR{+}r{-}t{\cdot}m)}\>\f_0^{\hat dR}
     \prod_{\a=1}^{M{-}1}{\vf_\a^{~m_\a}\over(p^\a m_\a)!}~. \cr}}
   \eqlabel{vp0Q}
$$

Now, to account for the possible integral-$\r$ poles, let
$$
  R_{D}~ \define ~{-}\Big[{-}{t{\cdot}m\over\hat d}\Big]~,
  \eqlabel{R2}
$$
mark the onset of candidate double poles, where $[x]$ denotes the largest
integer not bigger than $x$. Also, let
$R_1$ and $R_0$, respectively, denote the smallest and the second-smallest
integer among $\big\{{-}\big[{-}{s_i{\cdot}m\over\hat k_i}\big]\big\}$.
These mark, respectively, the onset of the first and the second band of
zeros; for $\r\leq{-}R_0$, all integral-$\r$ poles are canceled.
Clearly, $R_0 \geq R_1$. In addition, assume that
$$
   {-}R_0~ \leq ~{-}R_1~ \leq ~{-}R_D~ \leq ~{-}1~, \eqlabel{Assume}
$$
so that the integral-$\r$ residue contributions to~\eqref{MelBar} come
from simple poles for ${-}R_0<\r\leq{-}R_1$ and ${-}R_D<\r\leq{-}1$, but
double poles for ${-}R_1<\r\leq{-}R_D$. Straightforwardly then, we can
write
$$
  \vp_0(\f_0,\vf_\a)~ = ~\vp_0^{(Q)} + \vp_0^{(Z)}~,\qquad
  \vp_0^{(Z)}~        = ~\vp_0^{(1)} + \vp_0^{(2)} + \vp_0^{(3)}~,
  \eqlabel{WholeHog}
$$

\vbox{
\noindent
with
$$ \eqalignno{
 \vp_0^{(1)}
 &= -{1\over \hat d}\sum_{m_\a}
     \sum_{R=1}^{R_D{-}1} {\G(1{+}t{\cdot}m{-}\hat dR) \over
      \prod_{j=0}^N \G(1{+}s_j{\cdot}m{-}\hat k_jR)}\>\f_0^{\hat dR}
       \prod_{\a=1}^{M{-}1}{\vf_\a^{~m_\a}\over(p^\a m_\a)!}~,
                                                  \eqalignlabel{vp0Z1}\cr
\noalign{\vglue3mm}
 \vp_0^{(2)}
 &= -{1\over \hat d}\sum_{m_\a}
     \sum_{R=R_D}^{R_1{-}1} { (-1)^{t{\cdot}m{+}1} (-\f_0)^{\hat dR}
      \over \G(\hat dR -t\cdot m)
       \prod_{j=0}^N \G(1{+}s_j{\cdot}m{-}\hat k_jR)}
        \prod_{\a=1}^{M{-}1}{\vf_\a^{~m_\a}\over(p^\a m_\a)!}          \cr
 &\hskip15mm
   \times\Big[ \ln[{-}\f_0^{{-}\hat d}] + \hat d\J(\hat dR{-}t{\cdot}m)
              -\sum_{j=0}^N \hat k_j\J({-}\hat k_jR{+}s_j{\cdot}m{+}1)\Big]~,
                                                  \eqalignlabel{vp0Z2}\cr
\noalign{\vglue3mm}
 \vp_0^{(3)}
 &= -{1\over \hat d}\sum_{m_\a}
     \sum_{R=R_1}^{R_0{-}1} {\G(1{+}t{\cdot}m{-}\hat dR) \over
      \prod_{j=0}^N \G(1{+}s_j{\cdot}m{-}\hat k_jR)}\>\f_0^{\hat dR}
       \prod_{\a=1}^{M{-}1}{\vf_\a^{~m_\a}\over(p^\a m_\a)!}~,
                                                  \eqalignlabel{vp0Z3}\cr}
$$
}
\noindent where $\J(z)$ is the logarithmic derivative of the $\G$-function,
and $\vp_0^{(Q)}$ was given in~\eqref{vp0Q}. Clearly, sums with an inverted
range equal zero. For example, if $R_1=R_0$, then $\vp_0^{(3)}=0$.

When there is only one parameter the formula is very simple since, in this
case, there are no poles at all at $\r=-R$.  We obtain
\Ref\rBK{P.~Berglund and S.~Katz, in preparation.}
$$
  \vp_0(\f_0) = -{1\over {{\hat d}\p^N}}\sum_{r=1}^{{\hat d}{-}1}
    e^{{-}{\p ir\over{\hat d}}}\, ({-}\f_0)^r \,
    {{\prod_{j=0}^N \sin{{{\hat k}_jr\p}\over {\hat d}}}
        \over{\sin{{r\p}\over {\hat d}}}}\,
    \sum_{R=0}^{\infty} \f_0^{{\hat d}R}\,
     {{\prod_{j=0}^N
        \G\left({\hat k}_j\big(R{+}{r\over{\hat d}}\big)\right)}
         \over \G({\hat d}R{+}r)}~,   \eqlabel{onem}
$$
Using Eqs.~\eqref{UnFold}, this can now be written
in terms of a sum of generalized hypergeometric functions
 ${}_{{\hat d}} F_{{\hat d}-1}$. We obtain~\cite{\rBK}
$${\eqalign{
\vp_0(\f_0) = -{1\over {{\hat d}\p^N}}\sum_{r=1}^{{\hat d}{-}1}
    e^{{-}{\p ir\over{\hat d}}}\, ({-}\f_0)^r \,&
  {{\prod_{j=0}^N \sin{{{\hat k}_jr\p}\over {\hat d}}\,
           \G\left({{{\hat k}_jr}\over{\hat d}}\right)}
        \over{\sin{{r\p}\over {\hat d}}\, \G(r)}}\cr
 \noalign{\vglue3mm}
 &  \times {}_{{\hat d}}F_{\hat d{-}1}\left(\matrix{
      \smash{\hbox{$\overbrace{\ttt{r\over\hat d},\ldots,
                                   {r\over\hat d}}^{N+1~\rm times}$}},
                 a_0^1,\ldots, a_N^{{\hat k}_N{-}1}\cropen{3pt}
             {r{+}1\over{\hat d}},{{r{+}2}\over{\hat d}}
          \ldots,{{r{+}{\hat d}{-}1}\over{\hat d}}\cr};
  z~\right)~,\cr}}
  \eqlabel{XXX}
$$
where
$$
 a_i^{q_i} = \left( {r\over{\hat d}}{+}{q_i\over{\hat k}_i}\right)~,
 \quad i=0,\ldots,N~,\qquad q_i=0,\ldots,{\hat k}_i{-}1~. \eqlabel{XXX}
$$

So far, through \eqref{MelBar}, we are able to find a series
representation for the fundamental period for any value of the modulus
$\f_0$. The series converges for sufficiently small values of the
remaining moduli $\varphi_\a$.
The radius of convergence however, is determined by the values of these
parameters for which the \cym\ becomes singular. It is difficult to provide
any detail about this in general and precise information seems to require
study on a case by case basis.

\subsection{Convergence and analytic continuation: general $\vf_\a$}
Looking back at Eq.~\eqref{vp0Im}, it is clear that $\vp_0(\f_0,\vf_\a)$
will converge for small $\vf_\a$ if ${\cal I}_m(\f_0)$ does not grow too
fast as a function of $m_\a$. For particular cases, it may be even
possible to use known results on the behaviour of generalized
hypergeometric functions on their parameters. In general, however, this
approach does not seem to be fruitful.

We thus rewrite the fundamental period as
$$
  \vp_0~ = ~\sum_r {\f_0^{{-}{\hat d}r}\over r!}\,
                   {\G(r{+}1)\G(\hat dr{+}1)\over
                     \prod_{i=0}^N\G(\hat k_ir{+}1)}\>{\cal W}_r(\vf)~,
  \eqlabel{vp0Wr}
$$
where
$$
  {\cal W}_r(\vf)~ = ~\sum_{m_\a} {(\hat dr{+}1)_{t{\cdot}m}\over
                       \prod_{i=0}^N (\hat k_ir{+}1)_{s_i{\cdot}m}}
                       \prod_\a{\vf_\a^{\>m_\a}\over (p^\a m_\a)!}~.
  \eqlabel{Wr}
$$
The Mellin--Barnes integral representation of this is
$$
  \vp_0~ = ~{-}{1\over2\p i} \int_\g \rd\r~
   \f_0^{{-}\hat d\r}\, e^{i\p\r}\, {\p\over\sin\p\r}\, \G(\hat d\r{+}1)\,
            {{\cal W}_\r(\vf)\over\prod_{i=0}^N \G(\hat k_i\r{+}1)}~,
   \eqlabel{MelBarX}
$$
where the contour $\g$ is chosen as before, parallel to the imaginary
axis: $\r={-}\e{+}iy$, with $\e$ small, real and positive. In view of the
previous analysis, this integral converges for small $\vf_\a$ as long as
$0<\arg\f_0<{2\p\over\hat d}$.

The pole structure of the integrand in~\eqref{MelBarX} is of course the
same as the one in~\eqref{MelBar}.
 For $\Re e(\r)>{-}\e$, the only poles stem from $1/\sin\p\r$ and are
located at $\r=r$, $r=0,1,2,\ldots$.
 For $\r={-}1, {-}2,\ldots$, the poles stem both from
$\G(\hat d\r{+}1)$ and from $1/\sin\p\r$, and are double poles unless
canceled by the zeros of ${\cal W}_\r/\prod_i\G(\hat k_ir{+}1)$.
For $\r={-}(\hat dR{+}r)$,
with $R=0, 1,\ldots$ and $r=1,2,\ldots,(\hat d{-}1)$,
$\G(\hat d\r{+}1)$ has single poles.
 Indeed, although ${\cal W}_\r$ does have poles, the function
${\cal W}_\r/\prod_i\G(\hat k_ir{+}1)$ has {\it no poles whatsoever}.
This is easy to see from the fact that the Pochhammer symbol $(a)_n$ has no
poles for $n>0$ and vanishes when $a=0,-1,-2,\ldots$ but also $n{+}a>0$.
 Straightforwardly:
$$
  \hbox{zeros of~}{{\cal W}_\r\over\prod_i\G(\hat k_ir{+}1)}~:\quad
  \cases{\r={-}(Z{+}1)/\hat d   & $Z=0,1,2,\ldots,(t{\cdot}m{-}1)$, \cr
         \noalign{\vglue2mm}
         \r={-}(Z_i{+}s_i{\cdot}m{+}1)/\hat k_i
                                & $Z_i=0,1,2,\ldots$~,             \cr}
  \eqlabel{zeroWr}
$$
whence the same pole structure is recovered as in~\eqref{MelBar}.

Recall that the discussion in the previous section was done for small $\vf_\a$;
in fact to zeroth order in $\vf_\a$. In particular, the radius of convergence
only gives a condition on $\f_0$ which does not depend on the $\vf_\a$. In
order to find the radius of convergence of the Mellin-Barnes integral for
general $\vf_\a$ we need to know the asymptotic behaviour of $|W_\r|$ as
$|\r|\to\infty$; to close the integral in \eqref{MelBarX} to the left or right
we have to make sure that the integral over the semicircle vanishes as the
radius goes to infinity.

Unfortunately, it is  difficult to determine the
asymptotic behaviour of ${\cal W}_\r(\vf)$ as $|\r|\to\infty$.
We note however that the second formula in~\eqref{UnFold}, combined with
iterates of the identity
$$
  (a)_{m{+}n}~ = ~(a{+}m)_n\> (a)_m~, \eqlabel{XXX}
$$
allow the representation of ${\cal W}_\r(\vf)$ as
$$
  {\cal W}_\r(\vf_\a) = \sum_{m_\a} \prod_{\a=1}^{M-1}
   {\ddd \prod_{\ell_\a=0}^{t^\a-1}
     \Big({\hat d r{+}1{+}\sum_{\b=1}^{\a-1} t^\b m_\b{+}\ell_\a
            \over t^\a} \Big)_{m_\a} \over
    \ddd \prod_{\ell_\a=0}^{s_i^\a-1} \prod_{i=0}^N
      \Big({\hat k_i r{+}1{+}\sum_{\b=1}^{\a-1}s_i^\b m_\b{+}\ell_\a
             \over s_i^\a }\Big)_{m_\a}}\,
         {1\over\ddd\prod_{l_\a=1}^{p^\a{-}1}\left({l_\a\over
                            p^\a}\right)_{m_\a}}
         {\z_\a^{~m_\a}\over m_\a!}~,
  \eqlabel{ExpoWr}
$$
where
$$
  \z_\a~ \define ~{(t^\a)^{t^\a}\>\vf_\a \over (p^\a)^{p^\a}
                          \prod_{i=0}^N (s_i^\a)^{s_i^\a}}
  \eqlabel{newnewphi}
$$
are suitably rescaled variables. The admittedly bewildering
representation~\eqref{ExpoWr} of ${\cal W}_\r(\vf)$ however has the virtue
of making it clear that ${\cal W}_\r(\vf)$ is a multiple generalized
hypergeometric function.
The standard techniques~[Luke,~I:VII]~
 \Ref\rLuke{Y.L.~Luke, {\sl The Special Functions and Their
         Approximations, 2 Vols.},\Z (Academic Press, New York, 1969).}\
may thus be developed to study the asymptotic behaviour of
 ${\cal W}_\r(\vf)$ as $|\r|\to\infty$, but this is beyond
our present scope.
Here, we simply assume that the relevant integrals do converge and forego
determining the general, $\vf_\a$-dependent radii of convergence for the
large-$\f_0$ and small-$\f_0$ expansions.

There being only simple poles at $\r=r=0,1,2,\ldots$ when the contour
in~\eqref{MelBarX} is closed to the right, it is straightforward to verify
that the integral representation of $\vp_0(\f_0,\vf_\a)$ indeed reproduces
the small-$\f_0$ expansion~\eqref{vp0Wr}.

As for ${\cal I}_m(\f_0)$, the evaluation by residues of $\vp_0$
in~\eqref{MelBarX} may be separated into sums over nonintegral-$\r$:
$$
   \hbox{simple poles}~:~ \r^\sharp = {-}(R{+}{r\over\hat d})~,\qquad
    R=0,1,2\ldots~,\quad r=1, 2,\ldots, {\hat d}{-}1~,
 \eqlabel{SimpleP}
$$
and integral-$\r$ residues. The latter are again finite in number and
contain a possible interval of double poles, as indicated in~\eqref{Assume}.
The parametrization~\eqref{SimpleP} clearly avoids integral-$\r$, but
unlike in~\eqref{simpleP}, does not manifest that the poles are canceled
out down to $\r=-(t{\cdot}m)/\hat d$ by the zeros~\eqref{zeroWr}.
In this representation, we have
$$ \eqalignno{
   \vp_0^{(Q)}
 &= -{1\over\hat d\p^N} \sum_{r=1}^{\hat d{-}1} e^{{-}{i\p r\over\hat d}}
        {\prod_{j=0}^N \sin({\hat k_jr\p\over\hat d})
         \over \sin({\p r\over\hat d})}\>(-\f_0)^r                     \cr
 & \hskip25mm\times\sum_{R=0}^\infty
   {\prod_{j=0}^N \G(\hat k_j(R{+}{r\over\hat d}))
     \over\G(\hat dR{+}r)}\>\f_0^{\hat dR}\>
      {\cal W}_{-(R{+}{r\over\hat d})}(\vf)~,     \eqalignlabel{vp0WrQ}\cr
\noalign{\vglue3mm}
 \vp_0^{(1)}
 &= -{1\over \hat d}\sum_{R=1}^{R_D{-}1}
      \bigg[{\G(1{-}\hat dR)\> {\cal W}_{{-}R}(\vf)\over
       \prod_{j=0}^N \G(1{-}\hat k_jR)}\bigg]\>\f_0^{\hat dR}~,
                                                 \eqalignlabel{vp0WrZ1}\cr
\noalign{\vglue3mm}
 \vp_0^{(2)}
 &= +{1\over \hat d}\sum_{R=R_D}^{R_1{-}1} {{\cal W}_{{-}R}(\vf)\>
(-\f_0)^{\hat dR}
      \over \G(\hat dR) \prod_{j=0}^N \G(1{-}\hat k_jR)}               \cr
 &\hskip15mm
   \times\Big[\ln[{-}\f_0^{{-}\hat d}] + \hat d\J(\hat dR)
              -\sum_{j=0}^N \hat k_j\J(1-\hat k_jR) + w_{{-}R}(\vf)\Big]~,
                                                 \eqalignlabel{vp0WrZ2}\cr
\noalign{\vglue3mm}
 \vp_0^{(3)}
 &= -{1\over \hat d}\sum_{R=R_1}^{R_0{-}1}
      \bigg[{\G(1{-}\hat dR)\> {\cal W}_{{-}R}(\vf)\over
       \prod_{j=0}^N \G(1{-}\hat k_jR)}\bigg]\>\f_0^{\hat dR}~,
                                                 \eqalignlabel{vp0WrZ3}\cr}
$$
where
$$
  w_\r(\vf)~ \define
           ~{1\over{\cal W}_\r(\vf)}{\rd\over\rd\r}{\cal W}_\r(\vf)~.
$$
Despite appearances, the quantities in square brackets in~\eqref{vp0WrZ1}
and~\eqref{vp0WrZ3} are finite owing to the zeros~\eqref{zeroWr}.

\subsection{The remaining periods}
The analytic continuation of the fundamental period obtained
from \eqref{MelBar}\ very often suffices to find the other periods.
Let $\cal M$ denote the zero set of the defining polynomial~\eqref{eDefP}.
The $\f_\a$, $\a=0,\ldots,M{-}1$ parametrize the space of complex
structures for $\cal M$ centered around the reference polynomial $P_0$.
This reference manifold typically enjoys some symmetries
and those which act diagonally on the coordinates play a special
r\^ole~\cite{\rBriRon}; denote them by $G_{\cal M}$. Clearly, any two
deformations of the reference manifold which are related by a $G_{\cal M}$
transformation must be identified. Thus, there is a (lifted) action,
$\cal A$, of $G_{\cal M}$ on the parameter space, the proper moduli space
must be (at least) an $\cal A$-quotient of the parameter space and the
point represented by $P_0$ becomes an $\cal A$-orbifold point.  Although
the full modular group is in general not known, $\cal A$ is clearly
at least a subgroup and acts on the parameter space
$$
  {\cal A}~:~~ \f_\a \longmapsto \l_i^{a^i_\a} \f_\a\,\qquad
                \a=0,1,\ldots,M{-}1,\quad \l^{\hat d}=1,\qquad i=1,\ldots,r~,
\eqlabel{ModulG} $$
where ${\cal A}\sim \ZZ_{d_1}\times\ldots\times\ZZ_{d_r}$. In particular we
have  $d_1=\hat d$ where $\hat d$ is the degree of the transposed polynomial,
a quotient of which gives the mirror ${\cal W}$ of ${\cal M}$.
 We then define the set of periods $\vp_{j_1\ldots j_r}$ as
$$
 \vp_{j_1\ldots j_r}(\f_\a) \define
\vp_0(\l_1^{j\,a^1_\a}\f_\a,\ldots,\l_r^{j\,a^r_\a}\f_\a)~.\eqlabel{othervp}
$$

Assume first that $\vp_0^{(Z)}=0$; that is, only residues at
nonintegral-$\r$ contribute to the small-$\f_0$ expansion of $\vp_0$.
Since $\vp_0^{(Q)}$ does not contain $\hat d$'th powers of $\f_0$,
it follows that
$$
   \sum_{j=1}^{\hat d} \vp_0^{(Q)}(\l_1^j\f_0, \vf_\a)~ =
  ~\sum_{j=1}^{\hat d} \vp_0(\l_1^j\f_0, \vf_\a)~~ = ~0~.
   \eqlabel{relall}
$$
Thus, at most $(\sum_i d_i){-}1$ of the $\sum_i d_i$ periods
are linearly independent.
However, if at least one of $\vp_0^{(1)}, \vp_0^{(3)}$ is also nonzero,
this no longer holds. Clearly, $\vp_0^{(1)}$ and $\vp_0^{(3)}$ are
functions of $\f_0^{\hat d}$, rather than $\f_0$ and are consequently
invariant under the $\ZZ_{\hat d}$ action $\f_0\mapsto\l_1\f_0$, with
$\l_1^{\hat d}=1$. Then
$$
  \sum_{j=1}^{\hat d} \vp_0(\l_1^j\f_0, \vf_\a)~ =
  ~{\hat d}(\vp_0^{(1)} + \vp_0^{(3)})~ \neq ~0~.
  \eqlabel{noRel}
$$
Finally, because of the logarithmic term in~\eqref{vp0Z2}
and~\eqref{vp0WrZ2}, the presence of $\vp_0^{(2)}\neq0$, \ie\ the contribution
of double poles in~\eqref{MelBar} and~\eqref{MelBarX}, would change the
behaviour of $\vp_0(\f_0,\vf_\a)$ substantially.

We again assume that $\vp_0^{(Z)}=0$, so that $\vp_0=\vp_0^{(Q)}$.
Given the explicit factors of $\sin({\hat k_jr\p\over\hat d})$ in
Eqs.~\eqref{vp0Q} and~\eqref{vp0WrQ}, it should be clear that the terms
with $\hat k_jr = 0\,({\rm mod}\,\hat d)$ are also absent. This implies
additional conditions on the periods, corresponding to a modular
group action
$$
  (\f_0,\f_\a) \longmapsto
  (\l^{\hat k_j}\f_0,\l^{\hat k_j t^\a} \f_\a)~.
  \eqlabel{MoreModG}
$$
Of course, further relations may exist in specific models owing to
accidental cancellations.

With this we can now obtain a collection of complex structure periods
$\vp_j$ for general Calabi-Yau hypersurfaces in weighted projective spaces
and so also for the
corresponding Landau-Ginzburg models. The action of $\cal A$ in
Eq.~\eqref{othervp} produces a complete set of such periods for all models
described above. This provides a basis for a complete description of the
special geometry of the space of complex structures for these models, and
also the spaces of (complexified) K\"ahler classes for their mirror models.
In the subsequent sections, we will attempt various generalizations of
these results.

\newpage
\section{\cicys}{\cicy\ Manifolds}
We now turn to the calculation of the K\"ahler class
periods for some simple models. One difficulty here will be the
identification of suitable parameters and mirror symmetry will be
found most useful.
\subsection{Complete intersections in a single projective space}
We will compute here the periods for the K\"ahler class parameters for the
five families of manifolds:
 \CP4[5], \CP5[3,\,3], \CP5[2,\,4], \CP6[2,\,2,\,3] and
 \CP7[2,\,2,\,2,\,2].
Let us begin with the manifold \CP5[3,3]. Libgober and
Teitelbaum~\cite{\rLT}\ seem to have correctly guessed the mirror of this
manifold by considering the one parameter family, $\M_\j$, of \CP5[3,\,3]
manifolds given by the polynomials
$$ \eqalign{
  p_1 &= x_0^3 + x_1^3 + x_2^3 - 3 \j\, x_3 x_4 x_5~,  \cr
  p_2 &= x_3^3 + x_4^3 + x_5^3 - 3 \j\, x_0 x_1 x_2~.  \cr}
  \eqlabel{P533Eqs}
$$
Libgober and Teitelbaum observe that ({\it i})~there is a group of phase
symmetries $H$ such that these polynomials are in fact the most general
polynomials whose zero locus is invariant under $H$. ({\it ii})~The
manifold obtained by resolving $\M_\j/H$ has the values of $b_{1,1}$ and
$b_{2,1}$ exchanged relative to \CP5[3,\,3]. Moreover, ({\it iii})~the
numbers, $n_k$, of instantons of degree $k$ inferred by calculating the
Yukawa coupling yields the correct number in degree one and integral
values for all $k$ (the number of instantons has been checked also in
degree~two).

Underlying much of the present work is the important fact that the
holomorphic three-form may be realised as a residue where the manifold of
interest is a submanifold in a space of higher dimension.
For the manifolds of this section, which are given by $N$ transverse
polynomials $p^\a,~\a=1,\ldots,N$ in \CP{N+3}, the holomorphic three form
can be written as a residue of the $(N+3)$-form
$$
  {\e^{A_1A_2\cdots A_{N+4}}\, x_{A_1}\, \rd x_{A_2}\cdots \rd x_{A_{N+4}}
    \over p_1\, p_2\cdots p_N}~,
  \eqlabel{XXX}
$$
which is integrated around $N$ circles that enclose the loci $p_\a=0$. Let
$\deg(\a)$ denote the degree of $p_\a$ and note that both the numerator
and denominator of the $N+3$-form scale the same way under $x_A\to \l x_A$
in virtue of the \cy\ condition
$$
  \sum_{\a=1}^N \deg(\a) = N+4~. \eqlabel{XXX}
$$
In view of the above discussion, we may define a fundamental cycle $B_0$
for the mirror of \CP5[3,\,3] in analogy with \eqref{Bzero}. We
therefore arrive at the following expression for the fundamental period
that is the analogue of \eqref{vpzero}
$$
  \vp_0={1\over (2\p i)^6}\oint_{\g_0\times\cdots\times\g_5}
  {(3\j)^2\,\rd x_0\cdots \rd x_5 \over p_1\,p_2}~.
  \eqlabel{XXX}
$$
By extracting a factor of $x_0\cdots x_5$ from the denominator and
expanding in inverse powers of $\j$ we have
$$
  \vp_0={1\over (2\p i)^6}\int{\rd x_0\cdots \rd x_5 \over x_0\cdots x_5}\,
    \sum_{k=0}^\infty{\left[ x_0^3+x_1^3+x_2^3\right]^k \over
                             (3\j x_3 x_4 x_5)^k}
    \sum_{l=0}^\infty{\left[ x_3^3+x_4^3+x_5^3\right]^l \over
                             (3\j x_0 x_1 x_2)^l}~.
  \eqlabel{XXX}
$$
As in the case of the quintic the only terms that contribute are the terms
in the sums that are independent of the $x_A$. Such terms arise in the
product only when $k=l=3n$ and the contribution is then
$\left({(3n)!\over (n!)^3}\right)^2$. Thus, we find
$$
  \vp_0=\sum_{n=0}^\infty{\left( (3n)!\right)^2 \over (n!)^6\,(3\j)^{6n}}~
  = ~{}_4F_3\big({1\over3},{2\over3},{1\over3},{2\over3};
                                                  1,1,1; \j^{-6}\big)~.
  \eqlabel{vpP533}
$$
For the remaining manifolds the computations are analogous. The
fundamental period is given by the expression
$$
  \vp_0 = {\prod_{\a=1}^N \bigl(-\deg(\a)\j\bigr)\over (2\p i)^{N+4}}
          \int_{\g_1\times\cdots \times\g_{N+4}}
                  {\rd x_1 \cdots \rd x_{N+4}\over p_1\cdots p_N}~.
  \eqlabel{vpFund}
$$
The polynomials together with the resulting periods are displayed in
Table~\tabref{eqns}

\midinsert
\vbox{
$$
\vbox{\offinterlineskip\halign{
\strut#& \vrule#&\quad #\quad\hskip-3pt\hfil&#\vrule&\quad
$#$\quad\hskip-2pt\hfil&#\vrule
&\quad $\displaystyle{#}$\quad\hskip-2pt\hfil&#\vrule&\quad
$#$\quad\hfil&#\vrule\cr
\noalign{\hrule}
&height 15pt depth 10pt
  &\hfil Manifold
    &&\hfil\hbox{Equations}
      &&\qquad\quad\hbox{Period}
        &&\hbox{$a$-Indices}&\cr
\noalign{\hrule\vskip3pt\hrule}
&height 22pt depth 17pt
  &\CP4[5]
    &&\eqalign{p&=x_1^5{+}x_2^5{+}x_3^5{+}x_4^5{+}x_5^5\cr
           &\hskip50pt {-} 5\j\, x_1 x_2 x_3 x_4 x_5\cr}
     &&\sum_{n=0}^\infty{(5n)!\over (n!)^5\,(5^5\j^5)^n}
       &&{1\over 5},\,{2\over 5},\,{3\over 5},\,{4\over 5}&\cr
\noalign{\hrule}
&height 25pt depth 20pt
  &\CP5[3,\,3]
    &&\eqalign{
  p_1 &= x_1^3 {+} x_2^3 {+} x_3^3 {-} 3 \j\, x_4 x_5 x_6  \cropen{2pt}
  p_2 &= x_4^3 {+} x_5^3 {+} x_6^3 {-} 3 \j\, x_1 x_2 x_3  \cr}
     &&\sum_{n=0}^\infty{\big((3n)!\big)^2\over (n!)^6\,(3^6\j^6)^n}
       && {1\over 3},\,{2\over 3};\,{1\over 3},\,{2\over 3}&\cr
\noalign{\hrule}
&height 25pt depth 20pt
  &\CP5[2,\,4]
    &&\eqalign{
   p_1 &= x_1^2{+}x_2^2{+}x_3^2{+}x_4^2 {-}2\j\, x_5 x_6 \cropen{2pt}
   p_2 &= x_5^4 {+} x_6^4 {-} 4\j\, x_1 x_2 x_3 x_4   \cr}
     &&\sum_{n=0}^\infty{(2n)!\,(4n)!\over (n!)^6\,(2^8\j^6)^n}
       &&{1\over 2};\,{1\over 4},\,{2\over 4},\,{3\over 4}&\cr
\noalign{\hrule}
&height 35pt depth 30pt
  &\CP6[2,\,2,\,3]
    &&\eqalign{
  p_1 &= x_1^2 {+} x_2^2 {+} x_3^2 {-} 2 \j\, x_6 x_7 \cropen{2pt}
  p_2 &= x_4^2 {+} x_5^2 {-} 2 \j\, x_1 x_2 \cropen{2pt}
  p_3 &= x_6^3 {+} x_7^3 {-} 3\j\, x_3 x_4 x_5              \cr}
&&\sum_{n=0}^\infty{\big((2n)!\big)^2\,(3n)!\over (n!)^7\,
(2^5 3^2\j^7)^n}
    &&{1\over 2};\,{1\over 2};\,{1\over 3},\,{2\over 3}&\cr
\noalign{\hrule}
&height 45pt depth 40pt
  &\CP7[2,\,2,\,2,\,2]
    &&\eqalign{
  p_1 &= x_1^2 {+} x_2^2 {-} 2\j\, x_3 x_4 \cropen{2pt}
  p_2 &= x_3^2 {+} x_4^2 {-} 2\j\, x_5 x_6 \cropen{2pt}
  p_3 &= x_5^2 {+} x_6^2 {-} 2\j\, x_7 x_8 \cropen{2pt}
  p_4 &= x_7^2 {+} x_8^2 {-} 2\j\, x_1 x_2 \cr}
&&\sum_{n=0}^\infty{\big((2n)!\big)^4\over (n!)^8\,(2^8 \j^8)^n}
&&{1\over 2};\,{1\over 2};{1\over 2};\,{1\over 2}&\cr
\noalign{\hrule}}}
$$
\tablecaption{eqns}{The equations appropriate to the computation of the
K\"ahler class fundamental period, the fundamental period and the $a$-
indices for complete intersections in a single projective space.} }
\endinsert

One finds that the periods of Table~\tabref{eqns}\ may be related to
standard functions by means of the multiplication formula~\eqref{eProdG}.
We obtain in this way that the periods are given by generalized
hypergeometric functions ${}_pF_q$. Recall that these are defined by the
series
$$
 {}_pF_q\Big(\matrix{a_1,\ldots,a_p\cr
                     c_1,\ldots,c_q\cr};~z\Big)~
 = ~\sum_{k=0}^\infty {(a_1)_k,\ldots,(a_p)_k\over
                        (c_1)_k,\ldots,(c_q)_k}\,{z^k\over k!}~,
\eqlabel{eGHgF}
$$
where $(a)_r$ is the Pochhammer's symbol (see Eq.~\eqref{Poch}).
In fact, the periods of Table~\tabref{eqns}\ are all of the form
$$
  {}_4F_3\Big(\matrix{a_1,\ldots,a_4\cr 1,1,1\cr};~(C\j)^{-d}\Big)
  \eqlabel{XXX}
$$
with the $a$-indices $(a_1,a_2,a_3,a_4)$ taking the values shown in the
last column of Table~\tabref{eqns}. We see that each polynomial of degree
$d$ contributes the indices ${1\over d}, {2\over d},\cdots,{d-1\over d}$.

We will turn now to the analytic continuation for the fundamental periods
of these models.   We will mostly concentrate on $\CP5[3,\,3]$ to
illustrate the procedure since it is analogous for the other cases.  The
method is essentially the same as for $\CP4[5]$ \cite{\rCdGP}.

A very useful tool to understand the analytic continuation as well as the
behaviour of solutions near regular singular points is the Riemann symbol
for the $~{}_pF_q$ hypergeometric functions \eqref{eGHgF}.  For
$p=q+1=4$ it is given by
$$
  {\cal P} \left\{
     \matrix{0       &\infty   &1    &{}\cr
             0       &a_1      &0    &{}\cr
             1{-}c_1 &a_2      &1    &z\cr
             1{-}c_2 &a_3      &2    &{}\cr
             1{-}c_3 &a_4      &\sum c {-} \sum a    &{}\cr}
   \right\}~.\eqlabel{XXX}$$
The Riemann symbol gives the indices of the solutions for the regular
singular points\vadjust{\newpage}
of the space parametrized by $z$.
For the five cases  in Table \tabref{eqns} we have that
$$
  {\cal P} \left\{ \matrix{0     &\infty   &1    &{}\cr
                           0     &a_1      &0    &{}\cr
                           0     &a_2      &1    &(C\j)^{-d}\cr
                           0     &a_3      &2    &{}\cr
                           0     &a_4      &1    &{}\cr} \right\}~,
  \eqlabel{XXX}
$$
where $\sum a_i = 2$ for all the models, $C$ is a constant which is
$1$ for all models except $\CP5[3,\, 3]$ where $C=(2/3)^{1/7}$  and $d = \sum
{\rm degrees}$. The singularity at $\j = 1/C$ corresponds to a conifold point
and
the singularity at $\j = \infty$ corresponds to the large
complex structure limit.  For $\CP4[5]$, the singularity at $\j = 0$
corresponds to an orbifold point with a $\IZ_5$ symmetry and the
corresponding \cym\ is not singular.  For the
other cases in this series, this is not the case.  For example, for
$\CP5[3,\, 3]$ the space for $\j = 0$ has a singular curve, actually
a singular torus $\CP2[3]$.

We will use the information provided by the Riemann symbol to find a
complete basis for the periods in the region $|\j|<1/C$.  This basis,
which we denote $F_k, k= 1,\ldots,4\,$, is obtained directly
from the Riemann symbol by changing variables from $(C\j)^{-d}$ to
$(C\j)^d$ and exctracting factors of $\j^{a_i d}$.   We obtain
this way as many independent solutions as  there
are different indices $a_i$.     For example, by exctracting a factor of
$(C\j)^{a_1 d}$ we have a solution
$$ {\eqalign{F_1(\j) &\sim
 (C\j)^{a_1 d}\, {\cal P} \left\{
     \matrix{0             &\infty   &1    &{}\cr
             0             &a_1      &0    &{}\cr
             a_2{-}a_1     &a_1      &1    &(C\j)^d\cr
             a_3{-}a_1     &a_1      &2    &{}\cr
             a_4{-}a_1     &a_1      &1    &{}\cr}
   \right\}\cr
     &= (C\j)^{a_1 d}\, {}_4 F_3\left(a_1,a_1,a_1,a_1;
       1{-}a_2{-}a_1,1{-}a_3{-}a_1,1{-}a_4{-}a_1;(C\j)^d\right)~.\cr}}
  \eqlabel{XXX}
$$
The repetition of indices in the Riemann symbol for a given
regular singular point means that the remaining solutions will have
logarithms in the region of convergence around the singular point.  If the
indices $a_1$ and $a_2$ are equal, the other linearly independent solution
with index $a_1$ can be obtained, for example, by considering the
derivative of the solution $F_1$ with respect to the index $a_1$.   Except
for $\CP4[5]$, we do in fact find these logarithmic solutions near $\j=0$ thus
making the modular group intrinsically different from that of $\CP4[5]$.

Given the basis $\{F_k\}$,
we can find the periods $\vp_j$  as linear combinations
$$
  \vp_j = \sum_{k=1}^4 C_{jk} F_k~.\eqlabel{vpij}
$$
Most of the coefficients however can be found by using the symmetries of
the model.  All these spaces have a ``geometrical symmetry'' $\IZ_d$
corresponding to $\j\to\a\j,~\a^d=1$.  In the region $|C\j|<1$
we then define
$$
  \vp_j(\j) = \vp_0(\a^j\j),\qquad j = 0,\ldots,d-1~.
  \eqlabel{vpijsym}
$$
There are, of course, linear relations between these periods since
only four of them  are linearly independent.  Equation \eqref{vpijsym}
determines $C_{jk}$, for all $j\ne 0$.
The coefficients $C_{0k}$ can be found from the analytic continuation
of the period $\vp_0$ for $|C\j|>1$ to the region $|C\j|<1$.
The periods $\vp_j(\j)$ in the region $|C\j|>1$ can also be
obtained by analytic continuation to this region of the
representations in \eqref{vpij} for $|C\j|<1$.

Consider now $\CP5[3,\, 3]$; there will be two solutions with logarithms
in the region \hbox{$|\j|<1$} (with $C=1$).  The solutions without logarithms
are
$${\eqalign{
  F_k(\j) &= (3\j)^{2k}\, {\G({k\over 3})^4\over \G(k)^2}\,\,
  {}_4F_3\left({k\over 3},{k\over 3},{k\over 3},{k\over 3}
         ;\overbrace{
{{k{+}1}\over 3},{{k{+}2}\over 3},{{k{+}1}\over 3},{{k{+}2}\over 3}};\j^6
   \right)~,\cr
            &= (3\j)^{2k}\,
    \sum_{n=0}^{\infty} (3\j)^{6n}
            {\G^6(n{+}{k\over 3})\over\G^2(3n{+}k)}~,\qquad k = 1,2~,\cr}}
	    \eqlabel{XXX}
$$
where the overbrace means that an index that is equal to 1 must be deleted.
The other two linearly independent solutions can be chosen so that
$$\eqalign{
  F_{k+2}(\j) &= {{\rm d}\over{{\rm d} k}} \, F_k(\j)\cr
                &= \log(3\j)^2\, F_k \cr
           &\qquad +  2 (3\j)^{2k}\,\sum_{n=0}^{\infty}
         (3\j)^{6n}{\G^6(n{+}{k\over 3})\over\G^2(3n{+}k)}
            \left[\J\left(n{+}{k\over 3}\right) - \J(3n{+}k)\right]
     ~,\qquad k = 1,2~.\cr}\eqlabel{XXX}
$$

Let
$$
  \vp_0(\j) = \sum_{k=1}^2 (C_k F_k + C_{k+2} F_{k+2})~.\eqlabel{XXX}
$$
Since this theory has a geometrical symmetry $\IZ_6$, we define
$$
  \vp_j(\j) = \vp_0 (\a^j\j)~,\qquad\a^6 = 1~,\eqlabel{XXX}
$$
where not all six $\vp_j(\j)$ are linearly independent.
We obtain
$$
  \vp_j(\j) = \sum_{k=1}^2 \a^{2kj} \left[
      \left( C_k + {{2\p i}\over 3}\,j\, C_{k+2} \right) F_k +
            C_{k+2} F_{k+2}\right]~,\eqlabel{XXX}
$$
since
$$\eqalign{
  F_k(\a^j\j) &= \a^{2kj}\, F_k(\j)~,\cr
  F_{k+2}(\a^j\j) &= \a^{2kj} \left[F_{k+2}(\j) +
                     {{2\p i}\over 3}\,j\, F_k(\j)\right]~,
     \qquad k = 1,2~.\cr}\eqlabel{XXX}
$$
It is easily seen that, as expected, there are two relations between the
six periods:
$$
\eqalign{
  \vp_0 + 2\vp_1 + 3\vp_2 + 2 \vp_3 + \vp_4 &= 0\cr
  \vp_0 + \vp_1 + \vp_2
                 - \vp_3 - \vp_4 - \vp_5 &= 0~.\cr}\eqlabel{XXX}
$$

It is worthwhile at this point to contrast these solutions with the periods
for $\CP4[5]$ (and, in fact with any one modulus model embedded in a
weighted $\CP4$).  In the present cases, the monodromy transformation
${\cal A}$ around the point $\j=0$
in the moduli space {\it does not} satisfy the relation ${\cal A}^d=1$
associated to a relation between the periods
$\sum_{j=0}^{d-1}\vp_j=0$.  This is a consequence of the presence
of the logarithms in the solutions.  That is,  after transport
around $\j=0$ the period $\vp_0$ does not return to itself but
it goes to $\vp_6 = \vp_0 + 2 (\vp_3 - \vp_0)$.
Choosing a basis for the periods
$$
  \vp =
  \pmatrix{\vp_2 \cr \vp_1 \cr \vp_0 \cr \vp_5 \cr}~,
  \eqlabel{XXX}$$
the monodromy of the periods around $\j=0$ can be found to be
$$
  {\cal A}\vp(\j) = \vp(\a\j) = a\vp(\j)~,\eqlabel{XXX}
$$
where
$$
  a = \pmatrix{  -4    &-3    &-2    &\- 1\cr
               \- 1    &\- 0  &\- 0  &\- 0\cr
               \- 0    &\- 1  &\- 0  &\- 0\cr
                 -8    &-6    &-5    &\- 2\cr}~.\eqlabel{XXX}
$$

Returning to the analytic continuation of the periods to the region
$|\j|<1$, we still need to find the constants $C_k$.  To do this we write
the Mellin--Barnes integral for $\vp_0$
$$
  \vp_0 = {1\over {2\p i}}\int_{\gamma} {\rm d}\r\, e^{i\p\r}
               (3\j)^{-6\r}\,
     {{\G^2(3\r{+}1)\,\G(-\r)}\over \G^5(\r{+}1)}~,\eqlabel{XXX}
$$
which converges for $0<\arg\j<{{2\p}\over 6}$.
The contour of integration $\gamma$ is $\Ree(\r) = -\epsilon$ and
$-\infty < \imag(\r) < \infty$.  This integral has possible poles at
$$\eqalign{
   1.\quad \r &= n,\quad n = 0, 1,\ldots\cr
   2.\quad \r &= -{n\over 3},\quad n= 1, 2,\ldots\cr}
$$
Closing the contour to the right we
pick up the poles at $\r = n$, $n= 0, 1,\ldots\,$, and the sum over
residues gives back $\vp_0$ for $|\j| > 1$.  Closing the contour
to the left, we pick up the poles at $\r = -{n\over 3}$,
$n= 1, 2,\ldots$, which are actually {\it double} poles. We obtain
$$\eqalign{
  \vp_0 &= {1\over 9}\sum_{n=1}^{\infty} (3\j)^{2n} \a^{-n}
      {\G({n\over 3})\over{\G^2(n)\G^5(1{-}{n\over 3})}}\left\{
        \log[-(3\j)^{-6}] - \J\left({n\over 3}\right) -
           5 \J\left(1{-}{n\over 3}\right) + 6 \J(n)\right\}~,\cr
           &= -{{3^{3\over2}\over {(2\p)^5}}}\sum_{k=1}^2
        {\a^{-k}} \left\{
         \left[{\p\over 3}\left(-i +
             5\cot{k\p\over 3}\right)\right] F_k +
                F_{k+2}\right\}~,\qquad |\j|<1\cr}\eqlabel{XXX}
$$
Thus
$$\eqalign{
  C_k &= -{{3^{\half}}\over {2^5\p^4}} \a^{-k}
         \left(-i + 5\cot{k\p\over 3}\right)~,\cr
  C_{k+2} &= -{{3^{3\over 2}}\over {(2\p)^5}} \a^{-k} ~,
    \qquad k= 1,2~,\cr}\eqlabel{XXX}
$$
and
$$
   \vp_j(\j) = -{{3^{3\over 2}}\over {(2\p)^5}}
    \sum_{k=1}^2 \a^{k(2j-1)}  \left\{{\p\over 3}\left[
      (2j-1)i + 5 \cot{k\p\over 3}\right] F_k + F_{k+2}\right\}
  ~,\qquad |\j|<1~.\eqlabel{XXX}
$$

The reader may enjoy the straightforward exercise of finding the monodromy
of the periods around $\j = 1$ and $\j = \infty$.  Following \cite{\rCdGP},
the periods near $\j = 1$ are given by
$$
 \vp_j
 = - {c_j \over {2\p i}}(\vp_1 - \vp_0) \log(\j {-} 1) + g_j~,
 \eqlabel{XXX}
$$
with $g_j$ analytic for $|\j-1|<1$ and where the constants $c_j$ are
$$
  (c_j) = (1,1,-5,10,-8,-5)~.\eqlabel{XXX}
$$
The matrix $t$ corresponding to the monodromy ${\cal T}$ about $\j = 1$
is therefore
$$
  t = \pmatrix{\- 1  &\- 5  &  -5   &\- 0 \cr
               \- 0  &\- 0  &\- 1   &\- 0 \cr
               \- 0  & -1   &\- 2   &\- 0 \cr
               \- 0  &\- 5  &  -5   &\- 1 \cr}~.\eqlabel{XXX}
$$
The monodromy about $\j = \infty$ can be easily obtained from
the fact that a loop around
all the singularities ($\j=\a^j$ for $j=0,\ldots,5$, $\j=0$ and
$\j=\infty$) is contractible:
$$
  {\bf1} = t_{\infty} (a t)^6~.  \eqlabel{XXX}
$$
Therefore
$$ t_{\infty}^{-1} = \pmatrix{- 4 & -23 & \- 20  & - 1   \cr
                        \- 1 &\- 5 &  - 5   &\- 0  \cr
                         \- 0   &\- 0  &\- 1    &\- 0   \cr
                         - 8 & -25 & \- 14  & \- 2 \cr}~.
			 \eqlabel{XXX}
$$

A series representation for the periods in the region $|\j|>1$
can be similarly obtained ~\cite{\rCdGP}\ by considering the
following integral representation of $F_k$
$$
  F_k = - {(3\j)^{2k}\over{2\p i}}\int_{\g} {\rm d}s\, e^{i\p s}\,
      (3\j)^{6s}  {\p\over{\sin\p s}}\,
        {\G^6(s{+}{k\over 3})\over\G^2(3s{+}k)}\qquad k=1,2~,
	\eqlabel{XXX}
$$
but here we will not pursue this matter any further.

As a final comment, we would like to remark that the space
$\CP7[2,2,2,2]$ is a little different from the others in this series
in the sense that its periods for small-$\j$ behave very differently.
The singularity at $\j=0$ is of the same type as
the singularity at $\j=\infty$ and the reason behind this is that the
moduli space for this model has an extra symmetry $\j\to 1/\j$.

\subsection{Calabi-Yau hypersurfaces in products of projective spaces}
Consider next the five families of \cicys:
$$\vbox{
{\hsize=1in \parindent=0pt%
\valign{$$#$$\hfill\strut\vfil\cr
\Cnfg{\CP4\cr}{5\cr}~,                                        \cr
\Cnfg{\CP2\cr \CP2\cr}{3\cr 3\cr}~,                           \cr
\Cnfg{\CP1\cr \CP3\cr}{2\cr 4\cr}~,                           \cr
\Cnfg{\CP1\cr \CP1\cr \CP2\cr}{2\cr 2\cr 3\cr}~,              \cr
\Cnfg{\CP1\cr \CP1\cr \CP1\cr \CP1\cr}{2\cr 2\cr 2\cr 2\cr}~, \cr}}
      \vskip-20pt}
  \eqlabel{OneCol}
$$
that are defined by a single polynomial. We wish to obtain the fundamental
period for their space of K\"ahler classes using mirror symmetry. Each
family has one K\"ahler class parameter for each \CP{n}\ factor. The
generalization of the analysis presented in Section~2 to these models
requires several choices pertaining to the identification of the
deformations of the complex structure mirroring the K\"ahler classes of
each \CP{n} factor of the ambient space. We will return to discuss these
below. However, for the five families of spaces~\eqref{OneCol} the
fundamental period may be found by applying the construction of
Batyrev~\cite{\Baty}.

Consider for example the family of manifolds
$$
  \Cnfg{\CP2\cr \CP2}{3\cr 3\cr}~, \eqlabel{mnfld}
$$
and denote by $(x_0,x_1,x_2)$ and $(y_0,y_1,y_2)$ the homogeneous
coordinates of the two \CP2's. We restrict the coordinates to the set
\ca{T} on which no coordinate vanishes and take $x_0=y_0=1$ so that
$(x_1,x_2;y_1,y_2)$ are affine coordinates on the embedding space.
Let
$$
  f~ = ~{p\over x_1x_2y_1y_2} =
         \sum_{m_1,m_2;n_1,n_2}a_{m_1m_2n_1n_2}\,
          x_1^{m_1}x_2^{m_2}y_1^{n_1}y_2^{n_2}~.
  \eqlabel{XXX}
$$
The Newton polyhedron, $\D$, of the manifold is the convex hull of the
points $(m_1,m_2;n_1,n_2)$ for which the coefficients $a$ of the Laurent
polynomial $f$ are nonzero. For this elementary example these are the
points for which $$ \eqalign{
    -1\leq m_1~,~~~-1\leq m_2~&;~~~~ -1\leq n_1~,~~~-1\leq n_2\cropen{5pt}
               m_1 + m_2\leq 1~~~&;~~~n_1 + n_2\leq 1~.\cr}\eqlabel{Newton}
$$
The vertices of the polyhedron dual to $\D$ are the six points:
$$ \eqalign{
    (1,0,0,0)~,~~~(0,1,0,0)~&,~~~(0,0,1,0)~,~~~(0,0,0,1)~,\cr
       (-1,-1,0,0)~&,~~~(0,0,-1,-1)~.\cr}
  \eqlabel{XXX}
$$
These points are the coefficients in the inequalities \eqref{Newton} if
the inequalities are replaced by equalities and we multiply through if
necessary so that the constant term is $-1$ in each case. The Laurent
polynomial corresponding to the dual polyhedron is
$$
  \hat{f} = 1 + X_1 + X_2 + {\l_1\over X_1X_2} +
                Y_1 + Y_2 + {\l_2\over Y_1Y_2}
  \eqlabel{XXX}
$$
where the freedom to make linear redefinitions of the coordinates and to
multiply $\hat{f}$ by an overall factor has been used to reduce the number
of free parameters to two which is the correct number since $b_{1,1}=2$
for the original manifold~\eqref{mnfld}.

We may now revert to homogeneous coordinates by setting
$X_1={x_1\over x_0},~X_2={x_2\over x_0}$ etc. Then the mirror manifold is
given by the singular bicubic hypersurface $\hat{p}=0$ in $\CP2\times\CP2$
with
$$\eqalign{
  \hat{p} &= X_1X_2Y_1Y_2 \hat{f}\cr
          &= x_0 x_1 x_2\, y_0 y_1 y_2
            + (x_1^2 x_2 + x_1 x_2^2 + \l_1 x_0^3)\, y_0 y_1 y_2
            + (y_1^2 y_2 + y_1 y_2^2 + \l_2 y_0^3)\, x_0 x_1 x_2~.\cr}
  \eqlabel{eBatyP}
$$
The fundamental period for the mirror now follows in a manner that has
become familiar:
$$ \eqalign{
 \vp_0(\l_1,\l_2) &= {1\over (2\p i)^6}\int
   {\rd x_0 \rd x_1 \rd x_2\, \rd y_0 \rd y_1 dy_2 \over \hat{p}}
                                                              \cropen{5pt}
                  &=\sum_{n_1,n_2=0}^\infty {(3n_1 + 3n_2)!\over
                            (n_1!)^3 (n_2!)^3} \l_1^{n_1} \l_2^{n_2}~.\cr}
  \eqlabel{vpBaty}
$$
The other cases follow in precise analogy and we list the fundamental
K\"ahler class periods in Table~\tabref{cicys} to reveal the pattern.
These spaces would seem to furnish rather simple examples of
multiparameter spaces which might be easier to analyse than the
multiparameter cases studied thus far~\cite{{\rMorrison --\rCdFKM}}. Note also
that these spaces do not have a single `fundamental' parameter; rather
they have several, one corresponding to the K\"ahler class of each factor
space.

\midinsert
\vbox{
$$
\vbox{\offinterlineskip\halign{
\strut#& \vrule#&\hfil\quad $#$\quad\hfil&#\vrule
&\qquad $\displaystyle{#}$\qquad\hfil&#\vrule\cr
\noalign{\hrule}
&height 15pt depth 10pt
  &\hfil\hbox{Manifold}
    &&\hskip90pt\hbox{Period}&\cr
\noalign{\hrule\vskip3pt\hrule}
&height 22pt depth 17pt
  &\Cnfg{\CP4\cr}{5\cr}
      &&\hskip17pt\sum_{n=0}^\infty{(5n)!\over (n!)^5}\,\l^n &\cr
\noalign{\hrule}
&height 25pt depth 20pt
  &\Cnfg{\CP2\cr \CP2\cr}{3\cr 3\cr}
     &&\hskip10pt\sum_{n_1,n_2=0}^\infty {(3n_1 + 3n_2)!\over (n_1!)^3\,
(n_2!)^3}
\,\l_1^{n_1} \l_2^{n_2} &\cr
\noalign{\hrule}
&height 25pt depth 20pt
  &\Cnfg{\CP1\cr \CP3\cr}{2\cr 4\cr}
         &&\hskip10pt\sum_{n_1,n_2=0}^\infty
                      {(2n_1 + 4n_2)!\over (n_1!)^2\, (n_2!)^4}
                                      \,\l_1^{n_1} \l_2^{n_2} &\cr
\noalign{\hrule}
&height 30pt depth 25pt
  &\Cnfg{\CP1\cr \CP1\cr \CP2\cr}{2\cr 2\cr 3\cr}
      &&\hskip5pt\sum_{n_1,n_2,n_3=0}^\infty {(2n_1 + 2n_2 + 3n_3)!\over
(n_1!)^2
\,(n_2!)^2 \,(n_3!)^3}
\,\l_1^{n_1} \l_2^{n_2} \l_3^{n_3}&\cr
\noalign{\hrule}
&height 35pt depth 30pt
  &\Cnfg{\CP1\cr \CP1\cr \CP1\cr \CP1\cr}{2\cr 2\cr 2\cr 2\cr}
   &&\sum_{n_1,n_2,n_3,n_4=0}^\infty {(2n_1 + 2n_2 + 2n_3 + 2n_4)!\over
(n_1!)^2 \,(n_2!)^2 \,(n_3!)^2 \,(n_4!)^2}
\,\l_1^{n_1} \l_2^{n_2} \l_3^{n_3} \l_4^{n_4}&\cr
\noalign{\hrule}}}
$$
\tablecaption{cicys}{The K\"ahler class fundamental periods for Calabi-Yau
hypersurfaces in products of projective spaces.} }
\endinsert

\subsection{A natural conjecture}
We now know the K\"ahler class for the fundamental periods for \cicys\ for
which
the degree matrix consists either of a single row or a single column.
Based on this, it is natural to conjecture a form for a general weighted
\cicy---with
$F>1$ factor spaces
$$\CP{n_1}_{(k_1^{(1)},\ldots,k_{n_1{+}1}^{(1)})}\times\cdots
   \times\CP{n_F}_{(k_1^{(F)},\ldots,k_{n_F{+}1}^{(F)})}~, \eqlabel{XXX}
$$
and $N>1$ polynomials $p_\a,~\a=1,\ldots,N$ such that the degree of $p_\a$
in the variables of the $j$'th projective space is $\deg_j(\a)$ which, in
order to have a \cym, they must satisfy the relations
$$
  \sum_{\a} \deg_j(\a) = \sum_{i=1}^{n_j + 1} k_i^{(j)}~,\qquad
     \hbox{for all}~j~.  \eqlabel{XXX}
$$
The fundamental period is then given,
as a function of the parameters $\l_j$ corresponding to the K\"ahler
classes of each factor space, by
$$
  \vp_0(\l_1,\ldots,\l_F) = \sum_{m_1,\ldots,m_F=0}^\infty
  {\prod_{\a=1}^N\left( \sum_{j=1}^F \deg_j(\a)m_j \right)!
   \over \big(\prod_{i=1}^{n_1+1}(k_i^{(1)} m_1)!\big)
    \cdots \big(\prod_{i=1}^{n_F+1}(k_i^{(F)} m_F)!\big)}\,
   \l_1^{m_1}\cdots \l_F^{m_F}~.
  \eqlabel{Conj}
$$
We illustrate this conjecture below.  It is compatible with all other
examples for which the period has already been calculated~
\REF\rKT{A. Klemm and Theisen: `` Mirror Maps and Instanton Sums for Complete
Intersections in Weighted Projective Space'', Munich University preprint
LMU-TPW-93-08.}
\cite{{\rKT,\rBatnew}}.

By way of an example of this conjecture, consider the Schimmrigk
manifold~~\cite{\rROLF},
$$
  {\cal M} \in \Cnfg{\CP3\cr \CP2\cr}{3&1\cr 0&3\cr}~~ :
 ~~\left\{
\eqalign{
   p_{1,0}(x)   &= ~\sum_{i=0}^3 x_i^3\cr
   p_{2,0}(x,y) &= ~\sum_{i=1}^3 x_i y_i^3\cr}
\eqalign{
&\vphantom{\sum_{i=0}^3}= ~0~.  \cr
&\vphantom{\sum_{i=0}^3}= ~0~,  \cr}\right.
\eqlabel{cicyR}
$$
Identify first $x_1x_2x_3$ and $x_0y_1y_2y_3$ as the fundamental
deformations of $p^1_0$ and $p^2_0$ respectively. Thus, we want to find
the periods associated to a deformation of ${\cal M}$ given by
$$
  \eqalign{ p_1 &= p_{1,0} -\j_1 x_1 x_2 x_3~,     \cr
            p_2 &= p_{2,0} -\j_2 x_0 y_1 y_2 y_3~. \cr }
\eqlabel{cicyRdef}
$$

Note that $x_1x_2x_3$ and $x_0y_1y_2y_3$ are left invariant under action
of the quotient symmetry $H$ which produces the mirror manifold,
${\cal W}={\cal M}/H$. Therefore, by mirror symmetry, the fundamental
period may also be thought of as representing deformations of the K\"ahler
class of ${\cal M}$.
The period follows straightforwardly as in previous examples,
$$
  \vp_0~=~ (-\j_1)(-\j_2) {1\over (2\p i)^7} \oint_\G
              {\rd x_0 \ldots  \rd y_3 \over p_1\, p_2}~,
\eqlabel{XXX}
$$
which becomes
$$
 \hbox{}\mkern20mu
  \eqalign{ \vp_0~
 &=~{1\over (2\p i)^7}
    \int_\G{\rd x_0 \ldots \rd y_3\over x_0\ldots y_3}
     \sum_{n=0}^{\infty}\!
      \left({ x_0^3+\ldots +x_3^3\over\j_1 x_1x_2x_3 }\right)
     \sum_{m=0}^{\infty}\!
      \left({ x_1y_1^3+\ldots+x_3y_3^3\over\j_2 x_0y_1y_2y_3 }\right) \cr
 &= \sum_{m_1,m_2=0}^{\infty}{(3m_1)!\,(3m_2+m_1)!\over(m_1!)^4 (m_2!)^3}
              {1\over (\j_1)^{3m_2+m_1} (\j_2)^{3m_1}}~.              \cr}
\eqlabel{cicyRvp}
$$
Indeed, upon identifying $\l_1^{{-}1}=\j_1\j_2^3$ and
$\l_2^{{-}1}=\j_1^3$, this is seen to be the appropriate special case of
the conjecture~\eqref{Conj}. Note also that this is not the complete
answer; $b_{1,1}=8$ for this CICY manifold while we have here obtained the
fundamental period only as a function of two moduli, the ones
corresponding to the K\"ahler classes on \CP3 and \CP2. The dependence on
the remaining six parameters will be discussed in Section~\chapref{More}.

Another interesting example of the above is to consider the two-parameter
(Fermat) models which will be discussed in Section~4.1. For the first four
models in $(4.1)$ one can (for specific values of the moduli) relate them to
one-parameter families of complete intersections in a weighted projective
space. In particular, let us consider a two-parameter deformation of
$\CP4_{(1,1,2,2,6)}[12]^{128,2}_{-252}$ given by \eqref{pfer}.
{}From the discussion in Section~4.1 one finds that the fundamental period may
be written as
$$
\vp_0~=~\sum_{l=0}^{\infty}{\G(6l+1)\over \G^3(l+1)\G(3l+1)}\left({-1\over
(12\j)^6}\right) U_l(\f)
\eqlabel{birat}
$$
where $U_l(\f)$ is given by \eqref{Uhyp}. At the singular curve $\f=1$,
$U_l(\f=1)={\G(2l+1)\over \G^2(l+1)}$, and hence
\eqref{birat} becomes
$$
\vp_0~=~\sum_{l=0}{\infty}{\G(6l+1)\G(2l+1)\over
\G^5(l+1)\G(3l+1)}\left({-1\over
(12\j)^6}\right)^l~.
\eqlabel{ciwcy}
$$
But according to the conjecture, Eq.~\eqref{ciwcy} is the fundamental period
for
$\CP5_{(1,1,1,1,1,3)}[6,2]$. Indeed, it can be shown~\REF\rKatz{S.~Katz,
private communication.}
\cite{\rKatz}\
that this manifold is
birational to $\CP4_{(1,1,2,2,6)}[12]$.

\newpage
\section{twoparams}{Hypersurfaces with Two Parameters}
\subsection{Fermat hypersurfaces}
For Fermat hypersurfaces it is possible to derive some general results
from the explicit form of the fundamental period given in \eqref{PerM}.
There are five known examples of Fermat threefolds with $b_{11}=2$, and
they belong to the families
$$\matrix{
   \CP4_{(1,1,2,2,2)}[8]^{86,2}_{-168}~,~~~
   \CP4_{(1,1,2,2,6)}[12]^{128,2}_{-252}~,~~~
   \CP4_{(1,4,2,2,3)}[12]^{74,2}_{-144}~,\cropen{7pt}
   \CP4_{(1,7,2,2,2)}[14]^{122,2}_{-240}~,~~~
   \CP4_{(1,1,1,6,9)}[18]^{272,2}_{-540}~,\cr}  \eqlabel{twoexamp}
$$
respectively. As Fermat polynomials
are invariant under transposition, their mirrors are described by
$\{P=0\}/H$, for some appropriate group $H$, with $P$ taking the form
$$
  P = \sum_{j=0}^4 \ x_j^{d/k_j}   -  d\j\, x_0x_1x_2x_3x_4
     - {d\over q_0}\f\,  x_0^{q_0} x_1^{q_1} x_2^{q_2} x_3^{q_3} x_4^{q_4}
\eqlabel{pfer}
$$
Using the relations of the Jacobian ideal~\cite\rPhilip\ we can show that
in all cases $D\define d/q_0$ is an integer and furthermore that
$$
  {q_ik_i\over q_0}=\cases {0 & $i \geq D$\cr 1& $i < D$\cr}
  \eqlabel{emp}
$$
For $\CP4_{(1,1,1,6,9)}$, $D=3$, and for the remaining examples, $D=2$.

{}From \eqref{pert}, with $Q=m\,t=m\, q_0$ and realizing that $t,s_i\in\ZZ$ for
$p=1$, the fundamental period  is given by
$$
  \vp_0(\j, \f) = \sum_{n=0}^{\infty} \ \sum_{m=0}^{\infty}  \
  { (dn+q_0 m)!\,(d\j)^{-d n-q_0 m} (-D\f)^m\over m!\,n!\,
   \prod_{i=1}^4\left(k_i n + {k_i\over d}(q_0{-}q_i)m \right)!}
\eqlabel{pifer}
$$
which converges for sufficiently large-$\j$. Since $D$ is an
integer, it is convenient to set $l=Dn+m$ and sum over $l$ and $n$ to
obtain the alternative expression
$$
  \vp_0(\j,\f)=\sum _{l=0}^\infty{(q_0 l)!\,(d\j)^{-q_0 l}(-1)^l
   \over l!\,\prod _{i=1}^4\left({k_i\over d}(q_0{-}q_i)l\right)!}U_l(\f)
\eqlabel{piU}
$$
where
$$
  U_l(\f)= (D\f)^l \sum _{n=0}^{[{l\over D}]}{l!\prod_{i=1}^4
  \left({k_i\over d}(q_0{-}q_i)l\right)! (-D\f)^{-Dn}\over (l-Dn)!\, n!\,
  \prod_{i=1}^4 \left({k_iq_i\over q_0}n +
                      {k_i\over d}(q_0{-}q_i)l\right)!}
\eqlabel{Ufun}
$$
$U_l(\f)$ is a polynomial of degree $l$ but it can be extended to an
hypergeometric function for complex values $\nu$ of $l$. For instance,
when $D=2$ we find
$$
  U_\n (\f) = (2\f)^\n {}_2F_1\left(-{\n\over 2}, {1{-}\n\over 2};
  1{+}{\nu(k_1{-}1)\over 2}; {1\over \f^2}\right)\quad ,
  \quad \big|\f\big| > 1 \eqlabel{Uhyp}
$$
where we have used~\eqref{emp}.

$U_\n(\f)$ can be continued to the region $\big|\f\big| <1$
by means of contour integrals or by using well known
properties of the hypergeometric function. When $D=2$ the result is
$$
  U_\n(\f)~ = ~{ e^{i\p\n/2}\G \left(1{+}{\n(k_1{-}1)\over 2}\right)
                 \over 2\G(-\n)}
   \sum _{m=0}^\infty {e^{i\p m/2}\G\left({m{-}\n\over 2}\right)(2\f)^m
                        \over m!\,\G\left(1{-}{m{-}\n k_1\over2}\right)}
\eqlabel{Uzero}
$$
which can in turn
be written as a combination of hypergeometric functions around $\f=0$.

Following the discussion in Section~2, the region of convergence for the
fundamental period \eqref{pifer} can be found for the first two examples in
\eqref{twoexamp}\cite{\rCdFKM}.  For example, for the first case we obtain
that \eqref{pifer} converges in the region
$|(\phi\pm 1)/8\psi^4|$.  Unfortunately, the same analysis has proven more
difficult for the other examples with the main problem being the
calculation of the asymptotic behaviour of $U_\n(\f)$ as $\n\to\infty$.

In order to obtain a basis of independent periods as explained in
Section~2, we need to analytically continue $\vp_0(\j, \f)$ to a
small-$\j$ region.
Continuing the series given in~\eqref{piU} leads to an
equivalent result by virtue of~\eqref{Uzero}. Thus, when $D=2$, we find
$$
  \vp_0(\j, \f) = -{2\over d}\sum _{n=1}^\infty {\G \left({2n\over d}
  \right)(-d\j)^n\, U_{-{2n\over d}}(\f)\over
  \G (n) \G \left(1{-}{n\over d}(k_1{-}1)\right)
  \G\left(1{-}{k_2n\over d}\right)\G\left(1{-}{k_3 n\over d}\right)
  \G\left(1{-}{k_4n\over d}\right)} \eqlabel{pismall}
$$

A basis of periods can be obtained by the action of the phase symmetry of
the parameters as discussed in subsection~{\it2.5}. We define
$$
  \vp_j(\j, \f) = \vp_0 (\l^j\j, \l^{jq_0} \f)\quad ,\quad
  j=0,1\cdots, d-1
  \eqlabel{XXX}
$$
where $\l^d =1$. Note that since Fermat polynomials are invariant under
transposition, $\hat d = d$. Again, not all of the $\vp_j$ are independent.
In particular, the $\vp_j$ satisfy
$$
  \sum _{j=0}^{d-1} \vp_j (\j, \f) =0
  \eqlabel{XXX}
$$
Further relations among the $\vp_j(\j,\f)$ arise from the fact that
some terms in the series \eqref{pismall}
are absent due to the $\G$-functions in
the denominator. The resulting constraints on the $\vp_j$ depend on the
specific values of the weights $k_i$. In all cases we find that as
expected there are only six independent periods.

\subsection{A non-Fermat example}
We turn now to an example which is interesting because it is not of Fermat
type and it illustrates the important fact that the periods are most
naturally written in terms of the weights of the mirror manifolds. This is
not evident in Fermat examples precisely because the mirrors in these
cases are quotients of the original manifold.

Consider the following quasihomogenous polynomial,
$$
   P_0~~ = x_0^7 + x_1^7 x_3 +  x_3^3 + x_2^7 x_4 + x_4^3.\eqlabel{eP}
$$
$P_0=0$ defines a hypersurface ${\cal M}_0$ in a weighted projective space
$\CP4_{(3,2,2,7,7)}[21]^{50,11}_{-78}$. As it is, the space of complex
structures for this model is 50-dimensional and we wish to focus on a
smaller subspace. To this end, we consider a 2-parameter deformation
$$
   P_{\f_0,\f_1}~= ~x_0^7 + x_1^7 x_3 +  x_3^3 + x_2^7 x_4 + x_4^3
                         - \f_0\, x_0 x_1 x_2 x_3 x_4
                         + \f_1\, x_0^3 x_1^3 x_2^3~. \eqlabel{ePdef}
$$
This 2-parameter family of polynomials is invariant under the
symmetry
$$
  H~~ \define ~~(\IZ_{21}:12,2,0,7,0)~, \eqlabel{eOONE}
$$
and in fact is the most general subclass of such $\CP4_{(3,2,2,7,7)}[21]$
models; $P_{\f_0,\f_1}$ is the superpotential for the corresponding \LGO{s}.
Note that $H$ is {\it not} the quantum symmetry
$$
  Q_{\cal M}~ = ~(\IZ_{21}:3,2,2,7,7)~,
  \eqlabel{QSym}
$$
although both $Q_{\cal M}$ and $H$ are isomorphic to $\IZ_{21}$.

Related to this is the 2-parameter family ${\cal M}_{\f_0,\f_1}/H$,
for which $(b_{2,1},b_{1,1})=(2,95)$. Furthermore,
the orbifold ${\cal M}_0/H$ is the mirror model of a hypersurface
$\cal W$ in $\CP4_{(1,1,1,2,2)}[7]^{95,2}_{-186}$ for which the two
K\"ahler variations are mirrored in $\f_0,\f_1$. This can then be used to
study both the space of K\"ahler classes on $\cal W$ and the space of
complex structures for ${\cal M}/H$. Let us also point out that
the geometric symmetry of ${\cal M}/H$ is
$$
  G_{{\cal M}/H}~ = ~(\IZ_7 : 6,0,0,0,0)~. \eqlabel{GSymA}
$$
Now, $G_{{\cal M}/H}$ will leave $P_{\f_0,\f_1}$ invariant if also
$$
  G_{{\cal M}/H}~ : ~(\f_0,\f_1) \longrightarrow (\l\f_0,\l^3\f_1)~,
\qquad \l^7=1~.
  \eqlabel{GSymB}
$$
Note that $G_{{\cal M}/H}$ is isomorphic to the quantum symmetry of
$\cal W$, the mirror of ${\cal M}/H$.

Here, we are interested in obtaining explicit expressions for the
periods as functions of $\f_0,\f_1$. From our general discussion
of Section~2 we notice that using \eqref{ePOLY}\ and \eqref{eTheM}\
we have
$$
  q_0^0=q_1^0=q_2^0=3~,\qquad
  \big[a_{ij}\big]=\left[\matrix{ 7  &  0  &  0  &  0  &  0  \cr
                                  0  &  7  &  0  &  0  &  0  \cr
                                  0  &  0  &  7  &  0  &  0  \cr
                                  0  &  1  &  0  &  3  &  0  \cr
                                  0  &  0  &  1  &  0  &  3  \cr} \right]~.
  \eqlabel{FeMatP}
$$
Furthermore, from the transpose of $a_{ij}$ we obtain the weights of the
mirror manifold $(\hat k_1,\ldots,\hat k_5) = (1,1,1,2,2)$ and the
degree of its defining polynomial $\hat d = 7$.  From Eqs. \eqref{sumvar},
\eqref{dhat}\ and \eqref{relat}\ we have
$n_1=n_2=n_3=r$ and $n_4=n_5=2r+m$, and $s_{0,1,2}=0$, $s_{3,4}=1=p$ and
$t=3$. This yields the fundamental period
$$
  \vp_0~ = ~\sum_{m,r=0}^\infty
  {\G(7r{+}3m{+}1)\over\G^3(r{+}1)\G^2(2r{+}m{+}1)}\,
  {{1\over m!}\Big({\f_1\over{\f_0^3}}\Big)^{\!m}}\,
  {\f_0^{-7r}}~.
  \eqlabel{eEYPiO}
$$

Eq.~\eqref{eEYPiO}\ may be rewritten in a number of ways. Let us choose good
coordinates for the large complex structure limit~\cite{\rKatz},
$$
  X~={\f_1\over \f_0^3} \qquad;\qquad Y~=~{1\over \f_0\f_1^2}~.
  \eqlabel{XXX}
$$
Then, by changing the summation by setting $l=m+2r$, we find
$$
  \vp_0 = \sum_{l=0}^\infty {\G(3l+1) \over \G^3(l+1)} X^l U_l(Y)
  \eqlabel{pi0U}
$$
where
$$
  U_l(Y) = {\G(l+1) \over \G(3l+1)}
           \sum_{r=0}^{[l/2]} {\G(r+3l+1)\over \G^3(r+1) \G(-2r+l+1)}Y^r~.
    \eqlabel{Ul}
$$
 $U_l$ may be rewritten as a generalized hypergeometric function of type
${}_3F_2$. In fact since the series terminates we have a polynomial. The period
in the above
form  should prove useful in  studying the monodromy of the
periods around the singular curves in moduli space. This is however beyond the
scope of the present analysis, in particular because of the lack of an
asymptotic form for $U_l$ when $l\to \infty$, as discussed in Section~2.

The next step is to analytically continue $\vp_0$ to small-$\f_0$.
{}From our general analysis in Section~2, we find
$$
  \vp_0(\f_0,\f_1) = -{1\over 7} \sum_{n=1}^{\infty}
    e^{6i\p n/7}\>{\f_0^n \over \G(n) } \sum_{m=0}^{\infty}~
   {e^{-3i\p m/7}\> \G({n + 3m\over 7}) \over
    \G^2(1{-}{n{+}3m\over 7})\>\G^2(1{-}{2n{-}m\over 7})}
    \>{\f_1^m\over m!}
    \eqlabel{pi0sf}
$$
The remaining periods are then obtained from \eqref{pi0sf} by acting with
the modular group. Observe that terms with $(n+3m) = 0$ mod 7 are absent
in~\eqref{pi0sf} due to the $\G(1-{n+3m\over 7})$ in the denominator, and
so consider the action
$$
   {\cal A}~:~(\f_0,\f_1) \longmapsto (\l^j\f_0,\l^{3j}\f_1)~,
$$
and define
$$
  \vp_j(\f_0,\f_1) \define \vp_0(\l\f_0,\l^3\f_1)~,
  \qquad j=0, \cdots , 6~. \eqlabel{XXX}
$$
The $\vp_j$ are subject to the single relation
$$
   \sum_{j=0}^6 \vp_j = 0~. \eqlabel{XXX}
$$
Since $b_3=6$ for the 2-parameter model ${\cal M}/H$, these six linearly
independent periods form a complete set of periods for the space of
complex structures on ${\cal M}/H$ and also for the space of K\"ahler
classes on $\cal W$.

The appearance of the $\G(1{-}{2n{-}m\over7})$ may prompt considering
$$
   {\cal B}~:~(\f_0,\f_1) \longmapsto (\l^2\f_0,\l^{{-}1}\f_1)~,
$$
and define more periods using this action. However, in fact,
${\cal B} = {\cal A}^2$ (since $\l^6=\l^{{-}1}$) and no new periods or
conditions ensue.

Consider now briefly the complementary components of the complete moduli
space. To that end, we turn to the model
${\cal W} \in \CP4_{(1,1,1,2,2)}[7]$,
defined by the transposed defining equation
$$
  \hat{P}_0~ = ~y_1^7 + y_2^7 + y_2 y_4^3 + y_3^7 + y_3 y_5^3~,
  \eqlabel{eTrP}
$$
associated to the transposed matrix of exponents in~\eqref{FeMatP}.
Besides the fundamental monomial $\hat{M}_0 = y_1 y_2 y_3 y_4 y_5$, this
model has 94 independent deformations. The quantum symmetry being
$(\ZZ_7: 1,1,1,2,2)$, the geometric symmetry of $\cal W$ is
$$
  G_{\cal W}~ = ~(\ZZ_{21}: 0,18,0,1,0) \times (\ZZ_{21}: 0,0,18,0,1)~.
  \eqlabel{eGofW}
$$
Note that $G_{\cal W}$ is isomorphic to $Q_{{\cal M}/H}$, the quantum
symmetry of ${\cal M}/H$, as it should be since $\cal W$ and ${\cal M}/H$
are mirror models.

The techniques of Section~2 enable us to calculate the complex structure
periods $\hat{\vp}_0(\q_\b)$, $\b=0,1,\ldots,94$ for the 95-parameter
family of models with
$$
  \hat{P}_{\q_\b}~ = ~\hat{P}_0 - \q_0 y_1 y_2 y_3 y_4 y_5
                       + \sum_{\b=1}^{94} \q_\b \hat{M}_\b(y_i)~.
  \eqlabel{XXX}
$$
Once the small-$\q_0$ expansion of $\hat{\vp}_0$ is obtained, for example
upon analytic continuation of the large-$\q_0$ expansion as described
above, other periods will be obtained through the action of the modular
group. In the present case, the action of $G_{\cal W}$, defined
in~\eqref{eGofW}, is matched by the induced action $\cal A_W$:
$\q_0\to\l^2\m^2\q_0$, {\it etc}., so as to make $P_{\q_\b}(y)$ invariant.
Note that the action of $G_{\cal W}$ (and $\cal A_W$) is rather more
involved than the simple case in Eqs.~\eqref{GSymA}--\eqref{GSymB}, but is
of the general form discussed in subsection~{\it2.5}.

\subsection{The Picard--Fuchs equations}
The periods of $\W$ can also be determined as solutions of
certain Picard--Fuchs differential equations. These
equations can be derived from the defining polynomial $p$ by means of an
algebraic geometry construction~
\REF{\rCF}{A.~C.~Cadavid and S.~Ferrara, \plb{267} (1991) 193.}
\REF{\rBV}{B.~Blok and A.~Varchenko, Int. J. Mod. Phys. A{\bf 7},
      (1992) 1467.}
\REF{\rLSW}{W.~Lerche, D.~J.~Smit and N.~P.~Warner, \npb{372} (1992)
       87.}
\REF{\rCDFLL}{A.~Ceresole, R.~D'Auria, S.~Ferrara, W.~Lerche and
       J.~Louis, {\it Int. J. Mod. Phys.}{\bf A8} (1993) 79.}
\cite{{\rMorrison,\rCF-\rCDFLL}},
originally due to Griffiths~
\Ref{\Griffiths}{P.~Griffiths, Ann. Math. {\bf 90} (1969) 460.}.
We have analyzed particular examples of the Fermat
surfaces discussed above using the techniques explained in
Refs.~\cite{{\rCF,\rFont}}\
and found the linear system of partial differential equations
satisfied by the periods.

With the benefit of hindsight, we can describe how the Picard--Fuchs
equations follow from our fundamental period. For concreteness
consider $\CP4_{(1,1,2,2,6)}[12]$ and define
$$
  R(\j,\f) = {1\over \j} \vp_0(\j,\f)~. \eqlabel{XXX}
$$
$R(\j,\f)$ is given by a double series of the form
$$
  R(\j,\f) = \sum_{n,m=0}^{\infty} \ A_{n,m} \j^n \f^m~, \eqlabel{XXX}
$$
where the coefficients $A_{n,m}$ can be read off from \eqref{Uzero} and
\eqref{pismall}. The equations satisfied by $R(\j,\f)$ are equivalent to
recurrence relations verified by $A_{n,m}$.
 For instance, since $A_{n+q_0,m}/A_{n,m+1}$, with $q_0=6$, is a rational
function of $n$ and $m$, this relation derives from a partial differential
equation. Likewise, $A_{n,m+2}/A_{n,m}$ is a rational function and the
corresponding equation reflects the singularity at $\f^2=1$.
 The last equation is associated with the conifold type singularity
that exists at $(864\j^6 + \f)^2=1$.

\newpage
\section{More}{Other Important Generalizations}
In this section we show that the periods can be calculated even in the
cases where there is no known mirror. We show also that the fundamental period
can be obtained even when including non-polynomial deformations of the defining
polynomial.  In particular, in Ref.~\cite\rCfC\ it has been shown that some of
the parameters associated to the twisted sectors in \LG\ theories can
be represented by certain non-polynomial deformations.  We find however
that it is necessary to introduce
new contours associated to small values of the parameters.

\subsection{Manifolds with no known mirror}
 It was mentioned in Section~2 that the Ansatz~\eqref{ePOLY} for the form
of the reference polynomial does not cover all non-degenerate models and
that there are several classes which require more monomials than there are
coordinates. We now wish to illustrate how the above analysis applies to
such models.

Consider a model from the $\CP4_{(3,4,1,3,2)}[13]^{62,5}_{-114}$ class.
A little experimentation shows that there is no non-degenerate polynomial
consisting of five monomials in this class. This example therefore goes
beyond the lists of Ref.~\cite{\rBH}\ and in fact shows how to deal with
all the remaining models found in Refs.~\cite{{\rMaxSkI,\rMaxSkII}}.

In the case of $\CP4_{(3,4,1,3,2)}[13]$ it is enough to introduce an extra
monomial to obtain a transverse polynomial. We will calculate then the
period for the model
$$\eqalignno{
  P_0 &= ~x_0^3 x_1 + x_1^3 x_2 + x_2^{13} + x_2 x_3^4 + x_3 x_4^5~,
           \eqalignlabel{ePo}\cr
  P~ &= P_0 + ~\f x_1 x_3^3~ - ~\j x_0 x_1 x_2 x_3 x_4~,
           \eqalignlabel{ePoF}\cr}
$$
where, as explained, the reference polynomial $P_0$ is degenerate.
Proceeding as before, we obtain
$$ {\eqalign{
  \vp_0~ = ~\sum_{r,m=0}^\infty& {1\over\j^{45 r}}
  {\G(45r{+}15m{+}1) \over \G(15r{+}5m{+}1)\G(10r{+}2m{+}1)
                                    \G(2r{+}m{+}1)}     \cr
  &\times {1\over\G(9r{+}1)\G(9r{+}3m{+}1)}
           {(\f^4/\j^{15})^m\over (4m)!}~.    \cr}}
  \eqlabel{vpMore}
$$
Comparing with the general expression~\eqref{PerM}, we record (since the
index $\a$ takes only one value, we drop it):
$$
  \hat d=45~,\quad \hat k_i=15, 10, 2, 9, 9~,\quad p=4~,\quad
       s_i=5, 2, 1, 0, 3~,\quad t=15~, \eqlabel{NMparams}
$$
and introduce
$$
   \vf~ \define ~\f^4/\j^{15}~. \eqlabel{newphiNM}
$$
Note the coefficient of $r$ in $\G(45r{+}15m{+}1)$: 45 is indeed the degree
of the transpose,
$$
  \hat{P}_0~ = ~x_0^3 + x_0 x_1^3 + x_1 x_2^{13} x_3 + x_3^4 x_4 + x_4^5~,
  \eqlabel{eFoP}
$$
of the reference polynomial $P_0$ in~\eqref{ePo}. This transposed
polynomial~\eqref{eFoP} is also degenerate. Worse, there is no
 {\it polynomial} deformation which makes it transverse;
the entire family $\IP^4_{(15,10,2,9,9)}[45]$ consists of singular models!
The non-degeneracy of the reference polynomial (superpotential) is
important only from the Landau-Ginzburg point of view and, in fact, the
present study may well be useful in providing a sensible interpretation
to degenerate Landau-Ginzburg models.  We hope to return to this important
problem soon~
\Ref{\rCDK}{P.~Candelas, X.~C.~de la Ossa and S.~Katz, work in progress}.

Following the analysis in Section~2, from~\eqref{vpMore}
we write the Mellin--Barnes integral representation:
$$ {\eqalign{ \vp_0
 &= {-}{1\over2\p i} \int_\g\rd\r~ e^{i\p\r}\> \j^{{-}45\r}\>
            {\p\over\sin\p\r}\>\G(45\r{+}1)                    \cr
 &\mkern-60mu\hskip8mm\times \sum_{m=0}^\infty {(45\r{+}1)_{15m} \over
   \G(15\r{+}{5m}{+}1)\G(10\r{+}{2m}{+}1)\G(2\r{+}m{+}1)
   \G(9\r{+}1)\G(9\r{+}3m{+}1)} {\vf^m\over(4m)!}~,             \cr}}
   \eqlabel{MelBarNM}
$$
or
$$ {\eqalign{ \vp_0
 &= {-}{1\over2\p i} \int_\g\rd\r~ e^{i\p\r}\> \j^{{-}45\r}\>
            {\p\over\sin\p\r}\>\G(45\r{+}1)                    \cr
 &\hskip50mm\times {{\cal W}_\r(\vf) \over
   \G(15\r{+}1)\,\G(10\r{+}1)\,\G(2\r{+}1)\,\G^2(9\r{+}1)}~,   \cr}}
   \eqlabel{NMvp0Wr}
$$
where
$$
  {\cal W}_\r(\vf) \define
  \sum_{m=0}^\infty {(45\r{+}1)_{15m} \over
   (15\r{+}1)_{5m}\,(10\r{+}1)_{2m}\,(2\r{+}1)_m\,(9\r{+}1)_{3m}}\,
   {\vf^m\over4^{4m}({1\over4})_m({2\over4})_m({3\over4})_m\, m!}~.
   \eqlabel{WrNM}
$$
upon a little algebra, we obtain
$$ {\eqalign{ &{\cal W}_\r(\vf) =                                \cr
 &\mkern-40mu\hskip5mm {}_8F_7
  \left(\matrix{ 3\r{+}{1\over15}, 3\r{+}{2\over15},
                 3\r{+}{4\over15}, 3\r{+}{7\over15},
                 3\r{+}{8\over15}, 3\r{+}{11\over15},
                 3\r{+}{13\over15},3\r{+}{14\over15} \cropen{3pt}
                 5\r{+}{1\over2},~~ 5\r{+}1,~~ 2\r{+}1,~~ 3\r{+}1,~~
                {1\over4},~~ {1\over2},~~ {3\over4} \cr}\,;\>
                 \Big({3^{12}5^{10}\over2^2 4^4}\vf\Big)\right).\cr}}
   \eqlabel{WrNMghf}
$$
As discussed in Section~2, the convergence of the integral
in~\eqref{NMvp0Wr} depends on the asymptotic behaviour of ${\cal W}_\r$
for $|\r|\to\infty$ and ascertaining this is a difficult task~\cite{\rLuke}.
Instead, we assume that the relevant integrals converge for suitably small
values of $\vf$ and proceed with the evaluation of the
integral~\eqref{NMvp0Wr} by residues.

As usual, by closing the contour to the right, the (simple) poles of
$1/\sin\p\r$ at $\r=r=0,1,2,\ldots$ contribute and reproduce the
small-$\j$ expansion~\eqref{vpMore}.

Closing the contour to the left, the following poles are encircled:
 \item{1.} $\r={-}1,{-}2,{-}3,\ldots$~ from both $1/\sin\p\r$ and
           $\G(45\r{+}1)$,
 \item{2.} $\r={-}(R{+}{r\over45})$, $r=1,2,\ldots,44$ and
           $R=0,1,2,\ldots$ from $\G(45\r{+}1)$.

Now, in~\eqref{MelBarNM}, the zeros of $1/\G(9\r{+}1)$ include negative
integers, and since $s_4=0$, the position of these zeros does not shift
with growing $m$; think of these zeros as cancelling the poles from
$1/\sin\p\r$. It should also be clear from~\eqref{MelBarNM}, the
integral-$\r$ poles of $\G(45\r{+}1)$ are completely cancelled by the
combined zeros of
 $1/\G(10\r{+}2m{+}1)$ and of $(45\r{+}1)_{15m}$. Alternatively,
from~\eqref{vpMore}, the poles of $\G(45\r{+}15m{+}1)$ begin at
 $\r={-}(15m{+}1)/45$, while the zeros of $1/\G(10\r{+}2m{+}1)$ begin at
 $\r={-}(2m{+}1)/10$, whence the integral-$\r$ poles are all cancelled and
for all $m=0,1,\ldots$

The integral in~\eqref{NMvp0Wr}, with the contour closed to the left, may
thus be evaluated by summing the (simple pole) residues located at
$\r={-}(R{+}{r\over45})$, with $R=0,1,2,\ldots$ and $r=1,2,\ldots,44$.
Without further ado:
$$ {\eqalign{ \vp_0
 &= {\p\over45} \sum_{r=1}^{44} {e^{i\p r\over45}\over\sin({\p r\over45})}
 (-\j)^r \sum_{R=0}^\infty {(-\j)^{45R}\over\G(45R{+}r)}         \cr
 &\hskip18mm\times {{\cal W}_{{-}(R{+}{r\over45})}(\vf) \over
   \G(1{-}15R{-}{r\over3})\,\G(1{-}10R{-}{10r\over45})\,
   \G(1{-}2R{-}{2r\over45})\,\G(1{-}9R{-}{r\over5})}~,       \cr}}
   \eqlabel{smallNMvp}
$$
Note that we have {\it not} taken care explicitly of all cancellations,
hence many of the terms in this expansion actually vanish.

For a cursory search for other periods\Footnote{Eq.~\eqref{ePoF} does not
include the full complement of deformations and we cannot hope to recover
the $b_3=126$ periods.}, rewrite~\eqref{smallNMvp} more explicitly as
$$ {\eqalign{ \vp_0
 &= {\p\over45} \sum_{r=1}^{44} {e^{i\p r\over45}\over\sin({\p r\over45})}
    (-\j)^r \sum_{R=0}^\infty (-\j)^{45R} \sum_{m=0}^\infty
    {(-1)^{15m}\over\G(45R{+}r{-}15m)}                           \cr
 &\hskip5mm\mkern-60mu\times {\vf^m/(4m)! \over
   \G(1{+}5m{-}15R{-}{r\over3})\,\G(1{+}2m{-}10R{-}{10r\over45})\,
   \G(1{+}m{-}2R{-}{2r\over45})\,\G(1{+}5m{-}9R{-}{r\over5})}~.  \cr}}
   \eqlabel{sNMvp}
$$
Recall that $\vf=\f^4\j^{{-}15}$, so if $n$ is the exponent of $\j$, we
have $r=n+15m$. From the $\G$-functions in the denominator
in~\eqref{sNMvp} only the fractions provide non-trivial conditions and
$$ \threeqsalignno{
   r &= n+15m = 0\>({\rm mod}\,3)\quad
     &:~(\j,\f^4) &\mapsto (\l\j,\f^4)~,\quad
     &\l^3 &= 1~,                                \eqalignlabel{A} \cr
 10r &= 10n+15m = 0\>({\rm mod}\,45)\quad
     &:~(\j,\f^4) &\mapsto (\l^{10}\j,\l^{15}\f^4)~,\quad
     &\l^{45} &= 1~,                                              \cr
     &\quad
     &~\!\approx (\j,\f^4) &\mapsto (\l^2\j,\l^3\f^4)~,\quad
     &\l^9 &= 1~,                                \eqalignlabel{B} \cr
  2r &=2n+30m = 0\>({\rm mod}\,45)\quad
     &:~(\j,\f^4) &\mapsto (\l^2\j,\l^{30}\f^4)~,\quad
     &\l^{45} &= 1~,                             \eqalignlabel{C} \cr
   r &= n+15m = 0\>({\rm mod}\,5)\quad
     &:~(\j,\f^4) &\mapsto (\l\j,\f^4)~,\quad
     &\l^5 &= 1~.                                \eqalignlabel{D} \cr}
$$
It is easy to check that all of these actions may be generated
from
$$
  \vp_j(\j,\f^4)~ \define ~\vp_0(\l\j,\l^{15}\f^4)~,\qquad \l^{45}=1~.
$$
These 45 periods however satisfy the relation $\sum_{j=1}^{45}\vp_j=0$,
so that at most 44 of them are independent.

\subsection{Twisted moduli}
Recall the complete intersection space~\eqref{cicyR}
$$
  {\cal M} \in \Cnfg{\CP3\cr \CP2\cr}{3&1\cr 0&3\cr}^{35,8}_{-54}~~ :
 ~~\left\{ \twoeqsalign{
   p_{1,0}(x)   &= ~\sum_{i=0}^3 (x_i)^3~   & &= ~0~,  \cr
   p_{2,0}(x,y) &= ~\sum_{i=1}^3 x_i y_i^3~ & &= ~0~,  \cr}\right.
   \eqlabel{3103}
$$
for which we obtained the K\"ahler class fundamental period in
subsection~{\it3.3}. However, we obtained the period just as a function of
the two parameters corresponding to the K\"ahler classes of the two
projective spaces. In fact,
$b_{1,1}=8$ for $\cal M$ and so there are six more parameters.

It turns out that the additional six elements of $H^{1,1}({\cal M})$ may be
assigned explicit representatives, these however contain
square-roots~~\cite{\rCfC}. The six additional elements may be given as
`radical deformations' of the general form:
$$
   \tilde{M}_\a~~ \define ~~Q_\a(x_i)\sqrt{y_1 y_2 y_3}~,
   \eqlabel{Radical}
$$
where $Q_\a(x_i)$ are six suitable quadrics in the $x_i$. Such elements
cannot be added to either of the two defining equations in~\eqref{CICYR},
as they have bi-degree $(2,{3\over2})$.
However it is interesting that the complete intersection
space~\eqref{cicyR} has an alternative
description in terms of a single hypersurface, $\CP6_{(3,3,3,3,2,2,2)}[9]$,
which also has a Landau-Ginzburg model associated to it~\cite{\rGVW}. The
reference defining equation (superpotential) is then taken to be
$$
  P_0(x,y)~ = ~p_{1,0}(x) + p_{2,0}(x,y)~
            = ~x_0^3 + \sum_{i=1}^3 \big(\, x_i^3 + x_i y_i^3 \,\big)~.
  \eqlabel{CICYR}
$$
As the coordinates now have weights $k_x=3$ and $k_y=2$, the elements
in~\eqref{Radical} have the appropriate degree 9 and may appear as
deformations of the reference polynomial:
$$ \eqalign{
  P(x,y)~ &= ~x_0^3 + \sum_{i=1}^3 \big(\, x_i^3 + x_i y_i^3 \,\big)
           -\j_1\, x_1x_2x_3 -\j_2\, x_0y_1y_2y_3                   \cr
\noalign{\vglue-4mm}
          &\qquad~~~+ \sqrt{y_1 y_2 y_3}\, \sum_{i=1}^3 (\a_i x_i^2 +
\b_i x_{i+1}x_{i+2})         \cr}
  \eqlabel{CICYRdef}
$$
where the indices on $x_i$ are understood to be reduced modulo 3.

Now, for the complete intersection space~\eqref{cicyR}, it is natural to
realize the periods as integrals of the form
$$
  \int {\rd^4 x\> \rd^3 y \over p_1(x)\, p_2(x,y)}~,
  \eqlabel{XXX}
$$
where $p_1(x)$ and $p_2(x,y)$ were defined in Eq.~\eqref{cicyRdef}.
However, relative to the representation~\eqref{CICYRdef}, the periods are
most naturally
realised as
$$
  \int {\rd^4 x\> \rd^3 y \over \big(P(x,y)\big)^2 }~.
  \eqlabel{CICYRvp}
$$
The previous procedure of expanding the period in inverse powers of the
fundamental parameter is no longer applicable. The difficulty may be
stated in a number of different ways. One is that there is no fundamental
monomial for this case since the quantity $x_0x_1x_2x_3\,y_1y_2y_3$ does
not have the correct degree and so may not be added to $P$. Related to
this is the fact that attempts to define a fundamental contour in a manner
analogous to previous cases fail owing to the fact that the `fundamental
contour' so defined is trivial in homology. These facts seem to be true
generally for the generalised \cys. Fortunately we are able to proceed by
introducing contours analogous to those appropriate to small values of the
fundamental parameters.

We proceed now to calculate a period from \eqref{CICYRvp}.
To this end we take $x_0=1$ and let $\g_i$ be the contour in the
$y_i$-plane that connects the points $(e^{2\p i\over 3}\infty,0,\infty)$
and let $c_i$ be the contour in the $x_i$-plane that connects the points
$(-\infty + i\e, \infty + i\e)$.
Furthermore set $\G=\g_1{\times}\g_2{\times}\g_3$ and
$C=c_1{\times} c_2{\times} c_3$. The period~\eqref{CICYRvp} then becomes
$$
  \vp_0~\define~{1\over(2\p i)^6}\int_{\G\times C}
                                               {\rd^3x\,\rd^3y\over P^2}~.
  \eqlabel{Rpi}
$$
In order to perform the integrations we employ the identity
$$
  {1\over P^2}~ = ~-\int_0^\infty ds\, s\,e^{isP}        \eqlabel{std}
$$
and separate $P$ from~\eqref{CICYRdef} into a reference polynomial $p_0$
and a remainder
$${\eqalign{
  P(x,y)~ &= ~p + \j_1\, x_1x_2x_3 + \j_2\, y_1y_2y_3
 +  \sqrt{y_1 y_2 y_3}\, \sum_{i=1}^3 (\a_i x_i^2
+ \b_i x_{i+1} x_{i+2} ) \cr
p~ &= ~1 + \sum_{i=1}^3 (x_i^3 + x_i y_i^3)~.\cr}}
\eqlabel{REqs}
$$
On $\G\times C$, the quantity $y_i^3$ is real and positive and the
imaginary part of $x_i$ ensures the convergence of the integral
in~\eqref{std}\ and that of the $x$- and $y$-integrals when we
use~\eqref{std}\ in \eqref{Rpi}. To proceed we expand out the dependence
on the parameters. In order to avoid over-complicating the expressions
that follow it is useful to adopt a multi-index notation such that if
$x=(x_1,x_2,x_3)$ and $a=(a_1,a_2,a_3)$ then
$$ \matrix{
    x^a~ \define ~x_1^{a_1}\,x_2^{a_2}\,x_3^{a_3}~,\quad
    |a|~ \define ~a_1 + a_2 + a_3~,                    \cropen{3pt}
     a!~ \define ~(a_1)!\,(a_2)!\,(a_3)!~,             \cr}
\eqlabel{XXX}
$$
and so on. With this understanding we find the following expression for
$\vp_0$:
$$ {\eqalign{
  \vp_0 = {i\over (2\p i)^6} &\sum_{n,m,a,b}
          {\j_1^n\, \j_2^m\,\a^a\,\b^b \over n!\, m!\, a!\, b!}\,
  \int_0^\infty \rd s\,e^{is}\, (is)^{n+m+|a|+|b|+1}\,\times  \cropen{3pt}
& \int_C \rd^3x\,e^{is\sum_i x_i^3}\, x^{m+2a+|b|-b}
  \int_\G \rd^3y\,e^{is\sum_j x_jy_j^3}\, y^{n+\half |a| +\half |b|}~.\cr}}
  \eqlabel{XXX}
$$
The $x$- and $y$-integrals may be evaluated in virtue of the relations
$$
  \int_c \rd\x\,e^{is\x^3}\,\x^\s =
    -{ 2\p i\,e^{{\p i\over 2}(\s+1)}\over
       3\, s^{\s+1 \over3} \G\big({2-\s \over 3}\big)}~, \qquad
  \int_\g \rd\eta\, e^{is\x\eta^3}\,\eta^\r =
    -{ 2\p i\,e^{{\p i\over 2}(\r+1)}\over
       3\, (s\x)^{\r+1 \over 3} \G\big({2-\r \over 3}\big)}~.
  \eqlabel{InvG}
$$
After a little algebra we arrive at the expression:
$$
  \vp_0 = {1\over 3^6}\sum_{n,m,a,b}
  {\j_1^n\, \j_2^m\,\a^a\,\b^b \over n!\, m!\, a!\, b!}\,
   {\ddd e^{-{\p i\over 3}(n+1)}\,\G\Big({n+1\over 3}\Big)
   \over\ddd \G^3\left( {2{-}n{-}\half|a|{-}\half|b|\over 3} \right)
    \G_3\left( {2{-}m{-}2a{+}b{+}{1\over 6}|a|{-}{5\over 6}|b|{+}
                                       {n+1\over 3} \over 3}\right) }~.
  \eqlabel{Rvp}
$$
The suffix on $\G_3$ reminds us that, owing to the above multi-index
conventions, this term corresponds to a product of three $\G$-functions.
We remark that if we set the twisted parameters to zero we recover the
analytic continuation of the period defined by \eqref{cicyRvp}.

Clearly, there are many complete intersection spaces which do not have an
alternate description in terms of a single hypersurface and/or a
Landau-Ginzburg model, but this technique seems to present an important
extension of the general discussion in Section~2.

Indeed, such radicals also turn up in describing the complex structures.
Borrowing from Ref.~\cite{\rPDM}, we note that the complete intersection
space
$$
  {\cal M} \in \Cnfg{\CP4\cr\CP1\cr}{4&1\cr0&2\cr}^{86,2}_{-168}~~:
 ~~\left\{ \twoeqsalign{
   p_{1,0}(x)   &= \sum_{i=0}^4 x_i^4~    & &= 0~,  \cr
   p_{2,0}(x,y) &= x_0 y_0^2 + x_1 y_1^2~ & &= 0~,  \cr}\right.
  \eqlabel{4102}
$$
has $b_{1,1}=2$. The K\"ahler class periods can be obtained easily as
in Eqs.~\eqref{cicyRdef}--\eqref{cicyRvp}. The complex structure periods
however pose somewhat of a problem since only 56 of the 86 deformations
may be realised as polynomial deformations of the defining
equations~\eqref{4102}. The `missing' thirty deformations may however be
realised as `radical deformations' of the form~\cite{\rCfC}
$$
  \tilde{M}_\a(x,y)~ \define ~C_\a(x_i)\sqrt{x_0 x_1}~,
  \eqlabel{the30}
$$
where $\{C(x_i)\}$ are thirty suitable monomials in $x_i$. Note moreover
that this time the degree of these radical deformations is precisely right
to be added to the first of the defining equations~\eqref{4102}. In other
words, the defining equations~\eqref{4102} may be deformed in the manner
of Section~2, for example:
$$ {\eqalign{
  P_1(x)~  &=~\sum_{i=0}^4 x_i^4~ - ~\f_1 x_0 x_1 x_2 x_3~
            + ~\f_\a M_\a(x) + \f_{\a+56}C_\a(x_i)\sqrt{x_0 x_1}~, \cr
  P_2(x,y)~&=~x_0 y_0^2 + x_1 y_1^2~ - ~\f_2 x_4 y_0 y_1~
            + ~\f_\b M_\b(x,y)~, \cr }}
  \eqlabel{4102def}
$$
where $2<\a,\b\leq56$, so together with $x_0x_1x_2x_3$ and $x_4y_0y_1$,
$M_\a(x)$ and $M_\b(x,y)$ are the 56 conventional polynomial deformations.
Therefore, the 86 parameters $\j_1, \j_2, \f_\a$ then parametrize the
full 86-dimensional space of complex structures. The corresponding periods
may be realised as integrals of the type
$$
  \int {\rd^5 x\, \rd^2 y \over P_1(x)\, P_2(x,y)}~, \eqlabel{XXX}
$$
which may be calculated precisely as done in Sections~2 and~3.\bigskip

It seems that the radical complex structure deformations tend to be of the
correct degrees so they {\it can} be added as deformations of the defining
equations of
the original complete intersection model. By contrast, the radical
K\"ahler class variations tend to occur in such a way that they {\it
cannot} be added to the original defining equations of the complete
intersections but may be used as a deformation of the superpotential of
the corresponding Landau-Ginzburg orbifold, if such exists. In such cases,
this superpotential invariably corresponds to the defining equation of a
``generalized'' Calabi-Yau hypersurface, where~
\Ref\rolfII{R.~Schimmrigk: ``Critical Superstring Vacua from
Noncritical Manifolds: A Novel Framework'', Universit\"at Heidelberg report
HD-THEP-92-29.}
$$
   d~ \define ~\deg(P)~ = ~{1\over Q}\sum_{i=0}^N k_i~,\qquad
   Q=2,3~, \eqlabel{eCYGcond}
$$
holds instead of Eq.~\eqref{eCYcond}.

\subsection{A new look at the contours}
The purpose of this section is to repeat the calculation of
periods for quintics in \CP4, using contours that are
centered around the Fermat quintic.  Of course, these results are well
known but the idea is to illustrate the fact that these `new' contours are
the truly fundamental ones in the sense that there is generalization for
them, not only to the class of models embedded in weighted $\CP4$, but
to any \LG\ theory whose superpotential defines a hypersurface in
higher dimentional weighted projective spaces. We have already seen an
example for the importance of the existence of these contours for the
Schimmrigk manifold in subsection~{\it5.2}, where the `fundamental'
contour of Section~2 is trivial in homology.

Consider then, yet again, the model $\CP4[5]$.  In the affine coordinate
patch where $x_5=1$, the defining polynomial is
$$
  P~~ = ~~1 + \sum_{i=1}^4 x_i^{~5}~ - ~5\j x_1 x_2 x_3 x_4~.\eqlabel{XXX}
$$
We start by defining the periods as
$$ {{\eqalign{ \vp_j(\j)~
  &\define ~\j\Big({5\over2\p i}\Big)^4 \int_\G {\rd^4 x\over P}~,\cr
  &= ~\j\Big({5\over2\p i}\Big)^4 \int_\G \rd^4 x
         \int_0^\infty \rd s~ e^{-sP}~,\qquad \Re e(P)>0~, \cr}}}
  \eqlabel{XXX}
$$
Here, the poly-contours $\G_j$ must be chosen so that the convergence
criterion $\Re e(P)>0$ is satisfied. To this end, neglect temporarily the
$\j$-term and build the poly-contours $\G_j=\prod_{i=1}^4 \g_{ij}$ from
pieces where
$$
  x_i~ = ~\a^{k_i}\x_i~,\qquad \a~\define~e^{2\p i\over5}~,\quad
                                k_i\in\ZZ~,\quad 0\leq\x_i<\infty~,
  \eqlabel{XXX}
$$
so that $x_i^{~5}>0$, and the integrals converge.

Therefore
$$ {{\eqalign{ \vp_0(\j)~
  &= ~\j\Big({5\over2\p i}\Big)^4 \int_0^\infty\!\!\rd s\> e^{{-}s}
       \int_\G\!\!\rd^4 x\> e^{{-}s\sum_{i=1}^4 x_i^{~5}}~
        \sum_{n=0}^\infty {(5\j)^n\over n!}
                          s^n x_1^{~n} x_2^{~n} x_3^{~n} x_4^{~n}~, \cr
  &= ~\j\Big({5\over2\p i}\Big)^4 \sum_{n=0}^\infty {(5\j)^n\over n!}
       \int_0^\infty\!\!\rd s\,s^n\,e^{{-}s}
        \prod_{i=1}^4\int_{\g_{i,0}} \rd x_i\,x_i^{~n}\,e^{{-}s x_i^{~5}}
                                                                  ~.\cr}}}
  \eqlabel{XXX}
$$
The contours $\g_{i,0}$ are chosen to connect $(\a\infty, 0, \infty)$,
where $\a=e^{2\p i\over5}$, so that
$$
  \int_{\g_0}~~ = ~~\int_{\a\infty}^0 + \int_0^\infty~~
                = ~~\int_0^\infty - \int_0^{\a\infty}~. \eqlabel{XXX}
$$
Next, use
$$
  \int_0^{\a\infty}\!\!\rd\x\, \x^n\,e^{{-}s\x^{~5}}~
  = ~\a^{n+1}\int_0^\infty\!\!\rd\x\, \x^n\,e^{{-}s\x^{~5}}~
  = {\a^{n+1}\over5}{\G({n+1\over5})\over s^{n+1\over5}}~,
  \eqlabel{XXX}
$$
to obtain
$$ {{\eqalign{ \int_{\g_0}\!\!\rd\x\,\x^n\,e^{{-}s\x^{~5}}
  &= ~{1\over5}{\G({n+1\over5})\over s^{n+1\over5}}
       \Big(1-e^{{2\p i\over5}(n{+}1)}\Big)~,                       \cr
  &= ~{e^{i\p{(n{+}1)\over5}}\over5}{\G({n+1\over5})\over s^{n+1\over5}}
       \Big(-2i\sin{\ttt\big({n+1\over5}\p\big)}\Big)~,             \cr
  &= ~-{2\p i\over5}{e^{i\p{(n{+}1)\over5}}\over
                     s^{n+1\over5}\G\big(1-{n+1\over5}\big)}~.      \cr}}}
  \eqlabel{Int0}
$$

Thus,
$$ {{\eqalign{ \vp_0(\j)~
  &= ~\j\Big({5\over2\p i}\Big)^4 \sum_{n=0}^\infty {(5\j)^n\over n!}
       \int_0^\infty\!\!\rd s\,s^n\,e^{{-}s}
        \Big[-{2\p i\over5}{e^{i\p{(n{+}1)\over5}}\over
               s^{n+1\over5}\G\big(1-{n+1\over5}\big)}\Big]^4~,     \cr
  &= ~{1\over5}\sum_{n=0}^\infty {(5\j)^{n+1}\over\G(n+1)}
       {e^{{2\p i\over5}2(n{+}1)}\over\G^4\big(1{-}{n+1\over5}\big)}
       \int_0^\infty\!\!\rd s\,s^{{n+1\over5}-1}\,e^{{-}s}~,        \cr
  &= ~{1\over5}\sum_{m=1}^\infty {(5\a^2\j)^m\G({m\over5})
                           \over\G(m)\G^4\big(1{-}{m\over5}\big)}~, \cr}}}
  \eqlabel{vp0}
$$
which exactly recovers the well known result.

Now note that rotating the contour $\g_0$ in the integral~\eqref{Int0}
counterclockwise in units of $2\p\over5$, the result becomes
$$ \int_{\a^j\g_0}\!\!\rd\x\,\x^n\,e^{{-}s\x^{~5}}
  = ~-{2\p i\over5}{e^{i\p{(n{+}1)\over5}}\a^{j(n{+}1)}\over
                     s^{n+1\over5}\G\big(1-{n+1\over5}\big)}~.
  \eqlabel{Intj}
$$
Rotating only one of the four factor-contours in the poly-contour $\G$
then produces
$$ {{\eqalign{ \vp_j(\j)~
  &= ~\j\Big({5\over2\p i}\Big)^4 \sum_{n=0}^\infty {(5\j)^n\over n!}
       \int_0^\infty\!\!\rd s\,s^n\,e^{{-}s}~\a^{j(n{+}1)}
        \Big[-{2\p i\over5}{e^{i\p{(n{+}1)\over5}}\over
               s^{n+1\over5}\G\big(1-{n+1\over5}\big)}\Big]^4~,       \cr
  &= ~\vp_0(\a^j\j)~, \cr}}}
  \eqlabel{vpj}
$$
which are the same five periods found originally.

Notably, we have not derived anything new. However, the periods have been
now obtained by manifestly choosing different integration contours.
Although we have not proven {\it a priori} that these different
integration contours represent different cycles, this must be the case,
since four of the five periods in Eq.~\eqref{vpj} are linearly
independent, which is the correct count. Moreover, the relation
$\sum_{j=1}^5\vp_i(\j)=0$ acquires a convenient pictorial representation
in terms of adding the contours\Footnote{Of course, it is still important
that every fifth term in the summation in Eq.~\eqref{vpj} vanishes owing
to the poles of the $\G$-functions in the denominator.}.

\newpage
\section{concl}{Concluding Remarks}
We have described a method for computing the periods of Calabi-Yau models
which is applicable to the majority of currently known \cys\ and \LG\
models. We recall that the calculation of the periods provides the basis
for the explicit computation of the physically interesting quantities such
as the physically normalized Yukawa couplings of the effective low energy
theory and the threshold corrections to the gauge coupling
constants~
\Ref\rDixon{L.~Dixon, Talk Presented at the {\sl Mirror Symmetry
         Workshop}, MSRI (1991),\Z unpublished.}\
for $(2,2)$ models.
 Knowledge of the periods permits also the computation of the numbers of
instantons of the corresponding \cym~\cite{{\rCdGP--\rCdFKM}}. For
a recent advance concerning instantons of genus one, see Ref.~
\Ref\rBCOV{M.~Bershadsky, S.~Cecotti, H.~Ooguri and C.~Vafa,
        ``Holomorphic Anomalies in Topological Field Theories'', Harvard
         University preprint HUTP-93/A008.}.

Therefore, the present work brings us significantly closer to having a
complete control of the low energy effective action for any $(2,2)$
vacuum. It is illustrative to compare this with the conformal field theory
(CFT) techniques used in the case of free theories and coset spaces.
 Even though the CFT is well known in those cases, the calculation of the
effective couplings is always restricted to particular points
(Gepner--Kazama--Suzuki models) or a  subregion (orbifolds) of the whole
moduli space of the given $(2,2)$ model.
 Remarkably, the geometrical approach used here has now achieved the
same level of calculability as in the exact CFT's cases, even though the
underlying CFT's are known explicitly only for special subregions or
particular points of the moduli space. These achievements are due to the
methods developed during the last few years using topological and
algebro-geometrical techniques as well as special geometry and mirror
symmetry.
 In addition, this method has the advantage that the dependence of the
couplings on the whole moduli space is considered, whence the structure of
the full modular group can be obtained. An illustrative example to compare
both approaches is the mirror of the $Z_3$ orbifold~
\Ref\rCDP{P.~Candelas, E.~Derrick and L.~Parkes,
       ``Generalized Calabi-Yau Manifolds and the Mirror of a Rigid
        Manifold'', CERN preprint CERN-TH.6831/93 (1993).}.
In this case, the geometrical and the CFT methods
give identical results for the relevant quantities.
However, the explicit dependence of the couplings on the twisted moduli of
the corresponding orbifold model may now be obtained using geometrical
techniques---which goes beyond the common CFT practice.
It would be interesting to obtain a similar level of computability for
all couplings (including the $E_6$ {\bf1}'s) and eventually extend these
techniques to $(0,2)$ models, where the geometrical techniques are far
less well developed.

In principle the procedure presented here may be extended to other
cases that were not considered, such as complete intersection spaces in
products of weighted projective
spaces~
\Ref\rGRY{B.~Greene, S.-S.~Roan and S.-T.~Yau, \cmp{142} (1991) 245.}\
or more general Landau-Ginzburg models that do not have a standard
Calabi-Yau interpretation~\cite{\rCDP}.
 Also, the methods of Sections~2 and~3 cover only the dependence of the
K\"ahler class periods of \cicys\ on the parameters corresponding to the
K\"ahler classes of the embedding space or the mirror model. This appears
to be suitably generalised by the inclusion of non-polynomial deformations
and the alternate expansion method of Section~4. Another example of this
type concerns the mirror of the $Z_3$ orbifold~\cite{\rCDP}.\newpage

Finally, even though we can write the fundamental period for
very large classes of models, the explicit determination of
the full set of periods and the identification of the integral symplectic
basis requires an understanding of the singularity structure of each moduli
space and, at present, this has to be done model by model. Similar remarks
apply also to the determination of the modular group since this is also
largely determined by the singularity structure of the moduli space.
\vskip1in
\noindent{\bf Acknowledgments:}\ \
 % Help:
We are indebted to Paul Green and Sheldon Katz for discussions and
corrections.
 % Hospitality:
This work was supported by the American-Scandinavian Foundation, the
Fulbright Program, the NSF grants PHY 8904035 and PHY 9009850, the
Robert A.~Welch foundation and the  Swiss National Science Foundation.
P.B., X.D.\ and F.Q.\ would like to thank the Theory Division at CERN for
hospitality
and P.C.\ and X.D.\ wish to  acknowledge the
hospitality of the Institute for Theoretical Physics in Santa Barbara,
California, where some of this work was done.
A.F.\ thanks the Institut de Physique of the Universit\'e\:\ de\:\ Neuch\^atel
and the ICTP, Trieste for hospitality while part of this research was done as
well
as the Centro Cient{\'\ii}fico IBM-Venezuela for the use of its facilities.
\newpage
\immediate\closeout\referencewrite
                     \referenceopenfalse
                      \line{\bf\hfil References\hfil}\bigskip
                       \parindent=0pt\input referenc.texauxil
\bye